\documentclass[10pt]{article}
\pdfoutput=1
\usepackage{jheppub,amsmath,amssymb}
\usepackage{enumerate}
\usepackage{bm}
\usepackage{bbm}
\usepackage{enumitem}
\usepackage[usenames,dvipsnames]{xcolor}
\usepackage{amsmath}
\usepackage{amsfonts}
\usepackage{amssymb}
\usepackage{graphicx}
\usepackage{hhline}
\usepackage{subfig}
\usepackage{hyperref}
\usepackage{color}
\usepackage{verbatim}
\usepackage{amsmath,amsthm,amssymb}
\usepackage{lscape}
\usepackage{float}
\usepackage{caption}
\usepackage{lscape}
\usepackage{breqn}
\usepackage{multicol}
\usepackage{graphics}
\usepackage{tikz,todonotes}
\usepackage[utf8]{inputenc}
\usepackage{autobreak}
\usepackage{verbatim}
\usetikzlibrary{arrows,positioning,decorations.markings,decorations.pathmorphing,calc}
\pgfdeclarelayer{edgelayer}
\pgfdeclarelayer{nodelayer}
\usetikzlibrary{decorations.pathreplacing}
\pgfsetlayers{edgelayer,nodelayer,main}
\tikzset{none/.style={draw=none}}
\tikzset{new edge style 2/.style={black}}
\tikzset{new style 0/.style={black}}
\tikzset{rednode/.style={draw=none, scale=0.3pt,fill=red,circle, draw}}
\tikzset{redline/.style={line width=0.3mm,red}}
\tikzset{greyE/.style={line width=0.1mm,gray}}
\usepackage[utf8]{inputenc}
\usetikzlibrary{arrows,positioning,decorations.markings,decorations.pathmorphing,calc}

\definecolor{hyperref}{RGB}{026,028,087}

\newcommand{\beq}{\begin{equation}}
\newcommand{\eeq}{\end{equation}}
\newcommand{\bal}{\begin{aligned}}
\newcommand{\eal}{\end{aligned}}
\newcommand{\bea}{\begin{eqnarray}}
\newcommand{\eea}{\end{eqnarray}}
\def\be{\begin{equation}}
\def\ee{\end{equation}}

\def\beq{\begin{equation}}
\def\eeq{\end{equation}}

\newcommand{\mpl}{M_{\rm Pl}}

\newcommand{\K}{\mathcal K}

\renewcommand{\L}{\mathcal L}

\def\be{\begin{equation}}
\def\ee{\end{equation}}
\def\ba{\begin{eqnarray}}
\def\ea{\end{eqnarray}}

\def\d{\mathrm{d}}

\usepackage[normalem]{ulem}

\def\ba{\begin{eqnarray}}
\def\ea{\end{eqnarray}}

\def\L{\mathcal{L}}
\def\K{\mathcal{K}}
\def\X{\mathcal{X}}

\def\d{\mathrm{d}}
\def\mn{_{\mu \nu}}
\def\mnup{^{\mu \nu}}

\def\mupn{^\mu_{\phantom{\mu}\nu}}
\def\({\left(}
\def\){\right)}

\def\mpl{M_{\rm Pl}}
\def\p{\partial}

\begin{document}


\title{Cosmology of Extended Proca-Nuevo}

\author[a,b]{Claudia de Rham,}
\author[c,a]{Sebastian Garcia-Saenz,}
\author[d]{Lavinia Heisenberg,}
\author[a]{Victor Pozsgay}
\affiliation[a]{Theoretical Physics, Blackett Laboratory, Imperial College, London, SW7 2AZ, UK}
\affiliation[b]{CERCA, Department of Physics, Case Western Reserve University, 10900 Euclid Ave, Cleveland, OH 44106, USA}
\affiliation[c]{Department of Physics, Southern University of Science and Technology, Shenzhen 518055, China}
\affiliation[d]{Institute for Theoretical Physics, ETH Zurich, Wolfgang-Pauli-Strasse 27, 8093, Zurich, Switzerland}

\emailAdd{c.de-rham@imperial.ac.uk}
\emailAdd{s.garcia-saenz@imperial.ac.uk}
\emailAdd{lavinia.heisenberg@phys.ethz.ch}
\emailAdd{v.pozsgay19@imperial.ac.uk}

\abstract{Proca-Nuevo is a non-linear theory of a massive spin-1 field which enjoys a non-linearly realized constraint that distinguishes it among other generalized vector models. We show that the theory may be extended by the addition of operators of the Generalized Proca class without spoiling the primary constraint that is necessary for consistency, allowing to interpolate between Generalized Proca operators and Proca-Nuevo ones. The constraint is maintained on flat spacetime and on any fixed curved background.  Upon mixing extended Proca-Nuevo dynamically with gravity, we show that the constraint gets broken in a Planck scale suppressed way.
We further prove that the theory may be covariantized in models that allow for consistent and ghost-free cosmological solutions. We study the models in the presence of perfect fluid matter, and show that they describe the correct number of dynamical variables and derive their dispersion relations and stability criteria. We also exhibit, in a specific set-up, explicit hot Big Bang solutions featuring a late-time self-accelerating epoch, and which are such that all the stability and subluminality conditions are satisfied and where gravitational waves behave precisely as in General Relativity.}

\maketitle

\section{Introduction}
\label{sec:Intro}

Within the realm of charting consistent interacting effective field theories involving fields of different spins, the search for the most general theory of a self-interacting massive spin-1 is an interesting question that has enjoyed much progress over the past decade. With astrophysical and cosmological applications in mind, the embedding of these effective field theories in a fully gravitational framework is an exciting  problem connecting with  the ongoing program of classifying viable extensions of general relativity (GR). Similarly to their scalar-tensor counterparts, generalized vector-tensor theories have been shown to exhibit intriguing phenomenological properties in astrophysical systems \cite{Heisenberg:2017xda,Heisenberg:2017hwb,Kase:2017egk,Kase:2018owh,Kase:2020yhw,Garcia-Saenz:2021uyv,Brihaye:2021qvc} and cosmology \cite{Jimenez:2013qsa,DeFelice:2016yws,DeFelice:2016uil,Heisenberg:2016wtr,Nakamura:2017dnf,deFelice:2017paw,Nakamura:2018oyy,DeFelice:2020sdq,Heisenberg:2020xak,Garnica2021:2109.10154v1}. In the latter case, of particular interest, is the fact that a time-dependent vector condensate could behave as a dark energy fluid, driving the observed accelerated cosmic expansion in the present-day universe, with a technically natural vector mass and dark energy scale \cite{Heisenberg:2020jtr,deRham:2021mqq}.

An important milestone in this program was the discovery of the so-called Generalized Proca (GP) theory \cite{Tasinato:2014eka,Heisenberg:2014rta} (see \cite{Allys:2015sht,Allys:2016jaq,Jimenez:2016isa,Heisenberg:2018vsk} for related works). GP is an extension of the standard Proca theory of a spin-1 particle that includes self-interactions, with the virtue of maintaining the same constraint that renders the component $A_0$ of the field (in some frame) non-dynamical, thus ensuring the correct number of degrees of freedom at the non-linear level and, as a consequence, the absence of Ostrogradsky-type ghosts. It is worth emphasizing that the interactions of GP theory are non-trivial in that they are not simply constructed out of the field strength and the undifferentiated field, but includes derivative interactions that give rise to some unique properties, e.g.\ in relation to the screening mechanisms and the coupling to alternative theories of gravity \cite{DeFelice:2016cri,Heisenberg:2018acv,Garcia-Saenz:2021acj}.

While GP encompasses a broad class of models, there is no reason to expect it to be the unique non-linear completion of the free Proca theory. Indeed, vector-tensor theories that do not fall into the GP class have been found in \cite{Heisenberg:2016eld,Kimura:2016rzw}. There are two ways to see why GP is not necessarily the end of the story. The first is that the condition of having the desired number of degrees of freedom (i.e.\ three in four dimensions and ignoring gravity for the moment) only requires the existence of a constraint, which need not simply translate into the fact that a particular component of the field---$A_0$ in the case of GP---be non-dynamical. For instance, this constraint may be a non-linear functional of the vector field. The second insight into the question is provided by the decoupling limit of GP theory, in which the massive vector boson decomposes into massless spin-0 and spin-1 particles. A virtue of GP is that, in this limit, the equations of motion are second order, making the absence of extra unwanted degrees of freedom manifest. However, this feature of the equations of motion is sufficient but not necessary, since it is known that multi-field systems may in principle evade the Ostrogradsky theorem if the equations happen to be degenerate \cite{deRham:2016wji,Jimenez:2019hpl}.

Recently, an alternative extension of Proca theory, dubbed ``Proca-Nuevo'' (PN), was proposed in \cite{deRham:2020yet}. PN theory successfully exploits the above loopholes through a non-trivial realization of the primary constraint, motivated by the decoupling limit of massive gravity \cite{deRham:2010kj,Ondo:2013wka}. More in detail, if we denote by $V^{\mu}$ the vector spanning the null space of the Hessian matrix of time derivatives, then GP theory is characterized by having $V^{\mu}=\delta^{\mu}_0$, just like the linear theory, while in PN this vector is a non-linear function of the field itself. Crucially, this Hessian null eigenvector cannot be trivialized by performing a field redefinition and the two theories, GP and PN, are indeed dynamically inequivalent.

\begin{figure}
	\center{\includegraphics[width=6cm]{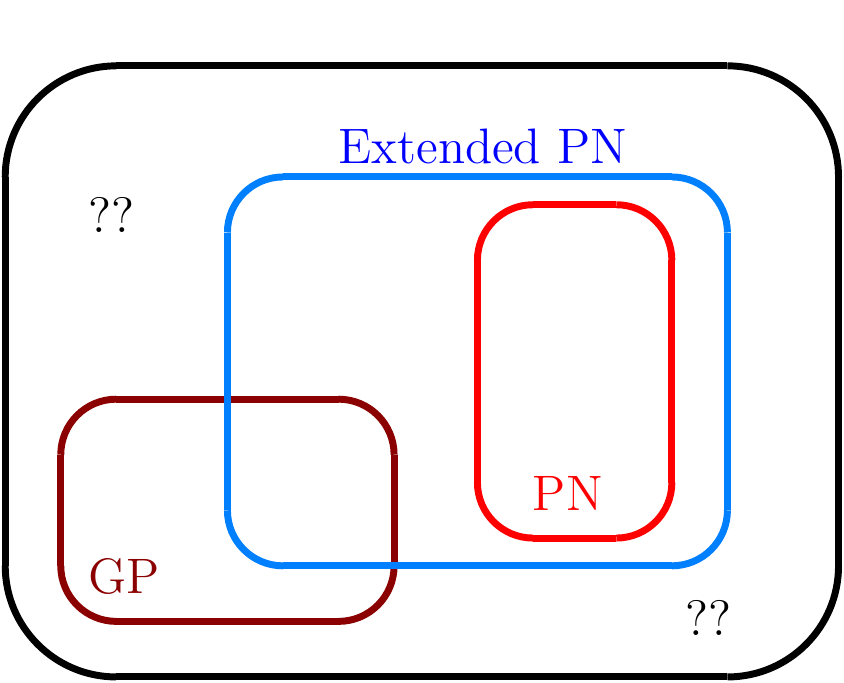}}
	\caption{Charting the space of massive spin-1 self-interacting theories that exhibit a constraint.}
	\label{fig:proca theory space}
\end{figure}

Given this inequivalence, it is natural to ask whether a still more general theory exists from which {\it both} GP and PN could arise as particular ``corners'' in the space of models, i.e.\ through particular choices (possibly in a limiting sense) of coupling constants. Exploring this question is the first aim of this paper. We shall show that such extension does exist, in a model that we imaginatively call ``Extended Proca-Nuevo''. While this proposal succeeds in furnishing a link between GP and PN, we should warn the reader of two caveats. First, in the ``GP limit'' of extended PN not all of the operators belonging to the GP class are obtained, although the whole PN class is included; this is represented artistically in Fig.\ \ref{fig:proca theory space}. Second, as it stands, extended PN is only complete when considered on a fixed background. While a full covariantization of the theory that maintains its constrained structure is still currently missing, as we shall see, since the breaking of the constraint is connected to the non-linear mixing between the gravitational degrees of freedom and the vector field, the ghost it implies always enters at a Planck-suppressed scale.

This last point regarding the coupling to gravity takes us to our second objective, namely to explore the cosmological implications of extended PN theory. Although not consistent in full generality, we will exhibit two alternative, partial covariantization schemes that successfully describe a massive spin-1 field coupled to Einstein gravity, with no additional degrees of freedom, for cosmological solutions at the levels of both the homogeneous and isotropic background and of general linear perturbations. Our main result is that, in each set-up, there exists a window of parameter values for which cosmological perturbations are free of ghost- and gradient-like instabilities and of superluminal propagation speeds. In particular, each scenario accommodates exactly luminal gravitational waves.

The first covariantization is particularly neat in that the coupling with gravity is minimal, unlike what occurs in GP theory. On the other hand, this model requires a technically-natural tuning of coefficients which has the advantage of providing a simple and tractable model with relatively few arbitrary functions. A particularly interesting property of this set-up is that, without any further tuning or special choices of coefficients, tensor fluctuations propagate exactly as in GR. As a consequence, observational bounds on the production and propagation of gravitational waves do not impose any extra constraints on the theory. After deriving the stability conditions for all types of perturbations---tensor, vector and scalar---for the model coupled to perfect fluid matter, we then analyze the resulting cosmological solutions. We will see that the model exhibits hot Big Bang solutions with epochs of radiation, matter and dark energy domination, with the latter corresponding to a ``self-accelerating'' phase, being driven by the vector field condensate and not a cosmological constant. We further show that perturbations within this model are fully under control, stable and causal.

The second covariantization is more general but requires non-minimal couplings between the vector field and the curvature. These non-minimal terms are precisely those of GP, so this model has the virtue of accommodating the covariant GP theory as a particular case, which is known to be free of pathologies for various choices of parameters. We will show however that this general set-up extends the cosmology of GP in interesting ways. For instance, we will prove that the dispersion relation of the Proca vector mode is non-linear, both in vacuum and when coupled to a perfect fluid. Similarly, the mixing of the perfect fluid with the extended PN sector results in a modification of the speed of propagation of the longitudinal fluctuation of the fluid, i.e.\ the phonon. As this effect is absent both in GR and in GP, it gives in principle a clean signature to test the theory and distinguish it from other vector-tensor models, although providing precise set-ups in which this signature would be detectable is beyond the scope of this work.

The rest of this paper is organized as follows, we begin in Sec.\ \ref{sec:ExtPN} with a brief review of the previously proposed PN theory before proceeding with its extension in flat spacetime, which includes most GP interactions as a subclass. The coupling to gravity is considered in Sec.\ \ref{sec:couplextPN}, where specific examples of covariantization are proposed, followed by the analysis of its constraint structure in the cases where the background spacetime is curved but non-dynamical and then when the metric is fully dynamical. We then present our analysis of cosmological solutions and perturbations for each of the covariantization schemes mentioned previously, first for the special case without non-minimal couplings in Sec.\ \ref{sec:SpecEx} followed by the general case in Sec.\  \ref{sec:fulltheory}. We summarize our main results in Sec.\ \ref{sec:discussion} and provide some final remarks. In order to ease the reading of the paper, we have omitted in the main text some technical derivations, which may be found in the Appendices.


\section{Extended Proca-Nuevo}
\label{sec:ExtPN}

\subsection{Review of Proca-Nuevo theory}
\label{ssec:RevPN}

We start with a vector field $A_\mu$ living on flat spacetime with Minkowski metric $\eta\mn$. The construction of PN theory follows the intuition drawn from the helicity decomposition of massive gravity \cite{deRham:2011qq} (see also \cite{deRham:2010kj,Ondo:2013wka}), beginning with the definition
\ba
	f\mn[A] = \eta\mn + 2 \frac{\p_{(\mu} A_{\nu)}}{\Lambda^2} + \frac{\p_{\mu} A^\rho \p_{\nu} A_\rho}{\Lambda^4} \,,
		\label{eq:deff2}
\ea
where $\Lambda$ is an energy scale that will ultimately control the strength of the vector self-interactions. Although reminiscent of the St\"uckelberg metric of massive gravity, we emphasize that we are ignoring gravity for the moment and therefore $f\mn$ here is simply a Lorentz tensor.

For later convenience we introduce
\ba
\label{eq:phia}
\phi^a=x^a+\frac{1}{\Lambda^2}A^a\,,
\ea
so that $f\mn$ may be written as
\ba
f\mn=\p_\mu \phi^a \p_\nu \phi^b \eta_{ab}\,.
\ea
The dependence of $\phi^a$ on the coordinates $x^a$ might naively suggest a breaking of Poincar\'e invariance. However, the quantity we shall use as a building block in the Lagrangian is $f\mn$, and this is manifestly a Poincar\'e-covariant object as is clear from \eqref{eq:deff2}.

Next we introduce the tensor $\K\mupn$ defined as \cite{deRham:2010kj,deRham:2014zqa}
\ba
\label{eq:defK1}
 \K\mupn &=& \X\mupn -\delta \mupn\\
 \text{with }	\quad \X\mupn[A] &=& \left( \sqrt{\eta^{-1}f[A]}  \right)\mupn \qquad \text{i.e.}\qquad \X^{\mu}_{\phantom{\mu} \alpha} \X^{\alpha}_{\phantom{\alpha} \nu} = \eta^{\mu \alpha}f_{\alpha \nu} \,.
		\label{eq:defK2}
\ea
In four dimensions, the PN theory for the vector field $A_\mu$ is then expressed as \cite{deRham:2020yet}
\begin{equation}
	\L_{\rm PN}[A] = \Lambda^4 \sum_{n=0}^4 \alpha_n(X) \L_n[\K]\,,
\label{eq:defLK}
\end{equation}
where the $n$th order PN term is defined by
\begin{equation}
	\L_n[\K] = - \frac{1}{(4-n)!} \epsilon^{\mu_1 \cdots \mu_n \mu_{n+1} \cdots \mu_4} \epsilon_{\nu_1 \cdots \nu_n \mu_{n+1} \cdots \mu_n} \K^{\nu_1}_{\phantom{\nu_1}\mu_1} \cdots \K^{\nu_n}_{\phantom{\nu_n}\mu_n}\,,
\label{eq:defLnK}
\end{equation}
or more explicitly
\begin{align}
	\L_0[\K] &= 1 \,, \label{eq:defL0K} \\
	\L_1[\K] &= [\K] \,, \label{eq:defL1K} \\
	\L_2[\K] &= [\K]^2 - [\K^2] \,, \label{eq:defL2K} \\
	\L_3[\K] &= [\K]^3 - 3[\K][\K^2] + 2[\K^3] \,, \label{eq:defL3K} \\
	\L_4[\K] &= [\K]^4 - 6[\K]^2[\K^2] + 3[\K^2]^2 + 8[\K][\K^3] - 6[\K^4] \label{eq:defL4K}\,,
\end{align}
and we use the notation $[\K] = \text{tr}(\K)$ for the trace. In \eqref{eq:defLK} the PN terms are multiplied by a set of coefficients $\alpha_n(X)$ which are arbitrary functions of
\beq
X=-\frac{1}{2\Lambda^2}\,A^{\mu}A_{\mu}\,.
\eeq
In our conventions, $X$, $\alpha_n$ and $\L_n$ are all dimensionless quantities. Note that $\L_0$ is just a constant, so that the product $\alpha_0(X)\L_0\equiv V(A^{\mu}A_{\mu})$ contains the standard potential of the vector field. In order for the trivial vacuum $\langle A_{\mu}\rangle=0$ to be a consistent state one should demand that $\alpha_0$ have a non-zero quadratic contribution, i.e.\ $\alpha_0\supseteq -\frac 12 (m^2/\Lambda^4) A^{\mu}A_{\mu}$.

\paragraph{Null Eigenvector.}

PN and GP are two inequivalent theories of a ghost-free massive vector field. It is natural to ask whether an extension of PN could be implemented in such a way that it would encompass both PN and GP. The two models realize the Proca constraint in very different ways, as can be seen at the level of the null eigenvector (NEV) of their respective Hessian matrices. As reviewed in the introduction, in GP theory the interactions are constructed in such a way that their NEV corresponds to the direction $(1, \vec{0})$, meaning that (in some frame) the component $A_0$ of the vector field is non-dynamical, just like in the linear theory. On the other hand, in PN theory the constraint is realized through a field-dependent NEV. Indeed, it was shown in \cite{deRham:2020yet} that the vector
\begin{equation} \label{eq:PN NEV}
	V_a^{\rm PN}(\Lambda) = (\mathcal{X}^{-1})^{0 \mu} \p_{\mu} \phi_a=(\mathcal{X}^{-1})^{0 \mu} \(\eta\mn+\frac{1}{\Lambda^2}\p_\nu A_\mu\) \,,
\end{equation}
is the non-perturbative normalized time-like NEV of the PN Lagrangian \eqref{eq:defLnK}, i.e.\ $V_a^{\rm PN}$ satisfies
\beq
\mathcal{H}^{ab} V^{\rm PN}_a = 0 \,,
\eeq
and $\eta^{ab}V^{\rm PN}_a V^{\rm PN}_b = -1$, with $\mathcal{H}^{ab}$ denoting the Hessian matrix of time derivatives,
\beq
\label{eq:Hessian}
\mathcal{H}^{ab}=\frac{\partial^2\L_{\rm PN}}{\partial\dot{A}_a\partial\dot{A}_b}\,.
\eeq

\subsection{Extended Proca-Nuevo theory}
\label{ssec:RevExtPN}

The PN model defined previously is special due to its link to massive gravity, but it is actually straightforward to include additional interactions within the same class. The main observation is that any operator that leaves the Hessian \eqref{eq:Hessian} fully invariant can be added to the theory without affecting the form of the NEV.
In 4D there exist precisely five operators  built out of the tensor $\partial^{\mu}A_{\nu}$ that respect that condition, namely the operators $d_n(X) \L_n[\p A]$, defined according to the rule in eq.\ \eqref{eq:defLnK}.

Although the $\L_n[\p A]$ operators are total derivatives and thus trivial in isolation, when added to the Lagrangian with a field-dependent coefficient, they produce non-trivial phenomenological effects, while being trivial at the level of the Hessian. These operators are precisely those that define the novel derivative interactions of GP theory (with the exception of $\L_4$, on which we will comment later). What we have uncovered here is that they may be added to the complete PN Lagrangian without thwarting the constraint structure. It is worth pointing out that the other members of the GP class, namely those that are not constructed solely in terms of elementary symmetric polynomials of $\partial^{\mu}A_{\nu}$, do in general contribute to the Hessian matrix and can therefore not trivially be added within this set-up.

We note that some redundancies are introduced through the construction we have just outlined: (i) $\L_0[\p A]$ is a constant, therefore its coefficient will contribute to the non-derivative potential and hence can be absorbed into $\alpha_0$; (ii) because of the identity
\begin{equation}
	\sum_{n=1}^4 \frac{\L_n[\K]}{n!} = \sum_{n=1}^4 \frac{\L_n[\p A]}{\Lambda^{2n} n!} \,,
\end{equation}
it follows that only three among the four remaining terms are linearly independent from the PN operators; (iii) moreover, it has been proved \cite{Jimenez:2019hpl} that $f(X) \L_4[\p A]$ is a total derivative for \textit{any} function $f$, therefore this term is always redundant. However, properties (ii) and (iii) hold only in flat spacetime, and are no longer true in a generic curved background upon replacing $\p A\to\nabla A$. Since our aim is to use this set-up as a starting point for building a covariant theory, we are thus led to consider all four GP terms $\L_1[\p A]$ through $\L_4[\p A]$.

With these considerations in mind, we now introduce the following Lagrangian,
\beq
	\L_{\text{EPN}} = \tilde{\Lambda}^4 \sum_{n=0}^4 \alpha_n(\tilde{X}) \L_n[\tilde{\K}[A]] + \Lambda^4 \sum_{n=1}^4 d_n(X) \frac{\L_n[\p A]}{\Lambda^{2n}} \,,
	\label{eq:LextPNflat}
\eeq
which we refer to as ``Extended Proca-Nuevo'' (EPN) theory. In four dimensions, it includes four additional arbitrary functions, $d_n(X)$, besides the original $\alpha_n$. Note that we have allowed for the possibility for the two families of operators to enter at different scales, namely at the scale $\Lambda$ and $\tilde \Lambda$, and denote as $\tilde \K$ and $\tilde X$ the same quantities as defined previously but suppressed with the scale $\tilde \Lambda$.

Obviously, the difference in scaling could be absorbed into the functions $\alpha_n$ (and we will do so later), but for now, we keep treating both scales independently as the relation between them can be used as a ``dialing'' parameter to interpolate between the respective PN and GP Lagrangians. In this sense, PN provides a perhaps unexpected link between the two previously known models.

This interpolation between PN and GP is seen most explicitly at the level of the NEV. Given that the additional  $d_n \L_n[\p A]$ operators do not affect the Hessian matrix, the NEV of EPN coincides with that of PN defined above in \eqref{eq:PN NEV}. Denoting as $V_a^{\rm EPN}(\tilde \Lambda, \Lambda)$ the NEV of EPN, it is straightforward to check that
\ba
V_a^{\rm EPN}(\tilde \Lambda, \Lambda)= V_a^{\rm PN}(\tilde \Lambda)\,,
\ea
where the NEV for PN is defined in \eqref{eq:PN NEV}. In the limit where $\tilde \Lambda \to \infty$, keeping the vector mass and the scale $\Lambda$ fixed, we have $\chi^\mu{}_\nu (\tilde \Lambda)\to \delta^\mu{}_\nu$, so it is clear that we recover the respective GP and PN null eigenvectors by taking the following limits for $\tilde \Lambda$
\begin{equation}
	\begin{cases}
		V_a^{\rm EPN}(\tilde \Lambda,\Lambda) \xrightarrow[\tilde \Lambda \rightarrow \infty]{} (1,\vec{0}) \,, \qquad &\text{(GP case)} \\
		V_a^{\rm EPN}(\tilde \Lambda,\Lambda) \xrightarrow[\tilde \Lambda \rightarrow \Lambda]{} V^{\rm PN}_a (\Lambda) \,, \qquad &\text{(PN case)}
	\end{cases}
\end{equation}
that is, the NEVs of GP and PN are obtained from the NEV of EPN in particular limits.

Let us emphasize that the GP limit is however non-trivial. Even though the limit $\tilde \Lambda\to \infty$ is well-defined and unambiguous at the level of the NEV, taking that limit at the level of the Lagrangian is on the other less trivial. Nevertheless, it is straightforward to see that one can indeed isolate all the $\L_n(\p A)$ GP operators via this procedure so long as the scale $\Lambda$ and the vector mass are both kept fixed in that limit. See Appendix \ref{sec:GPfromEPN} for details.

Having understood the relation with GP, we will focus on the full EPN theory for the remainder of the paper and without further loss of generality, we may set $\tilde{\Lambda}=\Lambda$.


\section{Coupling Extended Proca-Nuevo with gravity}
\label{sec:couplextPN}

As the theory of a Lorentz massive spin-1 field, the previous section naturally constructed the EPN theory on Minkowski, where the symmetries of the Lorentz and Poincar\'e groups make sense. However, in order to make contact with astrophysics and cosmology, we can also attempt to first extend the theory on arbitrary curved spacetime-dependent backgrounds and then further include the coupling with the gravitational dynamical degrees of freedom in the constraint analysis. We will address this by studying the existence of an NEV associated with the Hessian matrix of the covariantized version of the theory.

We will first prove that the suggested covariantized version of  EPN does possess an NEV on any arbitrary curved background metric no matter its spacetime dependence. However, this vector fails to be an eigenvector as soon as the metric is taken as a dynamical variable. By itself this simply indicates that the vector ought to be modified to include the non-trivial mixing with gravity, however, any modification of this vector would necessarily result in a non-vanishing eigenvalue and hence a breaking of the constraint. The presence of an additional degree of freedom is then inexorably linked to this loss of constraint, and standard analysis shows that when such additional degrees of freedom enter, they are always of ghostly nature. However since this loss of constraint is related to the mixing between the gravitational degrees of freedom and the Proca field, the resulting effects are Planck scale-suppressed. Moreover, we will see that the case of FLRW (Friedmann-Lema\^itre-Robertson-Walker) spacetime is special in that the additional degree of freedom is absent due to the isometries of the background so that the issue can be evaded at that level. This statement is strengthened by the fact that linear perturbations on cosmological backgrounds are free from any additional ghost degrees of freedom as will be shown explicitly.

\subsection{Covariant EPN theory}
\label{ssec:CovExtPN}

We define the covariant EPN theory by the action
\begin{equation}
	S = \int \d^4 x \sqrt{-g} \left(\frac{M_{\rm Pl}^2}{2}\,R + \L_{\rm EPN} + \L_M \right) \,,
	\label{eq:SextPN}
\end{equation}
where $R$ is the curvature scalar, $\L_M$ is the matter Lagrangian and
\begin{equation}
\label{eq:covEPN}
	\L_{\rm EPN} = - \frac14 \,F\mnup F\mn + {\Lambda}^4 \left( \L_0 + \L_1 + \L_2 + \L_3 \right) \,,
\end{equation}
with the definitions
\begin{align}
	 \L_0 &= \alpha_{0}(X) \label{eq:L0} \,,\\
	 \L_1 &=  \alpha_{1}(X) \L_1[\K] +  d_{1}(X) \frac{\L_1[\nabla A]}{\Lambda^2} \label{eq:L1} \,,\\
	 \L_2 &= \left( \alpha_2(X) + d_2(X) \right) \frac{R}{\Lambda^2} + \alpha_{2,X}(X) \L_2[\K] + d_{2,X}(X) \frac{\L_2[\nabla A]}{\Lambda^4} \label{eq:L2} \,,\\
	 \L_3 &= \left( \alpha_3(X) \K\mnup +d_3(X) \frac{\nabla^{\mu} A^{\nu}}{\Lambda^2} \right) \frac{G\mn}{\Lambda^2} - \frac16 \alpha_{3,X}(X) \L_3[\K] - \frac{1}{6} d_{3,X}(X) \frac{\L_3[\nabla A]}{\Lambda^6} \label{eq:L3} \,.
\end{align}
Here the subscript $X$ on the coefficient functions denotes differentiation w.r.t.\ $X$. While the Einstein-Hilbert term could be absorbed into the definition of $\alpha_2$ or $d_2$, we have chosen to write it independently in order to distinguish the Planck scale from the scale controlling the EPN interactions.

Some comments are in order regarding our definition of $\L_{\rm EPN}$. First, the Lagrangian includes non-minimal coupling terms, proportional to $R$ in $\L_2$ and to the Einstein tensor $G_{\mu\nu}$ in $\L_3$. These are motivated by the known non-minimal couplings of GP theory \cite{Heisenberg:2014rta}. Second, and also related to the question regarding non-minimal couplings, our Lagrangian omits the $\L_4$ term that was present in flat spacetime. As remarked before, this term does not belong to the GP class and neglecting this term  has the advantage of simplifying the analysis of cosmological perturbations, which is our main scope.

\subsection{EPN on an arbitrary background}
\label{ssec:ExtPNFix}

Before proceeding with the constraint analysis of the covariant EPN action defined in \eqref{eq:SextPN}, we make a brief digression here to point out that EPN admits a simpler covariantization in the case when the metric is non-dynamical (in the sense that we do not include the gravitational degrees of freedom in the counting of degrees of freedom), yet with a background spacetime that is arbitrarily curved. Indeed, in this situation, the minimal coupling prescription applied to the full flat-space Lagrangian, eq.\ \eqref{eq:LextPNflat}, is already enough to furnish a fully consistent theory. The proof is similar to that used in \cite{deRham:2020yet} to establish the consistency of PN theory in flat spacetime, i.e.\ through the explicit construction of the NEV of the Hessian matrix associated to the Lagrangian.

Unsurprisingly, this NEV is nothing but the minimal covariantization of the flat-space NEV. Starting with the vector field we define the auxiliary metric
\beq
f_{\mu\nu}=g\mn + 2 \frac{\nabla_{(\mu} A_{\nu)}}{\Lambda^2} + \frac{g^{\alpha \beta}\nabla_{\mu} A_\alpha \nabla_{\nu} A_\beta}{\Lambda^4} \,,
\eeq
where the covariant derivatives are taken with respect to the arbitrary metric $g\mn$ and the tensor
\beq
\X\mupn = \left( \sqrt{g^{-1}f}  \right)\mupn \,.
\eeq
The claim is that the vector
\beq
V_\mu = (\mathcal{X}^{-1})^{0}{}_{\alpha}\left(\delta^\alpha{}_\mu+\frac{g^{\alpha \beta}}{\Lambda^2}\,\nabla_{\beta}A_{\mu}\right) \,,
\label{eq:NEVcovariant}
\eeq
is the desired NEV. We will prove this here for the EPN term $\L_1$; the proof for the other terms can be found in Appendix \ref{ssec:NEVFix}.

As explained previously, the ``extended'' terms $\L_n[\nabla A]$ do not contribute to the Hessian matrix (again, in the absence of dynamical gravity), so it suffices to focus on $\L_1[\K]$. Actually, it is more convenient for the proof to consider instead $\L_1[\X]$, which entails no loss of generality given that the set $\L_n[\K]$ is linearly related to the set $\L_n[\X]$. Note that this statement is only true for the complete sets of operators $\L_n$, with $n$ going from $1$ to $4$. However, we prove in Appendix \ref{ssec:NEVFix} that the vector $V^{\mu}$ is the common NEV to each $\L_n[\X]$, including $\L_4[\X]$, and hence it is also the desired NEV for each $\L_n[\K]$.

We then define, for each $\L_n[\X]$, the associated canonical momentum conjugate to the vector field as
\begin{equation}
	p^{(n)}_{\alpha} = \Lambda^4 \frac{\p \L_n[\X]}{\p \dot{A}^{\alpha}} \,,
\end{equation}
and the corresponding Hessian matrix
\begin{equation}
	\mathcal{H}^{(n)}_{\alpha \beta} = \frac{\p p^{(n)}_{\alpha}}{\p \dot{A}^{\beta}} \,.
	\label{eq:HessianOrderN}
\end{equation}
For $\L_1[\X]$ we find $p^{(1)}_{\alpha} = {\Lambda}^{2} V_{\alpha}$, and therefore
\begin{equation}
	\mathcal{H}^{(1)}_{\alpha \beta} V^{\alpha} = {\Lambda}^2 \frac{\p V_{\alpha}}{\p \dot{A}^{\beta}} V^{\alpha} = \frac{{\Lambda}^2}{2} \frac{\p (V_{\alpha} V^{\alpha})}{\p \dot{A}^{\beta}} = 0 \,,
	\label{eq:HV1equals0}
\end{equation}
which follows because $V_{\alpha} V^{\alpha}=g^{00}$ is independent of the vector field velocity. A similar proof applies to the other $\L_n[\X]$ terms, see Appendix \ref{ssec:NEVFix} for details.

\subsection{Coupling with gravitational degrees of freedom}
\label{ssec:ExtPNDyn}

We now extend the previous analysis to accommodate a dynamical metric, in the sense where the dynamical degrees of freedom of the metric are included in the constraint analysis. While the degeneracy of the full Hessian matrix is preserved by the $\L_1$ EPN term upon minimal coupling to gravity, this property will be shown to fail for the other covariant EPN terms, $\L_2$ and $\L_3$, with or without the GP-inspired non-minimal couplings. Once again, because the GP-like contributions $\L_n[\nabla A]$ (with the appropriate non-minimal coupling terms) are known to be ghost-free, it suffices to focus on the PN terms, and without loss of generality we consider the set $\L_n[\X]$ instead of the set $\L_n[\K]$.

To proceed, we start by decomposing the metric in ADM variables,
\begin{equation}
	g_{00} = - N^2 + N^{i} N_i \,, \qquad g_{i0} = N_i \,, \qquad g_{ij} = \gamma_{ij} \,,
\end{equation}
where $N$ is the lapse, $N^i$ is the shift (defined with an upper index) and $\gamma_{ij}$ is the three-dimensional spatial metric, used to raise and lower indices on any spatial tensor. The vector field is parametrized by the time and spatial components of $A_{\mu}^{*}$, related to the original field via
\begin{equation}
	A_{\mu} = M_{\mu}^{\phantom{\mu} \nu} A_{\nu}^{\ast} \,,
\end{equation}
with
\begin{equation}
	M_{\mu}^{\phantom{\mu} \nu} \equiv
	\begin{pmatrix}
		N & N^i \\
		0 & \delta^i_j
	\end{pmatrix} \,.
\end{equation}
Even though $\dot{A}_{\mu}$ and $\dot{A}^*_{\mu}$ are not linearly related (because $M_{\mu}^{\phantom{\mu} \nu}$ is time-dependent), the corresponding conjugate momenta do satisfy a linear relation,
\beq
p^{\ast(n)}_{\mu} \equiv \Lambda^4 \frac{\p \L_n[\X]}{\p \dot{A}^{\ast\mu}} = M^{\nu}_{\phantom{\nu} \mu} p^{(n)}_{\nu} \,,
\eeq
and similarly for the Hessian submatrix
\beq
\mathcal{H}^{\ast(n)}_{\mu\nu} \equiv \frac{\p p^{\ast(n)}_{\mu}}{\p \dot{A}^{\ast\nu}} = M^{\rho}_{\phantom{\rho} \mu} M^{\sigma}_{\phantom{\sigma} \nu} \mathcal{H}^{(n)}_{\rho\sigma} \,.
\eeq

The full Hessian matrix of field velocities is now a $10\times 10$ matrix spanning the four components of the vector field $A^*_\mu$  and the six components of the spatial metric. In this analysis, we ignore the lapse and shift since it can be shown that no instance of $\dot{N}$ and $\dot{N}^i$ appears in the Lagrangian after performing the redefinition $A_\mu \mapsto A^*_\mu$ \cite{Langlois:2015skt}.

We claim that a natural candidate for the Hessian NEV of the covariant EPN theory is
\beq \label{eq:full EPN NEV}
\bm{\mathcal{V}} \equiv \left( V^*_\mu \,,\, 0\right) \,,
\eeq
where the null entry runs over the metric components and
\begin{equation}
	V^{\ast}_{\mu} \equiv \left( M^{-1} \right)_{\mu}^{\phantom{\mu} \nu} V_{\nu} \,.
	\label{eq:Vast}
\end{equation}
The vector $\bm{\mathcal{V}}$ indeed annihilates both the pure vector and pure metric subsectors. The latter property is trivial, while the former holds because
\beq
\mathcal{H}^{*(n)}_{\mu\nu}V^{*\nu} = M^{\rho}_{\phantom{\rho} \mu}\mathcal{H}^{(n)}_{\rho\sigma}V^{\sigma}=0 \,,
\eeq
where the last equality follows from the results of the previous subsection. Thus the outstanding question is whether $\bm{\mathcal{V}}$ annihilates the mixed vector-metric components of the Hessian.

It is easy to verify this for the $\L_1[\X]$ term. Defining
\beq
\mathcal{H}^{*(n)}_{\mu,ij} \equiv \frac{\p p^{*(n)}_{\mu}}{\p \dot{\gamma}^{ij}} \,,
\eeq
we have
\beq
\mathcal{H}^{*(n)}_{\mu,ij}V^{*\mu} = \frac{\p p^{(n)}_{\mu}}{\p \dot{\gamma}^{ij}}V^{\mu} \,,
\eeq
and in particular for $n=1$
\beq
\mathcal{H}^{*(1)}_{\mu,ij}V^{*\mu} = {\Lambda}^2 \frac{\p V_{\alpha}}{\p \dot{\gamma}^{ij}} V^{\alpha} = \frac{{\Lambda}^2}{2} \frac{\p (V_{\alpha} V^{\alpha})}{\p \dot{\gamma}^{ij}} = 0 \,.
\eeq
Therefore $\L_1[\X]$ defines a consistent ghost-free theory when coupled to dynamical gravity. Since $\bm{\mathcal{V}}$ is proven to be the null eigenvector for $\L_1[\X]$, we can directly infer that if $\bm{\mathcal{V}}$ fails to also be a null eigenvector for any  other $\L_n[\X]$, then irrespectively of what the appropriate eigenvectors would then be, it cannot be a null eigenvector for all $\L_n[\K]$ and thus the constraint will always be necessarily lost for generic theories given by \eqref{eq:covEPN}.
And indeed, as it turns out, when considering  $\L_2[\X]$ and $\L_3[\X]$, we can show that in the absence of any non-minimal couplings then $\mathcal{H}^{*(n)}_{\mu,ij}V^{*\mu}\neq0$ for $n=2,3$ (see Appendix \ref{ssec:NEVDyn} for the explicit expressions).

At this stage, this means that the covariant theory must necessarily include non-minimal couplings between the vector field and the curvature, unsurprisingly since we know this to be the case in the simpler GP theory. Our proposed covariant version of EPN theory, eq.\ \eqref{eq:SextPN}, contains the non-minimally coupled term
\beq \label{eq:L2 nonmin}
\L_2^{\rm(non-min)} = \alpha_{2,X}\,\L_2[\X]+\frac{\alpha_2}{\Lambda^2}\,R \,.
\eeq
As is shown in Appendix~\ref{ssec:NEVDyn}, while the vector $\bm{\mathcal{V}}$ fails to be a precise null eigenvector for the resulting Hessian matrix, our claim is that the addition of the curvature scalar operator allows us to consistently apply our model to cosmology.

The first virtue of the above choice \eqref{eq:L2 nonmin} of non-minimal coupling is that $\L_2^{\rm(non-min)}$ is indeed degenerate whenever the tensor $\nabla_{\mu}A_{\nu}$ is symmetric. For instance, this is the case for the cosmological backgrounds that we are interested in, namely the FLRW metric
\begin{equation}
	g_{\mu\nu}\d x^{\mu}\d x^{\nu} = -\d t^2 + a^2(t) \delta_{ij} \d x^i \d x^j \,,
	\label{eq:lineFLRW}
\end{equation}
and the vector field profile
\begin{equation} \label{eq:vector bkgd}
	A_{\mu}\d x^{\mu} = -\phi(t)\d t \,.
\end{equation}
In fact, for this background, the absence of additional degrees of freedom can be seen very directly by noting that\footnote{With some abuse of terminology, we will refer to the background defined by eqs.\ \eqref{eq:lineFLRW} and \eqref{eq:vector bkgd} as ``FLRW''.}
\begin{equation}
	\K^{\mu}_{\phantom{\mu}\nu} = \frac{1}{{\Lambda}^2}\,\nabla^{\mu}A_{\nu} \,,\quad  \text{on FLRW}\,.
\end{equation}
It follows that $\L_n[\K]=\L_n[\nabla A]/\Lambda^{2n}$ for this background, implying that the EPN theory actually reduces to a subclass of GP theory when restricted to FLRW, however the perturbations themselves differ quite significantly. Yet, we will confirm in Sec.\ \ref{sec:fulltheory} that the Lagrangian \eqref{eq:L2 nonmin} propagates the correct number of degrees of freedom also at the level of general linear perturbations about this background, where the equivalence between EPN and GP no longer holds. Although reassuring as an explicit check, let us emphasize that the presence of a constraint and absence of additional Ostrogradsky ghost was indeed expected given our proof that the NEV $\bm{\mathcal{V}}$ indeed annihilates the Hessian on the FLRW background.

We can also extend the derivations beyond cosmological backgrounds, and another virtue of the above choice of non-minimal coupling is that when expanded perturbatively in higher-dimensional operators, the matrix product $\bm{\mathcal{H}}^{(2)}_{\rm(non-min)}\bm{\mathcal{V}}$ correctly vanishes at leading order, but does not vanish when the two operators in \eqref{eq:L2 nonmin} are taken separately (see Appendix~\ref{ssec:NEVDyn}).
Since the constraint is present at leading order in an operator expansion and only gets broken at higher-order, it is in principle possible that the addition of new higher-order curvature terms could cancel the left-over, and so on in a perturbative fashion. Such precise constructions are however beyond the scope of this work and are kept to a future work.
Note however that the scale at which the vector $\bm{\mathcal{V}}$ ceases to be a null eigenvector is crucial.  The loss of constraint is related to the presence of operators that mix between the gravitational and vector degrees of freedom. Following the result presented in eq.~\eqref{eq:sumHcoupl}, and using the fact that at leading order the momentum is given by $p^*_\mu=\dot A^*_0 \delta^0_\mu$, we can infer that at the level of the Lagrangian, the loss of constraint is associated with an operator which behaves symbolically as
\ba
\L_{\rm ghost}\sim \frac{1}{\Lambda^2}\dot \gamma^{ij} \dot A_0^* F_{0i} A_j+\text{higher-dimensional operators}\,.
\ea
This term would be irrelevant if the gravitational degrees of freedom were not considered as dynamical, so including the gravitational tensor modes $h_T$ and the vector fluctuation $\delta A_0^*$,  this corresponds to a ghostly operator of the form
\ba
\L_{\rm ghost}\sim \frac{1}{\Lambda^2 \mpl}\dot h_{T} \delta\dot A_0^* \bar F_{0i} \bar A_j+\text{higher-dimensional operators}\,.
\ea
Remaining on the conservative side, this implies that a background configuration with vector vev $\bar A$ and field strength vev $\bar F$ would excite an additional ghost degree of freedom $\chi$, entering as $\dot A_0^*\sim \ddot \chi/\bar m$ at the symbolic cutoff scale
\ba
m_{\rm ghost}\sim \frac{\Lambda^2 \mpl \bar m}{\bar A_\perp \bar F}\,,
\ea
where $\bar m$ is the mass of the vector field on the background in question, and $\bar A_\perp$ is the dynamical part of the vector field. In particular on any background where the field strength tensor vanishes (i.e.\ where $\p_\mu \bar A_\nu$ is symmetric), we recover an absence of ghost, as is the case on the cosmological background we shall have in mind. Note that these considerations are meaningless on backgrounds where vector field happens to vanish $\bar m=0$ since the helicity-0 is then infinitely strongly coupled. On a background where $\bar A_\perp \sim \Lambda$, and $\bar F \sim \p \bar A_\perp \sim \bar m \Lambda$, the mass of the would-be ghost would be of order $\mpl$.

All the previous considerations also apply to the EPN term $\L_3$ and its associated non-minimal curvature coupling as given in \eqref{eq:SextPN}; details can be found in Appendix \ref{ssec:NEVDyn}. The upshot of this analysis is that our proposal for a covariantization of EPN theory, while not successful in complete generality, does indeed define a consistent cosmological model insofar as the number of degrees of freedom is concerned and as long as one is interested in linear perturbations about cosmological solutions defined by eqs.\ \eqref{eq:lineFLRW} and \eqref{eq:vector bkgd}.


\section{Special model without non-minimal couplings}
\label{sec:SpecEx}

We introduced in eq.~\eqref{eq:SextPN} what we have argued to be a natural first step in the covariantization of the flat-space EPN theory derived in Sec.\ \ref{sec:ExtPN}. We will refer to this Lagrangian as the ``general'' model, because it reproduces all the operators in \eqref{eq:LextPNflat} in the flat space limit (with the exception of $\L_4[\K]$, which we omit as previously explained). We will study this general model in the next section.

In the present section we focus instead on an alternative covariantization in which all non-minimal coupling terms are omitted. We recall that in our analysis of the Hessian matrix we found that the non-minimal couplings were in fact necessary for the NEV ansatz to succeed at leading order in a standard EFT operator expansion. While this statement is true generically, there remains the possibility that other covariantization schemes may exist when the theory is restricted by a specific choice of the coefficient functions $\alpha_n$ and $d_n$. In this section we show that this is indeed the case.

The ``special'' model we consider is defined by the action
\begin{equation}
	\hat{S} = \int \d^4 x \sqrt{-g} \left(\frac{M_{\rm Pl}^2}{2}\,R + \hat{\L}_{\rm EPN} + \L_M \right) \,,
	\label{eq:Shat}
\end{equation}
where we use a hat to distinguish it from the general model in eq.\ \eqref{eq:SextPN}. Here again $R$ is the curvature scalar, $\L_M$ is the matter Lagrangian, and the special EPN Lagrangian reads
\begin{equation}
\hat{\L} = - \frac14 \,F\mnup F\mn + \Lambda^4 \left( \hat{\L}_0 + \hat{\L}_1 + \hat{\L}_2 + \hat{\L}_3 \right) \,,
\end{equation}
where
\begin{align}
	 \hat{\L}_0 &= \alpha_{0}(X) \label{eq:L0example} \,,\\
	 \hat{\L}_1 &= \alpha_{1}(X) \L_1[\K] + d_{1}(X) \frac{\L_1[\nabla A]}{\Lambda^2} \label{eq:L1example} \,,\\
	 \hat{\L}_2 &= \alpha_{2,X}(X) \left(\L_2[\K] - \frac{\L_2[\nabla A]}{\Lambda^4}\right) \label{eq:L2example} \,,\\
	 \hat{\L}_3 &= - \frac16 \alpha_{3,X}(X) \left(\L_3[\K] - \frac{\L_3[\nabla A]}{\Lambda^6} \right)\label{eq:L3example} \,.
\end{align}

We remark that this Lagrangian can be formally obtained from that of the general model by setting $d_2=-\alpha_2$, $d_{2,X}=-\alpha_{2,X}$, $d_{3,X}=-\alpha_{3,X}$ and $\alpha_3=0=d_3$. But we emphasize that this is only a formal procedure, since the latter two conditions are obviously inconsistent as functional relations (except in the trivial case $\alpha_3(X)=0=d_3(X)$). With this choice of coefficients, the model has the advantage of being particularly simple, having no non-minimal couplings between the vector field and the metric and with comparatively few free coefficient functions. The precise constraint analysis for this special model is performed in details in Appendix~\ref{ssec:NEVspec}. It follows the precise same pattern as that discussed previously in the more general case in Section~\ref{sec:couplextPN}, and in particular the exact same conclusions as those of subsection~\ref{ssec:ExtPNDyn} hold here upon accounting for the dynamical mixing between the gravitational and vector degrees of freedom in this special example. Note in particular that this special model is free of ghost on cosmological backgrounds.

In the following subsections we derive the equations governing the dynamics of the FLRW background (defined by eqs.\ \eqref{eq:lineFLRW} and \eqref{eq:vector bkgd}) and of its linear perturbations. The matter sector is assumed to be a perfect fluid, although at this stage we do not specify its equation of state. As expected, our analysis recovers the correct number of propagating degrees of freedom on this cosmological background. In the last subsection we consider an admixture of pressureless matter and radiation, and then solve the background equations for a particular choice of the EPN coefficients. We further show that, for this particular choice, all the stability and subluminality conditions for the perturbations are satisfied. The proposed example thus provides a proof of principle that a heathy candidate for the Big Bang history of our Universe can be accommodated within EPN. This model does not rely on the presence of any cosmological constant, but rather with the presence of non-trivial Proca field self-interactions that enter at a technically natural scale \cite{deRham:2021mqq}.

\subsection{Background}
\label{ssec:SpecExbkgeom}

We proceed  by deriving the background cosmological equations of motion.
We focus on the FLRW metric
\begin{equation}
	\d s^2 = -N^2(t)\d t^2 + a^2(t) \delta_{ij} \d x^i \d x^j \,,
	\label{eq:lineFLRW2}
\end{equation}
with the vector field profile
\begin{equation} \label{eq:vector bkgd2}
	A_{\mu}\d x^{\mu} = -\phi(t)\d t \,.
\end{equation}
The equation obtained from varying the action with respect to the lapse yields the modified Friedmann equation
\begin{equation}
	H^2 = \frac{1}{3M_{\text{Pl}}^2} \left( \rho_M + \hat{\rho}_{\text{EPN}} \right) \,,
	\label{eq:Friedmannhat}
\end{equation}
where from now on we may set $N=1$ and where
\beq
	\hat{\rho}_{\text{EPN}} \equiv \Lambda^4 \left[ - \alpha_{0} + \alpha_{0,X} \frac{\phi^2}{\Lambda^2} + 3 \left( \alpha_{1,X} + d_{1,X} \right) \frac{H \phi^3}{\Lambda^4} \right] \label{eq:rhoExtPNhat} \,,
\eeq
is the effective energy density of the vector field. The Friedmann equation may be combined with the equation that follows from varying the action with respect to the scale factor $a(t)$ to produce the modified Raychaudhuri equation
\begin{equation}
	\frac{\ddot{a}}{a} = \dot{H} + H^2 = - \frac{1}{6 M_{\text{Pl}}^2} \left(\rho_M + \hat{\rho}_{\text{EPN}} + 3 P_M + 3 \hat{P}_{\text{EPN}}\right) \,,
\end{equation}
where
\begin{equation}
	\hat{P}_{\text{EPN}} \equiv \Lambda^4 \left[ \alpha_{0} - \left( \alpha_{1,X} + d_{1,X} \right)  \frac{\phi^2 \dot{\phi}}{\Lambda^4} \right] \,,
	\label{eq:PExtPNhat}
\end{equation}
is interpreted as the effective pressure of the vector condensate. Finally, variation with respect to $\phi(t)$ gives
\beq
	\alpha_{0,X} + 3 \left( \alpha_{1,X} + d_{1,X} \right) \frac{H \phi}{\Lambda^2} = 0 \,,
	\label{eq:eom3hat}
\eeq
which is however not independent of the other two as a consequence of the Bianchi identity. The fact that eq.\ \eqref{eq:eom3hat} is a constraint, enforcing an algebraic relation between $H$ and $\phi$, is no accident but follows from the precise form of the Lagrangian of the special model.

\subsection{Perturbations}
\label{ssec:SpecExPert}

\subsubsection{Definitions}
\label{sssec:SpecExIntroPert}

Next, we introduce perturbations about the FLRW background, following \cite{DeFelice:2016yws,DeFelice:2016uil}. Metric perturbations in the flat gauge are composed of two scalar modes $\alpha$ and $\chi$, one vector mode $V_i$ and one tensor mode $h_{ij}$. The line element reads
\begin{equation}
	g_{\mu\nu}\d x^{\mu}\d x^{\nu} = - \left(1+2 \frac{\alpha}{M_{\text{Pl}}} \right)\d t^2 + \frac{2}{M_{\text{Pl}}} \left(\frac{\p_i \chi}{M_{\text{Pl}}} + a V_i \right) \d t \d x^i + a^2(t) \(\delta_{ij} + \frac{h_{ij}}{\mpl}\) \d x^i \d x^j \,.
\end{equation}
The vector mode is transverse and the tensor mode is traceless and transverse, so that they each have two degrees of freedom. In here and what follows, spatial indices $i,j,\cdots$ are raised and lowered with respect to the spatial Euclidean metric $\delta_{ij}$.

The vector field $A_{\mu}$ is parametrized by two scalar perturbations $\delta \phi$ and $\chi_V$, together with a (transverse) vector mode $Z_i$. The perturbed vector field is then defined as
\beq\bal
A^0 &= \phi(t) + \delta \phi \,,\\
A^i &= \frac{1}{a^2} \delta^{ij}\left( aZ_j-\frac{a}{M_{\rm Pl}}\,\phi V_j +  \frac{\p_j \chi_V}{\Lambda} \right) \,.
\eal\eeq
The appearance of $V_i$ in the vector field perturbation may appear as unnecessary at this stage but will prove convenient later and prevent the need of additional field redefinitions.

For the perfect fluid matter we use the Schutz-Sorkin action \cite{SCHUTZ19771},
\begin{equation}
	S_M = - \int \d^4x \left[ \sqrt{-g} \, \rho_M(n) + J^{\mu} \left( \p_{\mu} l + \mathcal{A}_i \p_{\mu} \mathcal{B}^i \right) \right] \,.
	\label{eq:SMSS}
\end{equation}
Here
\begin{equation}
	n= \sqrt{\frac{J^{\mu} J_{\mu}}{g}} \,,
\end{equation}
is the fluid number density, whose background value is given by $\overline{n} = \mathcal{N}_0/a^3$, with $\mathcal{N}_0$ a constant. The current $J^{\mu}$ is decomposed as
\begin{align}
	J^0 &= \mathcal{N}_0 + M_{\text{Pl}}^2 \delta J \nonumber \,, \\
	J^i &= \frac{M_{\text{Pl}}}{a^2} \delta^{ik} \left( \p_k \delta j + M_{\text{Pl}} W_k \right) \,,
\end{align}
where $\delta J$ and $\delta j$ are scalars and $W_k$ is a transverse vector.

The scalar $l$ in \eqref{eq:SMSS} is such that on the background,
\begin{equation}
	\p_0 \overline{l} = - \rho_{M,n} \equiv - \frac{\p \rho_M}{\p n} \,,
\end{equation}
and we define its scalar perturbation $v$ by
\begin{equation}
	l = - \int^t \rho_{M,n} \d t' - \frac{\rho_{M,n} v}{M_{\text{Pl}}^2} \,,
\end{equation}
and note that on FLRW we have
\begin{equation}
	\rho_{M,n} = \frac{\rho_M + P_M}{\overline{n}} = a^3 \frac{\rho_M + P_M}{\mathcal{N}_0} \,.
	\label{eq:defrhoMn}
\end{equation}
Finally, the vectors $\mathcal{A}_i$ and $\mathcal{B}_i$ are also transverse. The canonical choice for their associated perturbations $\delta \mathcal{A}_i$ and $\delta \mathcal{B}_i$ reads
\begin{equation}
	\mathcal{A}_i = \frac{\delta \mathcal{A}_i}{M_{\text{Pl}}} \,, \qquad \mathcal{B}_i = M_{\text{Pl}} x_i + \frac{\delta \mathcal{B}_i}{M_{\text{Pl}}}  \,.
	\label{eq:dAidBi}
\end{equation}

For later use we note that the normalized $4$-velocity of the fluid can be found by varying the action \eqref{eq:SMSS} with respect to the current $J^{\mu}$, with the result
\begin{equation}
	u_{\mu} = \frac{J_{\mu}}{n \sqrt{-g}} = \frac{1}{\rho_{M,n}} \left( \p_{\mu} l + \mathcal{A}_i \p_{\mu} \mathcal{B}^i \right) \,,
	\label{eq:umu}
\end{equation}
and $u_i$ can be further split as
\begin{equation}
	u_i = - \frac{\p_i v}{M_{\text{Pl}}^2} + \frac{v_i}{M_{\text{Pl}}} \,,
	\label{eq:ui}
\end{equation}
where $v_i$ is transverse.

In the rest of this subsection we compute the quadratic part of the action for all perturbations, respectively for tensor, vector and scalar modes, and determine the conditions for every dynamical mode to be stable.

\subsubsection{Tensor perturbations}
\label{sssec:SpecExTens}

The quadratic action for the tensor perturbations is given by
\begin{equation}
	\hat{S}_T^{(2)} = \int \d^4 x \, a^3 \frac{1}{8} \left[ \dot{h}_{ij}^2 - \frac{1}{a^2} \(\p_k h_{ij}\)^2 \right] \,,
	\label{eq:SThat}
\end{equation}
We see that, in the special model, the EPN dynamics does not affect the quadratic action for the tensors, which are therefore entirely determined by the Einstein-Hilbert term. Thus not only is the tensor sector of \eqref{eq:Shat} free of instabilities, the speed of propagation of gravitational waves in this set-up is also exactly luminal. We find this to be a remarkable property considering the fact that the vector field, even though minimally coupled, still interacts non-trivially with the metric. Having the same dispersion relation as that of GR is of course also a phenomenological virtue given the recent experimental measurements on the speed of gravitational waves \cite{LIGOScientific:2017zic}.

\subsubsection{Vector perturbations}
\label{sssec:SpecExVec}

For the vector sector it is convenient to consider first the matter action \eqref{eq:SMSS}. Expanding to quadratic order in perturbations we find
\begin{equation}
	S_{M,V}^{(2)} = \int \d^4x \frac{1}{M_{\text{Pl}}^2} \left[ \frac{\rho_{M,n}}{2a^2 \mathcal{N}_0} \left( M_{\text{Pl}}^3 W_i + a \mathcal{N}_0 V_i \right)^2 - \frac12 a^3 \rho_M V^2 - \left( \mathcal{N}_0 \delta \dot{\mathcal{B}}_i + \frac{M_{\text{Pl}}^4}{a^2} W_i \right) \delta \mathcal{A}_i  \right] \,,
	\label{eq:SMV2}
\end{equation}
in agreement with \cite{DeFelice:2016uil}. We now proceed to eliminate the non-dynamical variables so as to identify the dynamical degrees of freedom. Varying \eqref{eq:SMV2} with respect to $W_i$ we have
\begin{equation}
	W_i = \frac{\mathcal{N}_0 \left( \delta \mathcal{A}_i - \rho_{M,n} a \frac{V_i}{M_{\text{Pl}}} \right)}{\rho_{M,n} M_{\text{Pl}}^2} \,.
\end{equation}
Plugging the definitions of $\delta \mathcal{A}_i$ and $\delta \mathcal{B}_i$ in eq.\ \eqref{eq:dAidBi} into eq.\ \eqref{eq:ui} we find
\begin{equation}
	\delta \mathcal{A}_i = \frac{\rho_{M,n} v_i}{M_{\text{Pl}}} \,,
	\label{eq:dAi1}
\end{equation}
and hence
\begin{equation}
	W_i = \frac{\mathcal{N}_0}{M_{\text{Pl}}^3} \left( v_i - a V_i \right) \,.
\end{equation}
Varying \eqref{eq:SMV2} with respect to $\delta \mathcal{A}^i$ we get
\begin{equation}
	v_i = a \left( V_i - a \frac{\delta \dot{\mathcal{B}}_i}{M_{\text{Pl}}} \right) \,.
	\label{eq:vi1}
\end{equation}
Combining these results we may integrate out $W_i$ and $\delta \mathcal{A}_i$ to obtain
\begin{equation}
	S_{M,V}^{(2)} = \int \d^4x \frac{a^3}{2} \frac{1}{M_{\text{Pl}}^2} \left[ (\rho_M + P_M) \left( V_i - a \frac{\delta \dot{\mathcal{B}}_i}{M_{\text{Pl}}} \right)^2 - \rho_M V_i^2 \right] \,.
	\label{eq:SMV2b}
\end{equation}
Collecting eq.\ \eqref{eq:SMV2b} with the expansion of the vector part of the EPN Lagrangian we arrive at
\begin{align}
	\hat{S}_V^{(2)} &= \int \d^4 x \frac{a^3}{2} \left[ \hat{q}_V \dot{Z}_i^2 - \frac{1}{a^2} \hat{\mathcal{C}}_1 (\p_i Z_j)^2 - H^2 \hat{\mathcal{C}}_2 Z_i^2 +\frac{1}{2a^2}(\p_i V_j)^2 + \frac{(\rho_M + P_M)}{M_{\text{Pl}}^2} \left( V_i - a \frac{\delta \dot{\mathcal{B}}_i}{M_{\text{Pl}}} \right)^2 \right] \,.
	\label{eq:SVhat2}
\end{align}
Note that to obtain this expression we made use of the background equations of motion. The coefficients appearing in \eqref{eq:SVhat2} are given by
\begin{align}
	\hat{q}_V &= 1 - \frac{1}{2 \left( 1 + \frac{\dot{\phi} + H \phi}{2 \Lambda^2} \right)} \left[ \alpha_1 - 2 \left( 1 - 2 \frac{H \phi}{\Lambda^2} \right) \alpha_{2,X} + \frac{H \phi}{\Lambda^2} \left( 2 - \frac{H \phi}{\Lambda^2} \right) \alpha_{3,X} \right] \label{eq:qVhat} \,,\\
	\hat{\mathcal{C}}_1 &= 1 - \frac{1}{2\left( 1 + \frac{H \phi}{\Lambda^2} \right)} \left[ \alpha_1 - 2 \left( 1 - \frac{H \phi + \dot{\phi}}{\Lambda^2} \right) \alpha_{2,X} + \left( \frac{H \phi}{\Lambda^2} + \left( 1 - \frac{H \phi}{\Lambda^2} \right) \frac{\dot{\phi}}{\Lambda^2} \right) \alpha_{3,X} \right] \label{eq:C1hat} \,, \\
	\hat{\mathcal{C}}_2 &= 2 \hat{q}_V + \frac{\p_t(\hat{q}_V H)}{H^2} + \frac{\dot{\phi}}{H^2} \left( \alpha_{1,X} + d_{1,X} \right) \label{eq:C2hat} \,.
\end{align}
The action \eqref{eq:SVhat2} describes two dynamical vector modes, since it is clear that $V_i$ is non-dynamical and may be integrated out (although the solution of its equation of motion involve non-linear instances of the 3-momentum). This integration could be performed formally but for what interests us here, namely the Proca vector mode $Z_i$, this degree of freedom is fully decoupled from $V_i$ and $\delta\mathcal{B}_i$, which are moreover independent of the parameters of the EPN model and thus evolve exactly as in GR.

Focusing then on the $Z_i$ mode, from \eqref{eq:SVhat2} we immediately infer the dispersion relation, with sound speed and effective mass being given by
\beq
\hat{c}_V^2 = \frac{\hat{\mathcal{C}}_1}{\hat{q}_V} \,,\qquad \hat{m}_V^2 = H^2 \frac{\hat{\mathcal{C}}_2}{\hat{q}_V} \,.
\eeq
Stability under ghosts and gradient modes then imposes the conditions
\beq
\hat{q}_V>0\,,\qquad \hat{\mathcal{C}}_1>0 \,.
\eeq
One may in addition ask for tachyon modes to be absent, which would then also require $\hat{\mathcal{C}}_2>0$.

\subsubsection{Scalar perturbations}
\label{sssec:SpecExScal}

Turning next to the scalar sector, we start again by expanding the matter action \eqref{eq:SMSS} to quadratic order. It proves useful to introduce
\begin{equation}
	\delta \rho_M \equiv \frac{\rho_{M,n}}{a^3} \delta J = \frac{\rho_M + P_M}{n_0 a^3} \delta J \,,
\end{equation}
in terms of which the scalar part of the action takes the form
\begin{align}
	S_{M,S}^{(2)} &= \int \d^4x \left[ M_{\text{Pl}}^2 \frac{\rho_{M,n}}{2a^5 \overline{n}} \left( \p_i \delta j + a^3 \frac{\overline{n}}{M_{\text{Pl}}^3} \p_i \chi \right)^2 + \frac{\rho_{M,n}}{a^2 M_{\text{Pl}}} \p_i \delta j \p_i v + a^3 \dot{v} \delta \rho_M - 3 \frac{a^3 \overline{n} \rho_{M,nn}}{\rho_{M,n}^2} H v \delta \rho_M \right. \nonumber \\
	& \quad \left. - \frac{a^3 M_{\text{Pl}}^4 \rho_{M,nn}}{2 \rho_{M,n}^2} \delta \rho_M^2 - a^3 M_{\text{Pl}} \alpha \delta \rho_M + \frac{a^3 \rho_M}{2 M_{\text{Pl}}^2} \left( \alpha^2 - \frac{(\p_i \chi)^2}{a^2 M_{\text{Pl}}^2} \right) \right] \,,
	\label{eq:SMS2}
\end{align}
where
\begin{equation}
	c_M^2 =  \frac{\overline{n} \rho_{M,nn}}{\rho_{M,n}} \,,
\end{equation}
is the squared sound speed of the fluid in pure GR. It also corresponds to the sound speed in GP theory and, as we will see, in the EPN special model, but not in the general model.

We may already integrate out at this stage the scalar mode $\delta j$. From its equation of motion we get
\begin{equation}
	\delta j = -a^3 \frac{\overline{n}}{M_{\text{Pl}}^3} \left( v + \chi \right) \,,
	\label{eq:eomdeltaj}
\end{equation}
and substituting into \eqref{eq:SMS2} furnishes
\begin{align}
	S_{M,S}^{(2)} &= \int \d^4x \, a^3 \left[ - \frac{\overline{n} \rho_{M,n}}{2M_{\text{Pl}}^4} \frac{(\p_i v)^2}{a^2} + \left( \frac{\overline{n} \rho_{M,n}}{M_{\text{Pl}}^4} \frac{\p^2 \chi}{a^2} - \delta \dot{\rho}_M - 3 H (1+ c_M^2) \delta \rho_M \right) v \right. \nonumber \\
	& \quad \left. - \frac{c_M^2 M_{\text{Pl}}^4}{2 \overline{n} \rho_{M,n}} (\delta \rho_M)^2 - M_{\text{Pl}} \alpha \delta \rho_M + \frac{\rho_M}{2 M_{\text{Pl}}^2} \left( \alpha^2 - \frac{(\p_i \chi)^2}{a^2 M_{\text{Pl}}^2} \right) \right] \,.
	\label{eq:SMS2b}
\end{align}
This result is to be combined with the expansion of the EPN Lagrangian. We eventually obtain (using again the background equations of motion)
\beq\bal \label{eq:SSnocoupl2}
	\hat{S}_S^{(2)} &= \int \d^4 x \, a^3 \left\lbrace  - \frac{\overline{n} \rho_{M,n}}{2M_{\text{Pl}}^4} \frac{(\p_i v)^2}{a^2} + \left[ \frac{\overline{n} \rho_{M,n}}{M_{\text{Pl}}^4} \frac{\p^2 \chi}{a^2} - \delta \dot{\rho}_M - 3 H (1+ c_M^2) \delta \rho_M \right] v   \right.  \\
	& \quad  - \frac{c_M^2 M_{\text{Pl}}^4}{2 \overline{n} \rho_{M,n}} (\delta \rho_M)^2 - M_{\text{Pl}}\, \alpha\, \delta \rho_M - \hat{\omega}_3 \frac{(\p_i \alpha)^2}{a^2 M_{\text{Pl}}^2} + \hat{\omega}_4 \frac{\alpha^2}{M_{\text{Pl}}^2}   \\
	& \quad - \left[ \left(3 H \hat{\omega}_1 - 2 \hat{\omega}_4 \right) \frac{\delta \phi}{\phi} - \hat{\omega}_3 \frac{\p^2 (\delta \phi)}{a^2 \phi} - \hat{\omega}_3 \frac{\p^2 \dot{\psi}}{a^2 \phi \Lambda} + \hat{\omega}_6 \frac{\p^2 \psi}{a^2 \Lambda} \right] \frac{\alpha}{M_{\text{Pl}}}   \\
	& \quad - \frac{\hat{\omega}_3}{4} \frac{(\p_i \delta \phi)^2}{a^2 \phi^2} + \hat{\omega}_5 \frac{(\delta \phi)^2}{\phi^2} - \frac12 \left[ \left( \hat{\omega}_2 + \hat{\omega}_6 \phi \right) \psi - \hat{\omega}_3 \dot{\psi} \right] \frac{\p^2 (\delta \phi)}{a^2 \phi^2 \Lambda}  \\
	& \quad \left. - \frac{\hat{\omega}_3}{4 \phi^2} \frac{(\p_i \dot{\psi})^2}{a^2 \Lambda^2} + \frac{\hat{\omega}_7}{2} \frac{(\p_i \psi)^2}{a^2 \Lambda^2} + \left( \hat{\omega}_1 \frac{\alpha}{M_{\text{Pl}}} + \hat{\omega}_2 \frac{\delta \phi}{\phi} \right) \frac{\p^2 \chi}{a^2 M_{\text{Pl}}^2} \right\rbrace   \,,\\
\eal\eeq
for the complete quadratic action of scalar perturbations in the special model. Here we introduced
\begin{equation}
	\psi \equiv \chi_V + \frac{\Lambda}{M_{\text{Pl}}^2} \phi \chi \,,
\end{equation}
and the (time-dependent) coefficients $\hat{\omega}_I$ are given in Appendix \ref{ssec:DefCoefsExScal}.

We observe that the action \eqref{eq:SSnocoupl2} has the same structure as the quadratic scalar action derived in GP theory \cite{DeFelice:2016uil}, only with different $\hat{\omega}_I$ coefficients. We emphasize that this is a non-trivial result since the special model \eqref{eq:Shat} is manifestly not of the GP class. Indeed, if one were to ``detune'' the operators in $\hat{\L}_2$ and $\hat{\L}_3$ in \eqref{eq:Shat} then additional operators would appear in \eqref{eq:SSnocoupl2}. These extra operators modify the degree of degeneracy of the equations of motion and, as a consequence, additional degrees of freedom become active.

To see that the action \eqref{eq:SSnocoupl2} propagates two dynamical modes one can simply analyse the resulting equations of motion,
\begin{align}
	&(3 H \hat{\omega}_1 -2\hat{\omega}_4) \frac{\delta \phi}{\phi} -2\hat{\omega}_4 \frac{\alpha}{M_{\text{Pl}}} + M_{\text{Pl}}^2 \delta \rho_M + \frac{k^2}{a^2 \Lambda^2} \left[ \hat{\mathcal{Y}} + \hat{\omega}_1 \frac{\Lambda^2}{M_{\text{Pl}}^2} \chi - \hat{\omega}_6 \Lambda \psi \right] = 0 \label{eq:SpecExeomalpha} \,,\\
	&\frac{(\rho_M + P_M)}{M_{\text{Pl}}} v + \hat{\omega}_1 \alpha + M_{\text{Pl}} \hat{\omega}_2 \frac{\delta \phi}{\phi} = 0 \label{eq:SpecExeomchi} \,,\\
	&(3 H \hat{\omega}_1 - 2 \hat{\omega}_4 )\frac{\alpha}{M_{\text{Pl}}} - 2 \hat{\omega}_5 \frac{\delta \phi}{\phi} + \frac{k^2}{a^2 \Lambda^2} \left[ \frac{1}{2}\hat{\mathcal{Y}} + \hat{\omega}_2 \frac{\Lambda^2}{M_{\text{Pl}}^2} \chi - \frac{\Lambda}{2} ( \hat{\omega}_2 + \hat{\omega}_6 \phi ) \frac{\psi}{\phi} \right] = 0 \label{eq:SpecExeomdphi} \,,\\
	&\frac{\dot{\hat{\mathcal{Y}}}}{H} + \left( 1 - \frac{\dot{\phi}}{H \phi} \right) \hat{\mathcal{Y}} + \frac{\Lambda^2}{H} \left\lbrace \hat{\omega}_2 \frac{\delta \phi}{\phi} + 2 \hat{\omega}_7 \frac{\phi \psi}{\Lambda} + \hat{\omega}_6 \left( 2 \frac{\alpha \phi}{M_{\text{Pl}}} + \delta \phi \right) \right\rbrace = 0 \label{eq:SpecExeomppsi} \,,\\
	&\dot{\delta \rho}_M + 3 H (1 + c_M^2) \delta \rho_M + \frac{k^2}{a^2} \frac{(\rho_M + P_M)}{M_{\text{Pl}}^4} ( v+ \chi) = 0 \label{eq:SpecExeomv} \,,\\
	&\alpha M_{\text{Pl}} + c_M^2 \left( 3 H v + \frac{M_{\text{Pl}}^4}{(\rho_M + P_M)} \delta \rho_M \right) - \dot{v} = 0 \label{eq:SpecExeomdrhoM} \,,
\end{align}
respectively for $\alpha$, $\chi$, $\delta \phi$, $\psi$, $v$ and $\delta \rho_M$, and we defined
\begin{equation}
	\hat{\mathcal{Y}} \equiv \frac{\Lambda^2}{\phi} \hat{\omega}_3 \left(  \delta \phi + 2 \frac{\alpha \phi}{M_{\text{Pl}}} + \frac{\dot{\psi}}{\Lambda} \right) \label{eq:SpecExY} \,.
\end{equation}
It is straightforward to show that these equations can be solved algebraically for $\alpha$, $\delta \phi$, $\chi$ and $v$ in order to be left with a system of two second-order differential equations for $\psi$ and $\delta \rho_M$. This completes the proof that the special EPN model exhibits the correct number of degrees of freedom in the tensor, vector and scalar sectors.

To study the stability of the propagating scalar modes we integrate out all the non-dynamical variables. The resulting action is formidably lengthy, but for the purpose of deciding whether the fields exhibit ghost- or gradient-type instabilities it suffices to focus on the short wavelength limit $k \rightarrow \infty$. In this regime, the action can be recast in the form
\begin{equation}
	\hat{S}_S^{(2)} = \int \d^4x \, a^3 \left[ \dot{\vec{\Omega}}^t \hat{\bm{K}} \dot{\vec{\Omega}} - \vec{\Omega}^t \left( \hat{\bm{M}} - \frac{k^2}{a^2} \hat{\bm{G}} \right) \vec{\Omega} - \vec{\Omega}^t \hat{\bm{B}} \dot{\vec{\Omega}} \right] \,,
	\label{eq:SpecExSMmatrix}
\end{equation}
where $\vec{\Omega}^t \equiv \left( \psi, \delta \rho_M / k \right)$ (note that $\delta \rho_M$ has mass dimension 2, hence the rescaling by $k$). For brevity we omit the explicit expressions for the matrices $\hat{\bm{K}}, \hat{\bm{M}}, \hat{\bm{G}}$ and $\hat{\bm{B}}$, but let us remark that they are independent of $k$ (again in the limit $k \rightarrow \infty$).

Absence of ghosts requires the kinetic matrix $\hat{\bm{K}}$ to be positive definite. Thanks to the special form of the Schutz-Sorkin action it turns out that $\hat{\bm{K}}$ is diagonal \cite{Kase:2014cwa, Kase:2014yya}, and we find
\begin{equation}
\hat{Q}_{S,\psi} = \frac{M_{\text{Pl}}^2 H^2}{\Lambda^2 \phi^2} \frac{3 \hat{\omega}_1^2 + 4 M_{\text{Pl}}^2 \hat{\omega}_4}{(\omega_1 - 2 \omega_2)^2} \,,\qquad \hat{Q}_{S,M} = \frac{a^2}{2} \frac{M_{\text{Pl}}^4}{(\rho_M + P_M)} \,,
\end{equation}
for the eigenvalues associated to $\psi$ and $\delta\rho_M$, respectively. Positivity of $\hat{Q}_{S,M}$ requires $\rho_M + P_M > 0$, i.e.\ the (strict) null energy condition, while the condition $\hat{Q}_{S,\psi}>0$ is equivalent to
\begin{equation}
	3 \hat{\omega}_1^2 + 4 M_{\text{Pl}}^2 \hat{\omega}_4 > 0 \,.
\end{equation}

Absence of gradient-unstable modes requires the sound speeds square to be positive. From the dispersion relation,
\begin{equation}
	\text{det} \left[ \hat{\omega}^2 \hat{\bm{K}} - \left( \hat{\bm{M}} - \frac{k^2}{a^2} \hat{\bm{G}} \right) \right] = 0 \,,
	\label{eq:SpecExdispSM}
\end{equation}
we obtain that the fluid propagates with speed $c_M^2$, as previously claimed, while the Proca scalar mode $\psi$ has
\begin{equation}
	\hat{c}_{S,\psi}^2 = \frac{1}{M_{\text{Pl}}^2 H^2 \phi^2} \frac{\hat{\Gamma}}{8 \hat{q}_V (3 \hat{\omega}_1^2 + 4 M_{\text{Pl}}^2 \hat{\omega}_4)} \,,
	\label{eq:cSpsihat}
\end{equation}
where
\begin{align}
	\hat{\Gamma} &\equiv 2 \hat{\omega}_2^2 \hat{\omega}_3 ( \rho_M + P_M ) - \hat{\omega}_3 ( \hat{\omega}_1 - 2 \hat{\omega}_2) ( \hat{\omega}_1 \hat{\omega}_2 + \phi ( \hat{\omega}_1 - 2 \hat{\omega}_2 ) \hat{\omega}_6 ) \left( \frac{\dot{\phi}}{\phi}- H \right) - \hat{\omega}_3 ( 2 \hat{\omega}_2^2 \dot{\hat{\omega}}_1 - \hat{\omega}_1^2 \dot{\hat{\omega}}_2 ) \nonumber \\
	&\quad + \phi ( \hat{\omega}_1 - 2 \hat{\omega}_2)^2 ( \hat{\omega}_3 \dot{\hat{\omega}}_6 + \phi (2 \hat{\omega}_3 \hat{\omega}_7 + \hat{\omega}_6^2)) + \hat{\omega}_1 \hat{\omega}_2 \left( \hat{\omega}_1 \hat{\omega}_2 + (\hat{\omega}_1 - 2 \hat{\omega}_2) \left( 2 \phi\hat{\omega}_6 - \hat{\omega}_3 \frac{\dot{\phi}}{\phi} \right) \right) \,.
\end{align}
Note that $\hat{q}_V>0$ is already required by the stability of vector perturbations (cf.\ eq.\ \eqref{eq:qVhat}), while $3 \hat{\omega}_1^2 + 4 M_{\text{Pl}}^2 \hat{\omega}_4 > 0$ from the above no-ghost condition. It therefore suffices to impose $\hat{\Gamma} > 0$ for Laplacian instabilities to be absent.

While tachyonic instabilities are less concerning---on the contrary, they are potentially interesting---later we will also examine the effective masses of the scalar modes. The expressions are somewhat lengthy and so we leave them for Sec.\ \ref{ssec:MassesExScal} in the Appendix.

Having derived the stability conditions for all the dynamical modes, the outstanding question is whether there exist choices of parameters of the special model Lagrangian such that all the criteria are satisfied while providing a consistent cosmological history. In the next subsection we show that this is the case.

\subsection{Cosmology of the special model}
\label{ssec:SpecExCosmo}

\subsubsection{Background}
\label{sssec:BkgSpec}

Having established that the simple model we analysed could be stable on FLRW, we can push analysis to whether it could be relevant for the cosmological evolution of our Universe. For this, we specify the matter perfect fluid to be a mixture of pressureless matter and radiation, respectively denoted by subscripts ``$m$'' and ``$r$'', i.e.\ $\rho_M = \rho_m + \rho_r$ and $P_M = P_m + P_r$ with
\begin{align}
	P_m = 0 \qquad &\Rightarrow\qquad  \dot{\rho}_m + 3H \rho_m = 0 \nonumber \,, \\
	P_r = \frac13 \rho_r \qquad &\Rightarrow \qquad \dot{\rho}_r + 4H \rho_r = 0 \,.
\end{align}
The effective energy density and pressure of the EPN field were defined previosuly in \eqref{eq:rhoExtPNhat} and \eqref{eq:PExtPNhat}.

We wish to recast the background equations as a dynamical system, again following the analysis of \cite{DeFelice:2016yws,DeFelice:2016uil}. As a first step solve for $\phi$ in terms of $H$ by using the constraint equation \eqref{eq:eom3hat}. Next it is convenient to introduce the density parameters
\begin{equation}
	\Omega_r \equiv \frac{\rho_r}{3 M_{\text{Pl}}^2 H^2}\,, \qquad \Omega_m \equiv \frac{\rho_m}{3 M_{\text{Pl}}^2 H^2}\,, \qquad \hat{\Omega}_{\text{EPN}} \equiv \frac{\hat{\rho}_{\text{EPN}}}{3 M_{\text{Pl}}^2 H^2} \,,
\end{equation}
so that the Friedmann equation reads
\begin{equation}
	\Omega_r + \Omega_m + \hat{\Omega}_{\text{EPN}} = 1 \,,
\end{equation}
which we use to solve for $H$ as a function of $\hat{\Omega}_{\text{EPN}}$ (or equivalently $\Omega_r + \Omega_m$). At this stage the scalar field $\phi$ and the Hubble parameter $H$ are now fully determined by the Lagrangian parameters, the Planck mass $M_{\text{Pl}}$, the mass scale $\Lambda$ and the density parameters.

We are interested in the time evolution of the density parameters $\hat{\Omega}_{\text{EPN}}$ and $\Omega_r$ ($\Omega_m$ being trivially determined from these). We employ the e-folding number $N=\log(a)$ as the time variable, with derivatives with respect to $N$ being denoted by a prime. In order to express $\hat{\Omega}_{\text{EPN}}'$ and $\Omega_r'$ solely in terms of $\hat{\Omega}_{\text{EPN}}$ and $\Omega_r$ we first use the Raychaudhuri equation to write the EPN pressure as
\beq
	\hat{P}_{\text{EPN}} = 3M_{\text{Pl}}^2 H^2 \left( \hat{w}_{\rm eff} - \frac{\Omega_r}{3} \right) \,,
\eeq
where
\begin{equation}
	\hat{w}_{\rm eff} \equiv - 1 - \frac{2\dot{H}}{3H^2} \,,
\end{equation}
is the effective equation of state parameter of the universe. For later use let us also introduce the effective equation of state parameter for the vector condensate,
\begin{equation}
	\hat{w}_{\text{EPN}} \equiv \frac{\hat{P}_{\text{EPN}}}{\hat{\rho}_{\text{EPN}}} = \frac{\hat{w}_{\rm eff} - \frac{\Omega_r}{3}}{\hat{\Omega}_{\text{EPN}}} \,.
\end{equation}
The only other ingredients we need are the time derivatives $\dot{\hat{\rho}}_{\text{EPN}}$ and $\dot{\rho}_{r}$ which are easily obtained from the respective continuity equations.

Collecting these preliminary results we get
\begin{equation}
\bal
	\hat{\Omega}_{\text{EPN}}' &= 3 \hat{w}_{\rm eff} ( \hat{\Omega}_{\text{EPN}} - 1 ) + \Omega_r = \mathcal{F}_1 ( \hat{\Omega}_{\text{EPN}}, \Omega_r)  \,,\\
	\Omega_r' &= (3\hat{w}_{\rm eff}-1) \Omega_r = \mathcal{F}_2 ( \hat{\Omega}_{\text{EPN}}, \Omega_r) \,.
\eal
\end{equation}
The precise form of the functions $\mathcal{F}_1$ and $\mathcal{F}_2$ can only be determined once $\hat{w}_{\rm eff}$ is known in terms of the density parameters, and for this we need to specify the model.

Before doing so let us comment on the size of the scales involved in the problem. For consistency we require the cutoff scale $\Lambda$ of the EPN sector to be parametrically smaller than the Planck scale, i.e.\ $\Lambda\ll M_{\text{Pl}}$. In line with our aim of using the Proca field as the dark energy fluid responsible for the late-time cosmic acceleration, we take
\begin{equation}
	\Lambda^4 \sim M_{\text{Pl}}^2 H_{\rm dS}^2 \,,
\end{equation}
where $H_{\rm dS}$ is the Hubble parameter of the late-time de Sitter fixed point, roughly of the order of the present-day Hubble constant. Note that this implies $H_{\rm dS}/\Lambda\sim \Lambda/M_{\text{Pl}} \ll 1$. Finally, we assume the bare mass of the vector field to be of order $m^2\sim H_{\rm dS}^2$, and for convenience we introduce
\begin{equation}
	c_m \equiv \frac{m^2 M_{\text{Pl}}^2}{\Lambda^4} \sim 1 \,.
\end{equation}
Although not necessary, we will eventually set $c_m=1$ for the sake of simplicity.

We now specify the parameters of the model by making the following choice:
\ba \label{eq:special model choice of coeffs}
&	\alpha_0 = - \frac{m^2}{\Lambda^2} X \,, \qquad
	\alpha_1 = - \frac{\Lambda^4}{M_{\text{Pl}}^4} b_1 X^2 - \frac{\Lambda^2}{M_{\text{Pl}}^2} c_1 X \,,\qquad   d_1 =- \frac{\Lambda^4}{M_{\text{Pl}}^4} e_1 X^2 - \frac{\Lambda^2}{M_{\text{Pl}}^2} f_1 X\\
&	\alpha_{2,X} = \frac{\Lambda^4}{M_{\text{Pl}}^4} b_2 X^2 + \frac{\Lambda^2}{M_{\text{Pl}}^2} c_2 X \,,\quad {\rm and}\quad
	\alpha_{3,X} = \frac{\Lambda^4}{M_{\text{Pl}}^4} b_3 X^2 + \frac{\Lambda^2}{M_{\text{Pl}}^2} c_3 X \,.
\ea
We leave the constants $b_I$, $c_I$, $e_1$ and $f_1$ unspecified for the time being. It is convenient to introduce
\begin{equation}
	y \equiv \frac{c_m}{3(b_1 + e_1)} \,,
\end{equation}
where just like $c_m$, $y$ is a dimensionless parameter which sets relations between the various scales in our system. With the benefit of hindsight we set
\beq \label{eq:choice y}
y = 4 \sqrt{\frac{6}{c_m}} \,.
\eeq
The reason behind this choice is that the effective squared mass of the Proca scalar mode, in this particular model, generically goes as $\propto -H_{\rm dS}^2 (y - 4 \sqrt{6/c_m})^2/(\hat{\Omega}_{\text{EPN}}-1)^2$ near the de Sitter attractor $\hat{\Omega}_{\text{EPN}}\to 1$. The tuning in \eqref{eq:choice y} then has the purpose of eliminating this pathological behavior.

From the background equation \eqref{eq:eom3hat} we obtain
\begin{equation}
\label{eq:H2y}
H^2 = \frac{y^2 \Lambda^4 M_{\text{Pl}}^4}{\phi^6} \,.
\end{equation}
From \eqref{eq:rhoExtPNhat} we can also evaluate the dark energy density in the Proca field and its associated density parameter,
\begin{equation}
\hat{\rho}_{\text{EPN}} = \Lambda^4 \frac{c_m y^{2/3}}{2} \left( \frac{\Lambda^4}{M_{\text{Pl}}^2 H^2} \right)^{1/3} \,,\qquad \hat{\Omega}_{\text{EPN}} =  \frac{c_m y^{2/3}}{6} \left(\frac{\Lambda^4}{M_{\text{Pl}}^2 H^2} \right)^{4/3} \,.
\end{equation}
Observe that $\hat{\rho}_{\text{EPN}} \sim \Lambda^4$ when approaching the de Sitter point $H \to H_{\rm dS}$, justifying the choice of scales made in \eqref{eq:special model choice of coeffs}.

From these results we obtain the following expressions for the effective equation of state parameters:
\begin{align}
	\hat{w}_{\text{eff}} &= \frac{- 4 \hat{\Omega}_{\text{EPN}}+ \Omega_r}{3 + \hat{\Omega}_{\text{EPN}}} \,,\\
	\hat{w}_{\text{EPN}} &= - \frac{12 + \Omega_r}{9 + 3 \hat{\Omega}_{\text{EPN}}} \,,
\end{align}
so that the autonomous system determining the evolution of $\hat{\Omega}_{\text{EPN}}$ and $\Omega_r$ reads
\beq\bal
		\hat{\Omega}_{\text{EPN}}' &= \frac{4 \hat{\Omega}_{\text{EPN}} ( 3(1-\hat{\Omega}_{\text{EPN}}) + \Omega_r) }{3 + \hat{\Omega}_{\text{EPN}}} \,,\\
		\Omega_r' &= - \frac{\Omega_r \left( 3 ( 1 - \Omega_r ) + 13 \hat{\Omega}_{\text{EPN}} \right)}{3 + \hat{\Omega}_{\text{EPN}}} \,.
	\label{eq:dynsys}
\eal\eeq
A straightforward analysis shows that this system admits three fixed points corresponding to radiation domination, matter domination and dark energy domination (de Sitter fixed point). The results are summarized in Table \ref{tab:fixedpoints}. In the last column we show the eigenvalues of the Jacobian matrix of the system evaluated at the respective fixed points, from which we can infer their stability. We conclude in particular that the de Sitter fixed point is an attractor.
\begin{table}[h!]
\begin{center}
\begin{tabular}{ c | c  c  c | c  c | c }
  & $\Omega_r$ & $\Omega_m$ & $\hat{\Omega}_{\text{EPN}}$ & $\hat{w}_{\rm eff}$ & $\hat{w}_{\text{EPN}}$ & eigenvalues \\ \hline
  radiation & $1$ & $0$ & $0$ & $\frac{1}{3}$ & $- \frac{13}{9}$ & $\{\frac{16}{3},1\}$: unstable \\
  matter & $0$ & $1$ & $0$ & $0$ & $-\frac43$ & $\{4,-1\}$: saddle point \\
  de Sitter & $0$ & $0$ & $1$ & $-1$ & $-1$ & $\{ -4, -3 \}$: stable
\end{tabular}
\caption{Fixed points of the autonomous system \eqref{eq:dynsys}, describing the cosmic density parameters carried respectively in radiation, matter and in the Proca field.}
\label{tab:fixedpoints}
\end{center}
\end{table}

\begin{figure}[!htb]
	\center{\includegraphics[width=8cm]{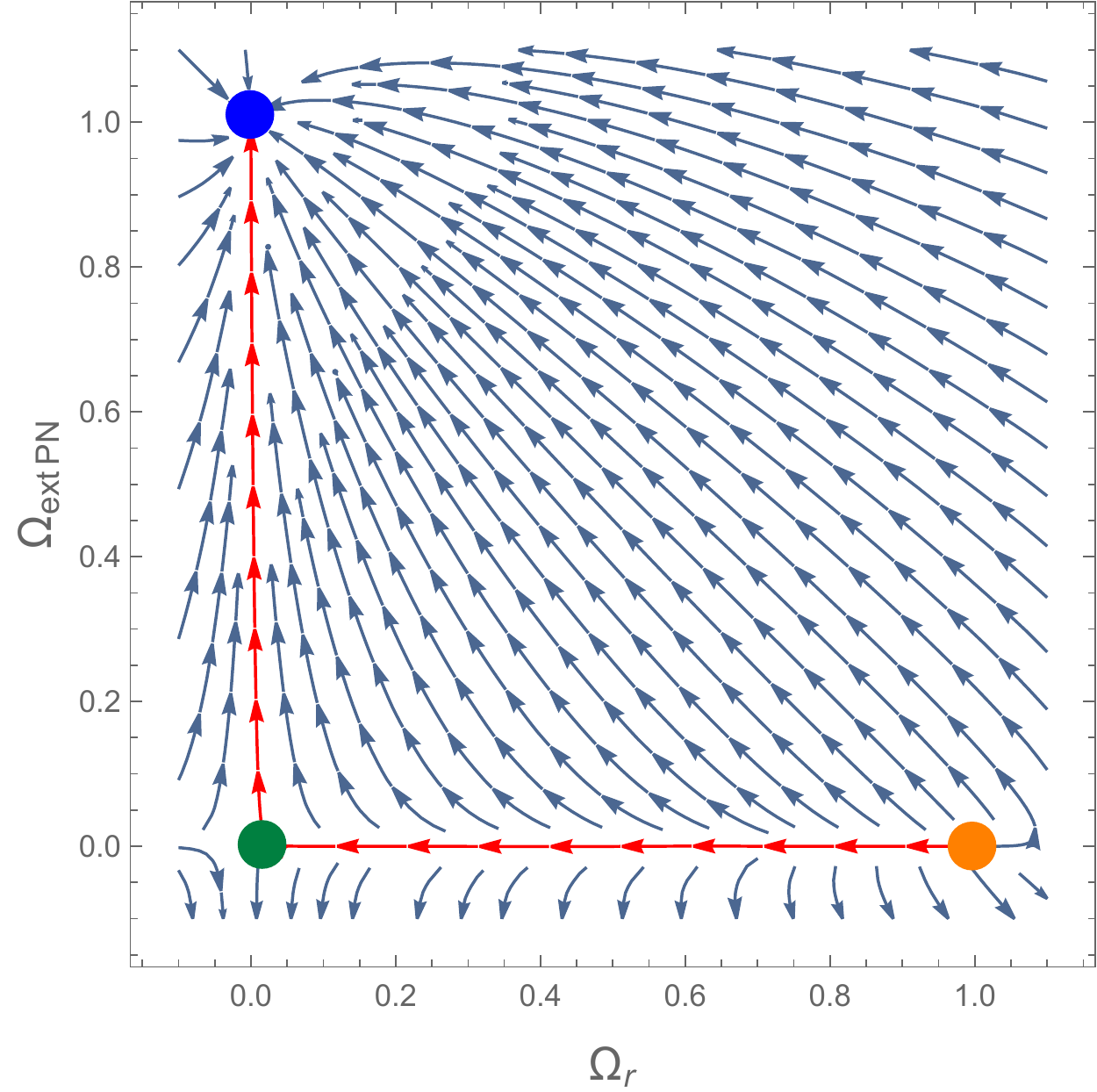}}
	\caption{\label{fig:plot_phase_portrait} Phase portrait associated to the autonomous system system \eqref{eq:dynsys}. The radiation, matter and dark energy fixed points are respectively indicated by the orange, green and blue dots. The red trajectory is a particular solution resembling the hot Big Bang phase of our universe with epochs of radiation, matter and dark energy domination.}
\end{figure}

In Fig.\ \ref{fig:plot_phase_portrait} we show the phase portrait of the dynamical system \eqref{eq:dynsys}, with the radiation, matter and dark energy fixed points shown as colored dots. The red trajectory is a particular solution that qualitatively mimics our universe's hot Big Bang phase, starting very close to the radiation point, flowing toward the matter point, and then asymptotically approaching the de Sitter point.

\begin{figure}[!htb]
	\center{\includegraphics[width=12cm]{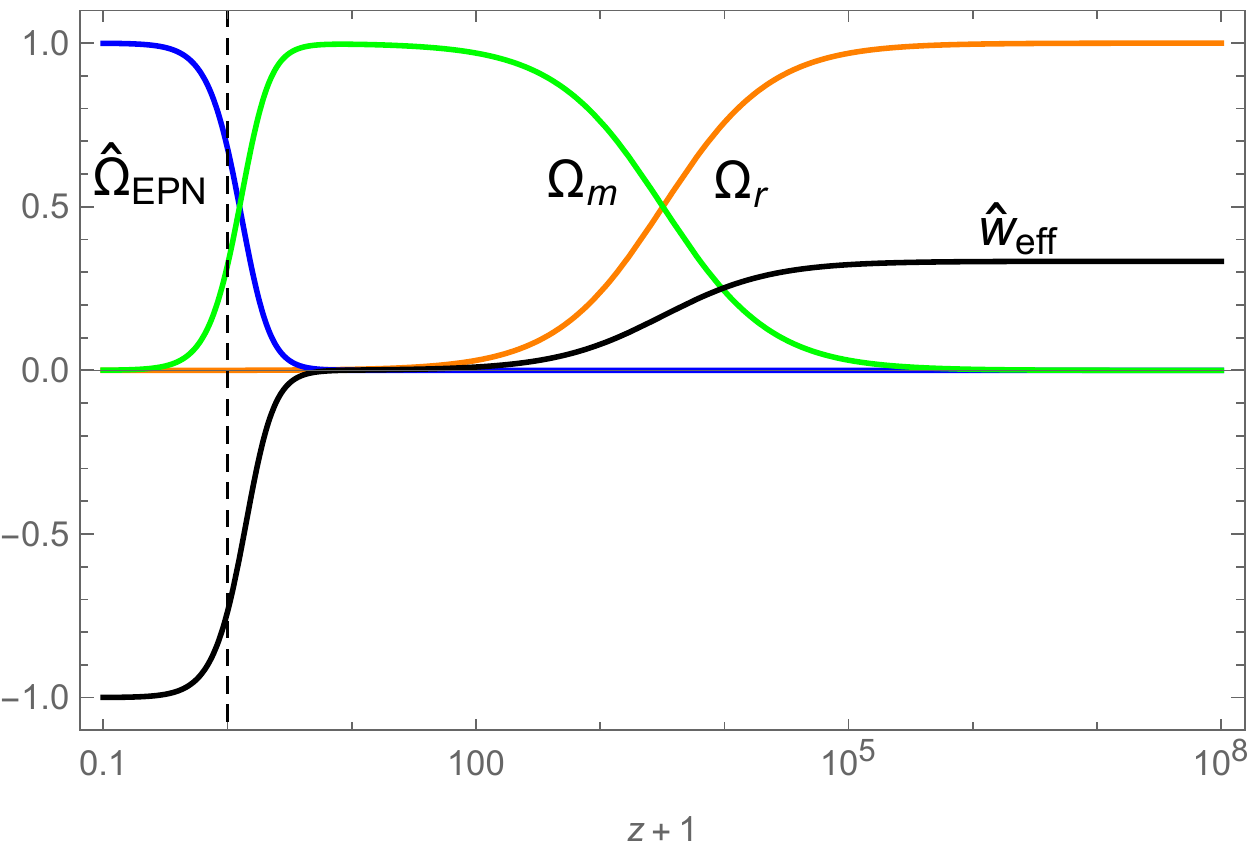}}
	\caption{\label{fig:plot_universe_content} Evolution of the density parameters and effective equation of state parameter $\hat{w}_{\rm eff}$ as functions of redshift, with initial conditions chosen such that $\hat{\Omega}_{\text{EPN}} = 0.68$ and $\Omega_r = 10^{-4}$ at $z=0$, indicated by the vertical dashed line.}
\end{figure}

The dynamical system \eqref{eq:dynsys} can also be solved numerically to obtain the time evolution of the density parameters. Rather than cosmic time we will show the results as functions of redshift $z$, setting $\hat{\Omega}_{\text{EPN}} = 0.68$ and $\Omega_r = 10^{-4}$ at $z=0$ (the present time), approximately the experimentally measured values. The solution for the three density parameters is shown in Fig.\ \ref{fig:plot_universe_content}, along with the effective equation of state parameter of the universe.

\subsubsection{Perturbations}
\label{sssec:PertSpec}

Finally we examine the stability  conditions for the perturbations as well as their speed. Even though all the stability conditions are time-dependent, we will evaluate them in the early- and late-time limits as way to derive a reduced set of algebraic constraints, and then verify numerically that there exists a choice of coefficients such that the constraints are satisfied at all times.

We already remarked that tensor perturbations propagate exactly as in GR. Starting then with the vector modes, we evaluate the kinetic term coefficient $\hat{Q}_V$ and squared sound speed $\hat{c}_V^2$ at both the radiation and dark energy fixed points:
\begin{equation}
	\hat{Q}_V = \begin{cases}
		1 - \frac{3 c_3 y}{10} + \mathcal{O}\left( 1 - \Omega_r, \hat{\Omega}_{\text{EPN}}^{1/4} \right) \qquad &\text{radiation} \\
		- \frac{y \left[ y (b_1 + 10b_2 + 8b_3) + 4 (c_1 + 10c_2 + 8c_3) \right]}{20 (1 - \hat{\Omega}_{\text{EPN}})}+ \mathcal{O}\left( \Omega_r, (1 - \hat{\Omega}_{\text{EPN}})^{0} \right) \qquad &\text{dS} \,,
	\end{cases}
\end{equation}
\begin{equation}
	\hat{c}_V^2 = \begin{cases}
		1 + \frac{4 c_3 y}{3 (10 - 3 c_3 y)} + \mathcal{O}\left( 1 - \Omega_r, \hat{\Omega}_{\text{EPN}}^{1/4} \right) \qquad &\text{radiation} \\
		0 + \frac{5y \left[ y (b_1 + 6b_2 + 2b_3) + 4 (c_1 + 6c_2 + 2c_3) \right] - 160}{8y \left[ y (b_1 + 10b_2 + 8b_3) + 4 (c_1 + 10c_2 + 8c_3) \right]} \left( 1 - \hat{\Omega}_{\text{EPN}} \right) + \mathcal{O}\left( \Omega_r, 1 - \hat{\Omega}_{\text{EPN}} \right) \qquad &\text{dS} \,.
	\end{cases}
\end{equation}
Notice that $\hat{Q}_V$ actually diverges while $\hat c_V$ asymptotes zero when one approaches the de Sitter point. Such a behaviour can be indicative of reaching strong coupling, however we analyze carefully the scale at which perturbative unitarity breaks down in Appendix \ref{sec:OperatordS} and show that the model becomes weakly coupled in the asymptotically de Sitter fixed point.

Recalling that we will later set $y$ as in \eqref{eq:choice y}, the conditions that $\hat{Q}_V>0$ and $0<\hat{c}_V^2\leq 1$ at early times\footnote{Note that since the tensor modes behave as in GR in this example and can be trivially decoupled, the relation between causality and subluminality of the other fields is more straightforward (see Ref.~\cite{deRham:2019ctd}).} imposes the constraint $c_3<0$, while positivity of $\hat{Q}_V$ at late times is clearly easy to achieve, for instance by taking all the coefficients $b_I$'s and $c_I$'s negative in \eqref{eq:special model choice of coeffs}. We will present a specific choice of values below. Note that this choice also ensures that $\hat{c}_V^2$ tends to $0$ from above about the dS point.

The Hubble-normalized effective mass of the vector mode is given, in our example, by
\begin{equation}
	\frac{\hat{m}_V^2}{H^2} = \begin{cases}
		0 + \left( \frac35 - \frac{1}{10 - 3 c_3 y} \right) (1 - \Omega_r ) + \mathcal{O}\left( 1 - \Omega_r, \hat{\Omega}_{\text{EPN}}^{1/4} \right) &\text{radiation} \\
		5 + \mathcal{O}\left( \Omega_r, 1 - \hat{\Omega}_{\text{EPN}} \right) &\text{dS}\,.
	\end{cases}
\end{equation}
Interestingly, the normalized mass of the vector mode approaches zero at early times and it acquires a Hubble-scale value at late times. Moreover, we will see below that it remains positive for all finite times, at least for this model and for a certain choice of coupling constants.

Continuing with the scalar sector, we focus on the Proca scalar mode $\psi$ as its stability has been shown to be independent of that of the matter fluid. The kinetic coefficient actually takes a very compact form with no need to evaluate at specific times,
\begin{equation}
	\hat{Q}_{S,\psi} = \frac{48 (3 + \hat{\Omega}_{\text{EPN}})}{y^2 (1 - \hat{\Omega}_{\text{EPN}})^2}\left(\frac{\Lambda}{M_{\rm Pl}}\right)^2 \,,
\end{equation}
while the expressions for the sound speed at early and late times read
\begin{equation}
	\hat{c}_{S,\psi}^2 = \begin{cases}
		\frac{11}{27} + \mathcal{O}\left( 1 - \Omega_r, \hat{\Omega}_{\text{EPN}}^{3/4} \right) \qquad &\text{radiation} \\
		0 + \frac{5}{24y^2}\frac{y^2 \left[ y (b_1 + 10b_2 + 8b_3) + 4 (c_1 + 10c_2 + 8c_3) \right] - 32}{y (b_1 + 10b_2 + 8b_3) + 4 (c_1 + 10c_2 + 8c_3)} \left( 1 - \hat{\Omega}_{\text{EPN}} \right) + \mathcal{O}\left( \Omega_r, 1 - \hat{\Omega}_{\text{EPN}} \right) \qquad &\text{dS}\,. \\
	\end{cases}
\end{equation}
We see that $\hat{Q}_{S,\psi}$ is always positive, although again we have a divergence about the de Sitter point. Similarly, $\hat{c}_{S,\psi}^2$ is manifestly positive and subluminal at early times, but tends to zero (while keeping positive values) at late times. As with the vector mode, it can be shown that this poses no problem, see Appendix \ref{sec:OperatordS}.

Finally, the effective mass of the scalar mode is given by
\begin{equation}
	\frac{\hat{m}_{S,\psi}^2}{H^2} = \begin{cases}
		0 - \frac{200}{y(10 - 3 c_3 y)} \hat{\Omega}_{\text{EPN}}^{3/4} + \mathcal{O}\left( 1 - \Omega_r, \hat{\Omega}_{\text{EPN}}^{3/4} \right) \qquad &\text{radiation} \\
		\frac{10}{y^5}\, \frac{\left[ 3y^3 (b_1 + 10b_2 + 8b_3) + 12 y^2 (c_1 + 10c_2 + 8c_3) + 64 \right]^2 - 96^2}{\left[ y(b_1 + 10b_2 + 8b_3) + 4(c_1 + 10c_2 + 8c_3) \right]^2} \frac{\Lambda^4}{M_{\text{Pl}}^2 H_{dS}^2} + \mathcal{O}\left( \Omega_r, 1 - \hat{\Omega}_{\text{EPN}} \right) \qquad &\text{dS} \,. \\
	\end{cases}
\end{equation}
Having $\hat{m}_{S,\psi}^2>0$ is clearly easy to achieve at late times, while at early times the Hubble-normalized effective mass approaches zero. It turns out that $\hat{m}_{S,\psi}^2$ actually tends to zero at early times from below, but being strongly suppressed relative to the time-dependent Hubble scale shows that the associated tachyonic instability is harmless.

\begin{figure}[!htb]
	\center{\includegraphics[width=7cm]{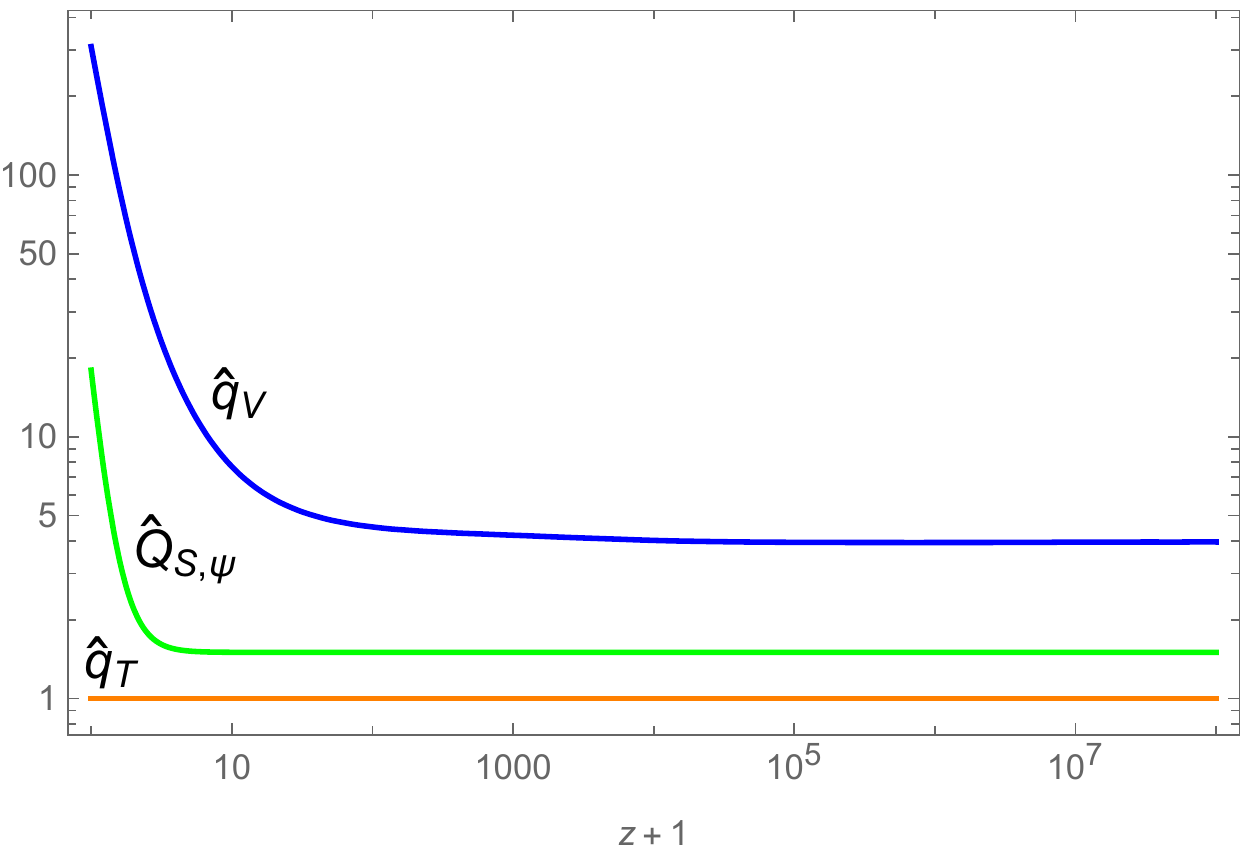}~~~~~~\includegraphics[width=7cm]{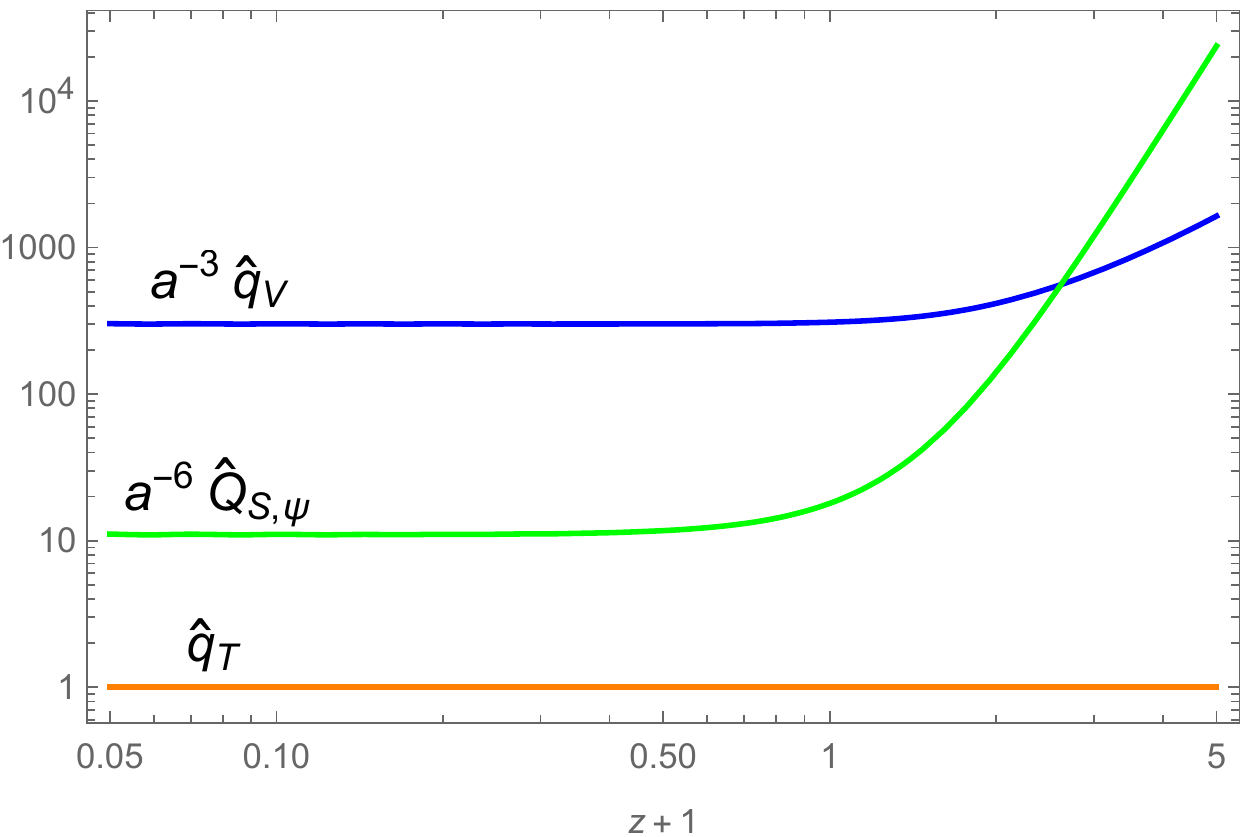}}
	\caption{\label{fig:plot_kinetic_terms} Left panel: Kinetic coefficients of the tensor, vector and scalar perturbations for $1 \leq z+1 \leq 10^8$. Right panel: The same kinetic coefficients rescaled by an appropriate power of the scale factor $a(t)$ in order to exhibit their scaling about the dS point. We observe they follow a power law scaling for small $z+1$.}
	\label{fig:kin}
\end{figure}

\begin{figure}[!htb]
	\center{\includegraphics[width=7cm]{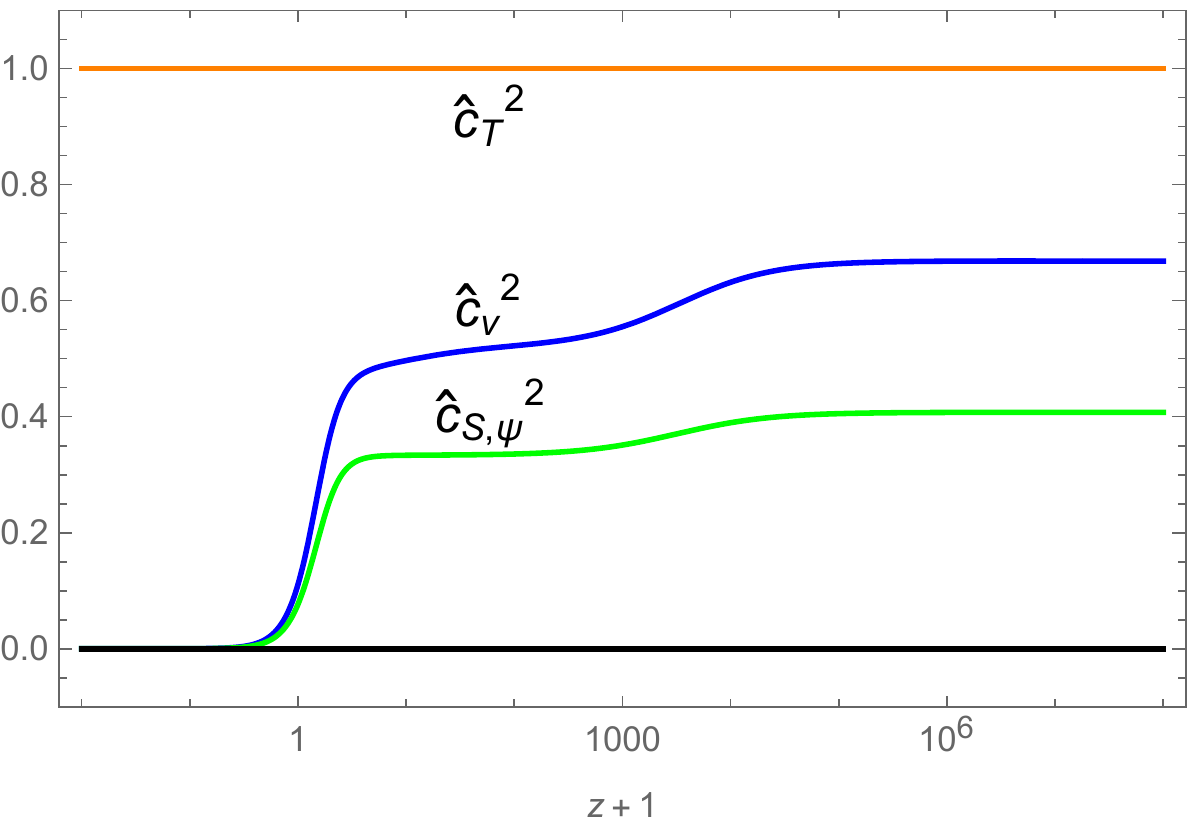}~~~~~~\includegraphics[width=7cm]{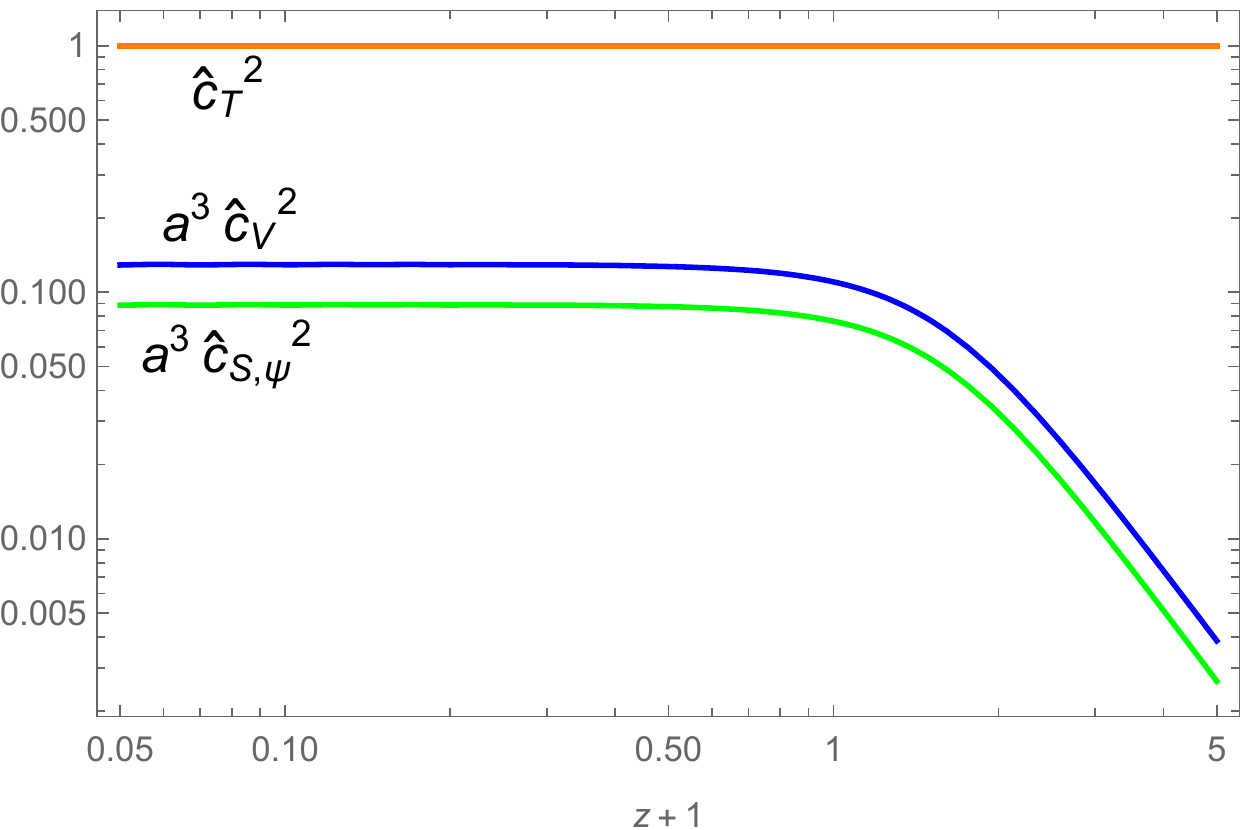}}
	\caption{\label{fig:plot_velocities} Left panel: Squared sound speeds of the tensor, vector and scalar perturbations. Right panel: The same speeds rescaled by an appropriate power of the scale factor $a(t)$ in order to exhibit their scaling about the dS point. We observe they follow a power law scaling for small $z+1$.}
	\label{fig:velocities}
\end{figure}

\begin{figure}[!htb]
	\center{\includegraphics[width=8cm]{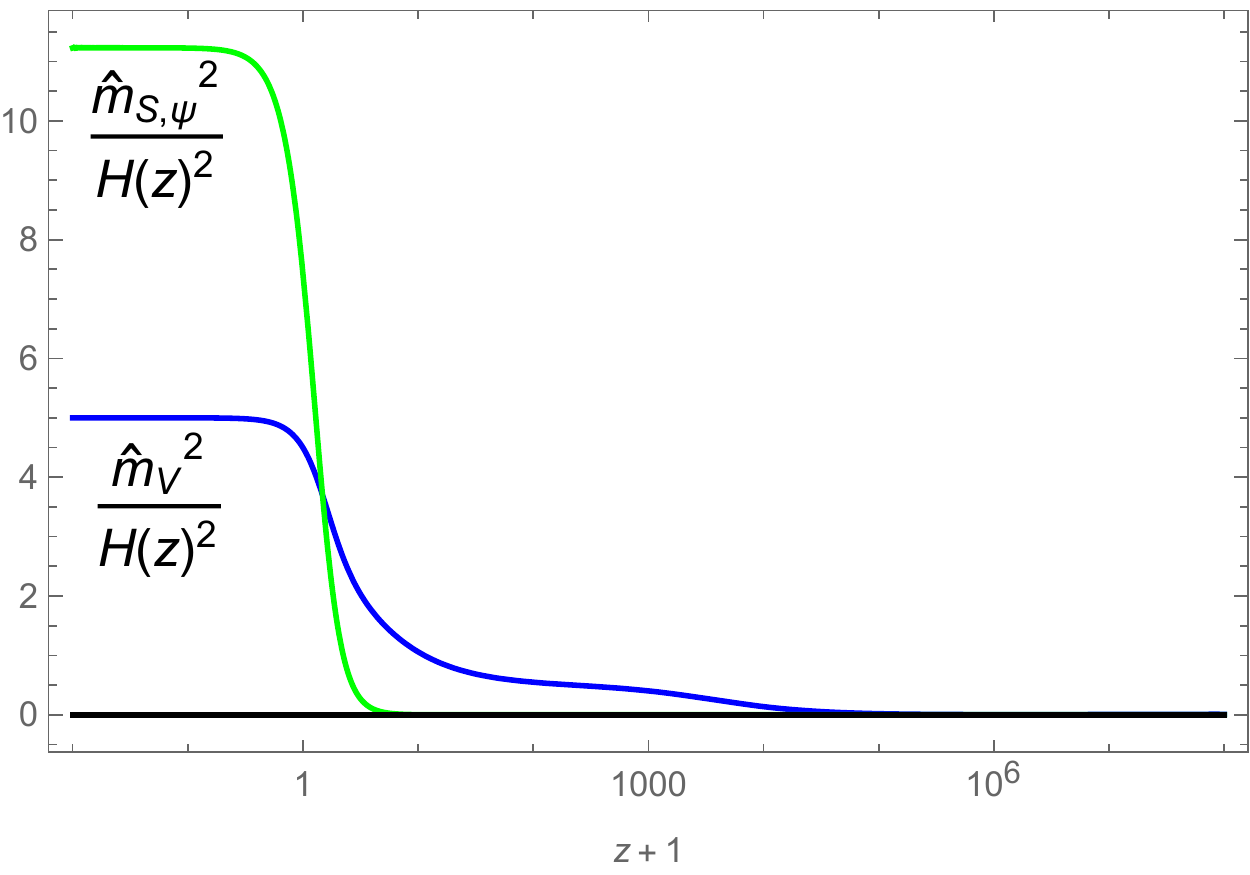}}
	\caption{\label{fig:plot_masses} Effective squared masses of the vector and scalar perturbations, normalized by the time-dependent Hubble parameter $H(z)$. Note that $\hat{m}_{S,\psi}^2/H^2$ is proportional to the ratio $M_{\rm{Pl}}^2 H_{\rm dS}^2 / \Lambda^4$, which we have kept generic in our analysis but have set to 1 in this particular plot.}
\end{figure}

We have argued that all the stability conditions for the dynamical modes can be met, at least in the epochs of radiation and dark energy domination and the fields are then all also subluminal. We can further show numerically that there exist coefficients such that no pathologies arise at any times. One simple explicit choice is
\begin{equation}
	c_m = 1\,, \qquad c_I = b_I = -1, \quad \text{for } I=1,2,3 \,.
\end{equation}
The results for the time evolution (plotted as functions of redshift) of the kinetic coefficients, squared sound speeds and effective squared masses are shown respectively in Figs.\ \ref{fig:plot_kinetic_terms}, \ref{fig:plot_velocities} and \ref{fig:plot_masses}. The cosmological history is the same as that of the previous subsection, with $\hat{\Omega}_{\text{EPN}} = 0.68$ and $\Omega_r = 10^{-4}$ at $z=0$.

To summarize, we have demonstrated that the EPN special model, with the particular choice of coefficient functions given in \eqref{eq:special model choice of coeffs}, admits a window of parameters such that all perturbations are free of ghost- and gradient-type instabilities and propagate subluminally. Although the velocities $\hat{c}_V^2$ and $\hat{c}_{S,\psi}^2$ approach zero in the late-time de Sitter limit, we have given an argument in Appendix \ref{sec:OperatordS} which shows that this is not a pathology. Furthermore, $\hat{c}_V^2$ and $\hat{c}_{S,\psi}^2$ are finite and positive for all $z\geq0$, so all the degrees of freedom behave in a smooth, stable and subluminal way throughout the cosmological history. While gravitational waves behave identically as in GR, the presence of the vector and scalar modes could have intriguing signatures for instance at the level of structure formation. The study of those is beyond the scope of this work, and saved for future considerations.


\section{General model}
\label{sec:fulltheory}

Having focused on a specific model for sake of concreteness, we now  return to the more generic covariant EPN theory given by eq.\ \eqref{eq:SextPN}, what we refer to as the general model. Since the procedure follows a very similar pattern to what was given in the previous section, in what follows, we will  omit intermediate steps in most cases and highlight only the final results. We also refer the reader to the previous section for our parametrization of perturbations and other conventions.

\subsection{Background}
\label{ssec:bkgeom}

The Friedmann and Raychaudhuri equations take the same form as before,
\beq
H^2 = \frac{1}{3M_{\text{Pl}}^2} \left( \rho_M + {\rho}_{\text{EPN}} \right) \,,\qquad \dot{H} + H^2 = - \frac{1}{6 M_{\text{Pl}}^2} \left(\rho_M + \rho_{\text{EPN}} + 3 P_M + 3 P_{\text{EPN}}\right) \,,
\eeq
with effective density and pressure for the dark energy fluid given by
\beq\bal
	\rho_{\text{EPN}} &= \Lambda^4 \left\{ - \alpha_{0} + \alpha_{0,X} \frac{\phi^2}{\Lambda^2} + 3 \left( \alpha_{1,X} + d_{1,X} \right) \frac{H \phi^3}{\Lambda^4} \right.  \\
	& \quad \left. + 6 \left[ - \left( \alpha_2 + d_2 \right) + 2 \left( \alpha_{2,X} + d_{2,X} \right) \frac{\phi^2}{\Lambda^2} + \left( \alpha_{2,XX} + d_{2,XX} \right) \frac{\phi^4}{\Lambda^4} \right] \frac{H^2}{\Lambda^2} \right.  \\
	& \quad \left. - \left[ 5 \left( \alpha_{3,X} + d_{3,X} \right) + \left( \alpha_{3,XX} + d_{3,XX} \right) \frac{\phi^2}{\Lambda^2} \right] \frac{H^3 \phi^3}{\Lambda^6} \right\} \,,\label{eq:rhoExtPN}
\eal\eeq
\beq\bal
	P_{\text{EPN}} &= \Lambda^4 \left\{ \alpha_{0} - \left( \alpha_{1,X} + d_{1,X} \right) \frac{\phi^2 \dot{\phi}}{\Lambda^4} + 2 \left(\alpha_2 + d_2 \right) \frac{3H^2 + 2\dot{H}}{\Lambda^2} \right.  \\
	&\quad - 2 \left( \alpha_{2,X} + d_{2,X} \right) \frac{\phi \left( 3 H^2 \phi + 2 H \dot{\phi} + 2 \dot{H} \phi \right)}{\Lambda^4} - 4 \left( \alpha_{2,XX} + d_{2,XX} \right) \frac{H \phi^3 \dot{\phi}}{\Lambda^6}  \\
	& \quad \left. + \left[ \left( \alpha_{3,X} + d_{3,X} \right) \frac{2 H^2 \phi + 3 H \dot{\phi} + 2 \dot{H} \phi}{\Lambda^3} + \left( \alpha_{3,XX} + d_{3,XX} \right) \frac{H \phi^2 \dot{\phi}}{\Lambda^5} \right] \frac{H \phi^2}{\Lambda^3} \right\} \,.
	\label{eq:PExtPN}
\eal\eeq

Finally, from the variation of the action with respect to $\phi$ one infers
\begin{align}
	&\alpha_{0,X} + 3 \left( \alpha_{1,X} + d_{1,X} \right) \frac{H \phi}{\Lambda^2} + 6 \left[ \left( \alpha_{2,X} + d_{2,X} \right) + \left( \alpha_{2,XX} + d_{2,XX} \right) \frac{\phi^2}{\Lambda^2} \right] \frac{H^2}{\Lambda^2} \nonumber \\
	& - \left[ 3 \left( \alpha_{3,X} + d_{3,X} \right) + \left( \alpha_{3,XX} + d_{3,XX} \right) \frac{\phi^2}{\Lambda^2} \right] \frac{H^3 \phi}{\Lambda^4} = 0 \,,
	\label{eq:eom3}
\end{align}
after discarding the trivial solution $\phi=0$. We observe that \eqref{eq:eom3} is again a ``constraint'' equation, relating $H$ and $\phi$ algebraically. While for the simple model this property was a consequence of the specific tuning of coefficients, in the general model this follows from the particular form of the non-minimal couplings and their coefficients.

We also note that this set of background equations is equivalent to those of GP theory \cite{DeFelice:2016uil}. As mentioned previously, this is simply because PN and GP coincide at the level of the FLRW background, and therefore so does the general EPN model.

\subsection{Perturbations}
\label{ssec:Perturbations}

\subsubsection{Tensor perturbations}
\label{sssec:TensPert}

Interestingly, in the presence of tensor perturbations the relation
\begin{equation}
	\K_{\mu\nu} = \frac{1}{\Lambda^2}\,\nabla_{\mu} A_{\nu} \qquad\qquad \mbox{(tensor modes, quadratic order)}\,,
\end{equation}
remains true up to quadratic order in the tensor modes. As for the background, it follows as an immediate result that the quadratic action for tensor perturbations in the general model will match that of GP:
\begin{equation}
	S_T^{(2)} = \int \d^4 x \, a^3 \frac{q_T}{8} \left[ \dot{h}_{ij}^2 - \frac{c_T^2}{a^2} \(\p_i h_{jk} \)^2 \right] \,,
	\label{eq:ST2}
\end{equation}
with
\begin{align}
	q_T &= 1 + 2 \frac{\Lambda^2}{M_{\text{Pl}}^2} \left[ (\alpha_2 + d_2) - \frac{\phi^2}{\Lambda^2}( \alpha_{2,X} + d_{2,X} ) \right] + \frac{H \phi^3}{M_{\rm{Pl}}^2 \Lambda^2} \left( \alpha_{3,X} + d_{3,X} \right) \label{eq:qT} \,,\\
	c_T^2 &= \frac{1 + 2 \frac{\Lambda^2}{M_{\rm{Pl}}^2} (\alpha_2 + d_2) + \frac{\phi^2 \dot{\phi}}{M_{\text{Pl}}^2 \Lambda^2} \left( \alpha_{3,X} + d_{3,X} \right)}{q_T} \,. \label{eq:cTsq}
\end{align}
These results  imply that the general model describes the expected two degrees of freedom in the tensor sector. Stability of tensor perturbations then dictates $q_T,c_T^2>0$. Subluminality of the tensor modes would also require $c_T^2<1$ but we refer to Refs.~\cite{deRham:2019ctd,deRham:2020zyh,deRham:2021fpu} for a word of caution on applying generic subluminality criteria to gravitational waves propagation without other further considerations. Imposing the speed of gravitational waves to be exactly luminal requires setting $\alpha_{2,X}+d_{2,X}=\alpha_{3,X}+d_{3,X}=0$ at all times, meaning for all values of the argument $X$ of those functions (unless $\dot \phi$ is constant). Such a choice would correspond to the example explored in details in the previous section.
We point out however that there may exist some subtleties related to the frequency at which the existing constraints on the speed of gravitational waves are satisfied \cite{deRham:2018red}, and in principle one would only require $\alpha_{2,X}+d_{2,X}=\alpha_{3,X}+d_{3,X}=0$ for a given range of arguments.

\subsubsection{Vector perturbations}
\label{sssec:VecPert}

Continuing with the vector perturbations, combining the expansion of \eqref{eq:SextPN} with the matter action derived before in \eqref{eq:SMV2b}, we find
\begin{align}
	S_V^{(2)} &= \int \d^4 x \, \frac{a^3}{2} \Bigg[ q_V \dot{Z}_i^2 - \frac{1}{a^2} \mathcal{C}_1 (\p_i Z_j)^2 - H^2 \mathcal{C}_2 Z_i^2 + \frac{1}{a^2} \mathcal{C}_3 \p_i V_j \p_i Z_j  + \frac{1}{a^2} \mathcal{C}_4 \p_i V_j \p_i \dot{Z}_j \nonumber \\
	& \quad + \frac{q_T}{2a^2}(\p_i V_j)^2 + \frac{(\rho_M + P_M)}{M_{\text{Pl}}^2} \bigg( V_i - a \frac{\delta \dot{\mathcal{B}}_i}{M_{\text{Pl}}} \bigg)^2 \Bigg] \,,
	\label{eq:SV2}
\end{align}
where the coefficients entering in this result are given in Appendix \ref{ssec:DefCoefsVec}. The structure of \eqref{eq:SV2} matches that of GP \cite{DeFelice:2016uil} except for the presence of the operator proportional to $\mathcal{C}_4$. Nevertheless, we see that this extra term does not spoil the counting of degrees of freedom, since the mode $V_i$ is still non-dynamical. This establishes that the general EPN model propagates the correct number of vector modes, namely one.

The extra operator proportional to $\mathcal{C}_4$ is interesting in that it modifies the dispersion relation of the dynamical field $Z_i$ in a way that is qualitatively different from GP. In order to highlight this effect we will ignore matter for the moment and return to the general case at the end.
Taking $\rho_M,P_M=0$ in \eqref{eq:SV2} and integrating out $V_i$ we obtain, after Fourier transforming and performing a partial integration,
\beq
S_V^{(2)}=\int \d t\,\frac{\d^3k}{(2\pi)^3}\,\frac{a^3}{2}\bigg\{q_V\bigg[1-\frac{k^2}{a^2}\,\frac{\mathcal{C}_4^2}{2q_Tq_V}\bigg]|\dot{Z}_i(k)|^2-\bigg[\mathcal{C}_2H^2+\frac{k^2}{a^2}\left(\mathcal{C}_1+\frac{\mathcal{C}_3^2}{2q_T}-a^{-1}\partial_t\left(a \frac{\mathcal{C}_3\mathcal{C}_4}{2q_T}\right)\right)\bigg]|Z_i(k)|^2\bigg\} \,.
\eeq
For a localized sub-Hubble perturbation we then infer the dispersion relation
\beq
\omega_V^2=\frac{\mathcal{C}_2H^2+\frac{k^2}{a^2}\left(\mathcal{C}_1+\frac{\mathcal{C}_3^2}{2q_T}-a^{-1}\partial_t\left(a \frac{\mathcal{C}_3\mathcal{C}_4}{2q_T}\right)\right)}{q_V\left(1-\frac{k^2}{a^2}\,\frac{\mathcal{C}_4^2}{2q_Tq_V}\right)} \qquad\qquad \mbox{(no matter)}\,.
\eeq
We see that the presence of the new coefficient $\mathcal{C}_4$ makes the dispersion relation non-linear. Expanding at small momenta (more precisely for $k^2/a^2 \ll |q_T q_V| / \mathcal{C}_4^2$) we have the linear approximation
\beq
\omega_V^2\simeq m_V^2+c_V^2\,\frac{k^2}{a^2} \,,
\eeq
with effective mass and speed of sound
\beq \label{eq:gen model mass speed vector without matter}
m_V^2\equiv \frac{\mathcal{C}_2}{q_V}\,H^2 \,,\qquad c_V^2\equiv \frac{1}{q_V} \left( \mathcal{C}_1+\frac{\mathcal{C}_3^2}{2q_T}- a^{-1}\partial_t\left(a \frac{\mathcal{C}_3\mathcal{C}_4}{2q_T}\right) + \frac{\mathcal{C}_2 \mathcal{C}_4^2}{2q_T q_V}\,H^2 \right) \,.
\eeq
In this approximation and remembering that we are neglecting matter, absence of gradient instabilities requires  $q_V, c_V^2>0$. Similarly one may also wish to demand the absence of tachyonic modes, which is achieved if $m_V^2>0$. Note that while the coefficient of the kinetic term is also modified by the $\mathcal{C}_4$ coupling, at low energies we still have the simple no-ghost condition $q_V>0$.

Returning to the general set-up with matter present, we proceed again to integrate out $V_i$ from its equation of motion. The resulting action is non-diagonal in the fields $Z_i$ and $\delta\mathcal{B}_i$,
\beq\bal \label{eq:SV2 after int out}
S_V^{(2)}&=\int \d t\,\frac{\d^3k}{(2\pi)^3}\,\frac{a^3}{2}\bigg\{q_V\bigg[1-\frac{\frac{k^4}{a^4}\,\mathcal{C}_4^2}{2q_V\left(\frac{k^2}{a^2}\,q_T+M^2\right)}\bigg]|\dot{Z}_i|^2+\frac{1}{2}\,\frac{\frac{k^2}{a^2}\,q_T}{\frac{k^2}{a^2}\,q_T+M^2} \frac{M^2}{M_{\text{Pl}}^2}\,|a\delta\dot{\mathcal{B}}_i|^2 \\
&\quad -\bigg[\mathcal{C}_2H^2+\frac{k^2}{a^2}\,\mathcal{C}_1+\frac{\frac{k^4}{a^4}\,\mathcal{C}_3^2}{2\left(\frac{k^2}{a^2}\,q_T+M^2\right)}-a^{-3}\partial_t\bigg(a^3\frac{\frac{k^4}{a^4}\,\mathcal{C}_3\mathcal{C}_4}{2\left(\frac{k^2}{a^2}\,q_T+M^2\right)}\bigg)\bigg]|Z_i|^2 \\
&\quad +\frac{1}{2}\,\frac{\frac{k^2}{a^2}\,\mathcal{C}_3}{\frac{k^2}{a^2}\,q_T+M^2}\frac{M^2}{M_{\text{Pl}}}\left(a\delta\dot{\mathcal{B}}_i^{*}Z_i+{\rm c.c.}\right)+\frac{1}{2}\,\frac{\frac{k^2}{a^2}\,\mathcal{C}_4}{\frac{k^2}{a^2}\,q_T+M^2}\frac{M^2}{M_{\text{Pl}}}\left(a\delta\dot{\mathcal{B}}_i^{*}\dot{Z}_i+{\rm c.c.}\right)\bigg\} \,,
\eal\eeq
and we introduced
\beq
M^2\equiv 2 \frac{\rho_M+P_M}{M_{\rm Pl}^2} \,.
\eeq
Observe that the scale $M$ acts as a sort of infrared regulator modifying the long wavelength behavior of the coefficients in the action. Fourier transforming with respect to time in the sub-Hubble limit we find the following dispersion relation for the Proca vector mode:
\beq \label{eq:gen model vec disp rel}
\omega_V^2=\frac{\mathcal{C}_2H^2+\frac{k^2}{a^2}\left(\mathcal{C}_1+\frac{\mathcal{C}_3^2}{2q_T}\right)-\frac{1}{2}a^{-3}\partial_t\left(a^3\frac{\frac{k^4}{a^4}\,\mathcal{C}_3\mathcal{C}_4}{\frac{k^2}{a^2}\,q_T+M^2}\right)}{q_V\left(1-\frac{k^2}{a^2}\,\frac{\mathcal{C}_4^2}{2q_Tq_V}\right)} \,.
\eeq
Expanding at small momenta, assuming $k^2/a^2 \ll |q_T q_V| / \mathcal{C}_4^2$ and $k^2/a^2 \ll M^2/q_T$, we find the same effective mass as before (cf.\ eq.\ \eqref{eq:gen model mass speed vector without matter}) and a speed of sound
\beq
c_V^2 = \frac{1}{q_V} \left( \mathcal{C}_1+\frac{\mathcal{C}_3^2}{2q_T} + \frac{\mathcal{C}_2 \mathcal{C}_4^2}{2q_T q_V}\,H^2 \right) \,,
\eeq
which curiously is a simpler expression than in the case without matter, as a consequence of the modified infrared behavior mentioned before.

The dispersion relation for the matter perturbation $\delta\mathcal{B}_i$ is $\omega^2=0$. We emphasize that this result only assumes that the fluctuation is localized on sub-Hubble scales but is otherwise exact. This degenerate dispersion relation may seem pathological but was in fact expected. The variable $\delta\mathcal{B}_i$ corresponds physically to the vorticity field of the fluid, which is indeed gapless and has no gradient energy (see \cite{Endlich:2010hf} for a discussion of this aspect in an EFT context).

The condition for the vector mode not to be ghostly is less immediate because of the non-trivial derivative couplings appearing in \eqref{eq:SV2 after int out}. To determine the norm of the propagating field we compute the residue matrix (see for instance \cite{Garcia-Saenz:2021uyv} for a review of this method),
\beq
\lim_{\omega^2\to\omega_V^2}(\omega^2-\omega_V^2) \mathcal{P}(\omega,k) \,,
\eeq
where $\mathcal{P}$ is the matrix of propagators that we read off from \eqref{eq:SV2 after int out}. By construction the residue matrix has a single non-zero eigenvalue, which we find to be
\beq\bal
\frac{1}{Q_V}&\equiv \frac{1}{q_V}\left[\frac{1+M_{\text{Pl}}^2\frac{\mathcal{C}_4^2}{q_T^2}}{1-\frac{k^2}{a^2}\,\frac{\mathcal{C}_4^2}{2q_Tq_V}}+\frac{M_{\text{Pl}}^2\frac{\mathcal{C}_3^2q_V}{q_T^2}}{\mathcal{C}_2H^2+\frac{k^2}{a^2}\left(\mathcal{C}_1+\frac{\mathcal{C}_3^2}{2q_T}\right)-\frac{1}{2}a^{-3}\partial_t\left(a^3\frac{\frac{k^4}{a^4}\,\mathcal{C}_3\mathcal{C}_4}{\frac{k^2}{a^2}\,q_T+M^2}\right)}\right] \\
&\simeq \frac{1}{q_V}\left[1+M_{\text{Pl}}^2 \left( \frac{\mathcal{C}_4^2}{q_T^2}+\frac{\mathcal{C}_3^2q_V}{\mathcal{C}_2q_T^2H^2} \right) \right] \,,
\eal\eeq
where in the second line we have neglected $k$-dependent corrections. Absence of ghosts in the vector sector then implies the condition $Q_V>0$. Note that this does not necessarily imply $q_V>0$ as one might have naively inferred from the action in the form \eqref{eq:SV2}.

\subsubsection{Scalar perturbations}
\label{sssec:ScalPert}

The analysis of scalar perturbations in the general model proceeds very analogously to that of the special model. Expanding the full action including matter we find
\beq\bal
	S_S^{(2)} &= \int \d^4 x \, a^3 \bigg[  - \frac{\overline{n} \rho_{M,n}}{2M_{\text{Pl}}^4} \frac{(\p_i v)^2}{a^2} + \left( \frac{\overline{n} \rho_{M,n}}{M_{\text{Pl}}^4} \frac{\p^2 \chi}{a^2} - \delta \dot{\rho}_M - 3 H (1+ c_M^2) \delta \rho_M \right) v \\
	& \quad  - \frac{c_M^2 M_{\text{Pl}}^4}{2 \overline{n} \rho_{M,n}} (\delta \rho_M)^2 - M_{\text{Pl}}\, \alpha\, \delta \rho_M  - \left( \omega_3 -2 \omega_8 +2 \omega_9 \right) \frac{(\p_i \alpha)^2}{a^2 M_{\text{Pl}}^2} + \omega_4 \frac{\alpha^2}{M_{\text{Pl}}^2}  \\
	& \quad - \bigg( (3 H \omega_1 - 2 \omega_4) \frac{\delta \phi}{\phi} - \left( \omega_3 - 3\omega_8 + \omega_9 \right) \frac{\p^2 (\delta \phi)}{a^2 \phi} - \left( \omega_3 - \omega_8 + \omega_9 \right) \frac{\p^2 \dot{\psi}}{a^2 \phi \Lambda} + \omega_6 \frac{\p^2 \psi}{a^2 \Lambda} \bigg) \frac{\alpha}{M_{\text{Pl}}}  \\
	& \quad  -  ( \omega_3 - 4 \omega_8 ) \frac{(\p_i \delta \phi)^2}{ 4a^2 \phi^2} + \omega_5 \frac{(\delta \phi)^2}{\phi^2} - \frac12 \left( \left( \omega_2 + \omega_6 \phi \right) \psi -  (\omega_3 - 2\omega_8) \dot{\psi} \right) \frac{\p^2 (\delta \phi)}{a^2 \phi^2 \Lambda} - \frac{\omega_3}{4} \frac{(\p_i \dot{\psi})^2}{a^2 \phi^2 \Lambda^2} \\
	&  \quad + \frac{\omega_7}{2} \frac{(\p_i \psi)^2}{a^2 \Lambda^2} + \left( \omega_1 \frac{\alpha}{M_{\text{Pl}}} + \omega_2 \frac{\delta \phi}{\phi} \right) \frac{\p^2 \chi}{a^2 M_{\text{Pl}}^2} \bigg] \,.
	\label{eq:SS2}
\eal\eeq
The coefficients $\omega_I$ are given in Appendix \ref{ssec:DefCoefsScal}. We have defined them in a way that highlights the differences with the result of GP theory \cite{DeFelice:2016uil}, in which case the coefficients $\omega_8$ and $\omega_9$ vanish. Although these parameters do not introduce any new operators (as it occurred in the vector sector), they do have the effect of ``detuning'' the relative coefficients among some of the terms. In \eqref{eq:SS2} we introduced
\begin{equation}
	c_M^2 \equiv \frac{\overline{n} \rho_{M,nn}}{\rho_{M,n}} \,,
\end{equation}
which we recall is the GR value of the matter fluid speed of sound. As anticipated previously, the actual speed of sound in EPN will turn out to be different.

The counting of degrees of freedom is again most easily performed by examining the equations of motion. Varying the action with respect to $\alpha$, $\chi$, $\delta \phi$, $\p \psi$, $v$ and $\delta \rho_M$, respectively, we derive
\begin{align}
	&(3 H \omega_1 -2\omega_4) \frac{\delta \phi}{\phi} -2\omega_4 \frac{\alpha}{M_{\text{Pl}}} + M_{\text{Pl}}^2 \delta \rho_M + \frac{k^2}{a^2 \Lambda^2} \left[ \mathcal{Y}_1 + \omega_1 \frac{\Lambda^2}{M_{\text{Pl}}^2} \chi - \omega_6 \Lambda \psi \right] = 0 \,, \label{eq:eomalpha} \\
	&\frac{(\rho_M + P_M)}{M_{\text{Pl}}} v + \omega_1 \alpha + M_{\text{Pl}} \omega_2 \frac{\delta \phi}{\phi} = 0 \,, \label{eq:eomchi} \\
	&(3 H \omega_1 - 2 \omega_4 )\frac{\alpha}{M_{\text{Pl}}} - 2 \omega_5 \frac{\delta \phi}{\phi} + \frac{k^2}{a^2 \Lambda^2} \left[ \frac{1}{2} \mathcal{Y}_2 + \omega_2 \frac{\Lambda^2}{M_{\text{Pl}}^2} \chi - \frac{\Lambda}{2} ( \omega_2 + \omega_6 \phi ) \frac{\psi}{\phi} \right] = 0 \,, \label{eq:eomdphi} \\
	&\frac{\dot{\mathcal{Y}}_3}{H} + \left( 1 - \frac{\dot{\phi}}{H \phi} \right) \mathcal{Y}_3 + \frac{\Lambda^2}{H} \left\lbrace \omega_2 \frac{\delta \phi}{\phi} + 2 \omega_7 \frac{\phi \psi}{\Lambda} + \omega_6 \left( 2 \frac{\alpha \phi}{M_{\text{Pl}}} + \delta \phi \right) \right\rbrace = 0 \,, \label{eq:eomppsi} \\
	&\dot{\delta \rho}_M + 3 H (1 + c_M^2) \delta \rho_M + \frac{k^2}{a^2} \frac{(\rho_M + P_M)}{M_{\text{Pl}}^4} ( v+ \chi) = 0 \,, \label{eq:eomv} \\
	&\alpha M_{\text{Pl}} + c_M^2 \left( 3 H v + \frac{M_{\text{Pl}}^4}{(\rho_M + P_M)} \delta \rho_M \right) - \dot{v} = 0 \,, \label{eq:eomdrhoM}
\end{align}
where
\begin{align}
	\mathcal{Y}_1 &\equiv \frac{\Lambda^2}{\phi} \left[  \left(\omega_3 - 3\omega_8 + \omega_9 \right) \delta \phi + 2 \left(\omega_3 - 2\omega_8 + \omega_9 \right) \frac{\alpha \phi}{M_{\text{Pl}}} + \left(\omega_3 - \omega_8 + \omega_9 \right) \frac{\dot{\psi}}{\Lambda} \right] \,, \label{eq:Y1} \\
	\mathcal{Y}_2 &\equiv \frac{\Lambda^2}{\phi} \left[  \left(\omega_3 - 4\omega_8 \right) \delta \phi + 2 \left(\omega_3 - 3\omega_8 + \omega_9 \right) \frac{\alpha \phi}{M_{\text{Pl}}} + \left(\omega_3 - 2 \omega_8 \right) \frac{\dot{\psi}}{\Lambda} \right] \,, \label{eq:Y2} \\
	\mathcal{Y}_3 &\equiv \frac{\Lambda^2}{\phi} \left[  \left(\omega_3 - 2\omega_8 \right) \delta \phi + 2 \left(\omega_3 - \omega_8 + \omega_9 \right) \frac{\alpha \phi}{M_{\text{Pl}}} + \omega_3 \frac{\dot{\psi}}{\Lambda} \right] \,. \label{eq:Y3}
\end{align}
Note that $\mathcal{Y}_1=\mathcal{Y}_2=\mathcal{Y}_3$ when $\omega_8=\omega_9=0$. The equations for the variables $\alpha$, $\delta \phi$, $\chi$ and $v$ can be solved algebraically in terms of $\psi$ and $\delta \rho_M$. These expressions can be plugged back into \eqref{eq:eomppsi} and \eqref{eq:eomdrhoM} leading to a system of two second-order differential equations for $\psi$ and $\delta \rho_M$. This concludes the proof that the covariant EPN theory is completely free from unwanted degrees of freedom at the level of linear perturbations about the FLRW background.

To determine the dispersion relations and stability conditions we proceed as in Sec.\ \ref{sssec:SpecExScal}. After integrating out the non-dynamical modes, and focusing from the outset on the long wavelength approximation, we may recast the resulting action in the same form as in eq.\ \eqref {eq:SpecExSMmatrix} for the propagating fields $\psi$ and $\delta \rho_M$. Recall that $\bm{K}$, $\bm{M}$, $\bm{G}$ and $\bm{B}$ are all independent of $k$ in this approximation. Moreover, we find that the kinetic matrix $\bm{K}$ is still diagonal. The no-ghost conditions are therefore immediately inferred from its entries, which we denote by $Q_{S,\psi}$ and $Q_{S,M}$. We find
\beq\bal
Q_{S,\psi}&=\frac{1}{\Lambda^2 \phi^2 \left[ ( \omega_1 -2 \omega_2)^2 \omega_3 - 4 ( \omega_1 - \omega_2) ( ( \omega_1 - 2 \omega_2 ) \omega_8 + \omega_2 \omega_9 ) - 2 ( \rho_M + P_M) ( (\omega_8 - \omega_9 )^2 + \omega_3 \omega_9) \right]^2} \\
&\quad\times \bigg\{ \big(3H \omega_1^2 - 2(\omega_1-\omega_2) \omega_4\big) \big( (\omega_1 - 2 \omega_2) \omega_3 - 2(\omega_1 - \omega_2) \omega_8 - 2 \omega_2 \omega_9 \big)  \\
&\quad \Big[ 2(\rho_M + P_M) (\omega_3 - \omega_8 + \omega_9) \omega_9 - \frac12 (\omega_1 - \omega_2) \left( (\omega_1 - 2 \omega_2) \omega_3 - 2(\omega_1 - \omega_2) \omega_8 - 2 \omega_2 \omega_9 \right) \Big] \\
&\quad + \Big[ 6H \big( (\omega_1-\omega_2) \omega_8^2 + ((\omega_1 + \omega_2)\omega_3 - 2\omega_2 \omega_8)\omega_9 + (\omega_1 + \omega_2) \omega_9^2 \big) \big(\omega_8^2 + 2\omega_8 \omega_9 - (\omega_3 + \omega_9) \omega_9 \big) \\
&\quad + 4 \omega_4 \omega_9^2 (\omega_3 - \omega_8 + \omega_9)^2 \Big] (\rho_M + P_M)^2 \bigg\} \,,
\eal
\eeq
\beq
Q_{S,M}=\frac{a^2}{2} \frac{M_{\text{Pl}}^4}{(\rho_M + P_M)}\,\frac{1}{1-\Delta} \,,
\eeq
where
\begin{equation}
	\Delta \equiv 2 (\rho_M + P_M) \frac{(\omega_8 - \omega_9)^2 + \omega_3 \omega_9}{\omega_3 ( \omega_1 - 2 \omega_2)^2 - 4 ( \omega_1-\omega_2) \left[ (\omega_1 - 2 \omega_2) \omega_8 + \omega_2 \omega_9 \right]} \,.
\end{equation}
We observe that the ``new'' coefficients $\omega_8$ and $\omega_9$ have the interesting effect of inducing a modification of the kinetic term of the Proca scalar mode $\psi$ that depends on the matter density and pressure. These coefficients similarly affect the matter fluid's kinetic term through the parameter $\Delta$. In particular, we see that $Q_{S,M}$ now depends on the EPN Lagrangian parameters, whereas in GP the result would coincide with that of pure GR.

From the long-wavelength expansion of the dispersion relations we obtain the sound speeds $c_{S,\psi}^2$ and $c_{S,M}^2$ respectively for the Proca scalar and the fluid. The fluid speed of sound can be written as
\beq \label{eq:gen model phonon speed}
c_{S,M}^2 = (1 - \Delta) c_M^2 \,,
\eeq
showing that the parameter $\Delta$ has the interesting effect of modifying the GR (and also GP) value of the sound speed. On the other hand, the precise expression for $c_{S,\psi}^2$ is  not particularly illuminating, so we choose to omit it.
However, one can get insight on the difference between the GP and the EPN predictions by going to a minimal example where we set some of the coefficients to $0$ for simplicity's sake. A particularly simple example that is consistent with the GP constraints \cite{DeFelice:2016yws,DeFelice:2016uil} is reached when taking $\omega_2 = \omega_4 = \omega_6 = 0$. As a result, $\omega_1$ and $\omega_4$ are written solely in terms of $q_T$, and hence the problem is fully described by the variables $\{q_T, q_V, \omega_7,\omega_8,\omega_9\}$. Furthermore, we will redefine the variables $\omega_8$ and $\omega_9$ into the dimensionless $W_8 \equiv \omega_8/(q_V \phi^2)$ and $W_9 \equiv \omega_9/(q_V \phi^2)$. With these definitions, we can write
\begin{equation}
\label{eq:cpsi}
	c_{S,\psi}^2 = c_{S,\psi}^{\text{(GP)}2} \left[ 1 +2 W_8 - \frac{q_V}{4q_T^2} \frac{\phi^2}{M_{\text{Pl}}^2} \frac{\rho_M + P_M}{M_{\text{Pl}}^2 H^2} \left( 2W_9 - (W_8-W_9)^2 \right) \right]^2  \Upsilon^{-1} \,,
\end{equation}
where
\begin{align}
 c_{S,\psi}^{\text{(GP)}2}  =& - \frac{\omega_7 \phi^2}{6 M_{\text{Pl}}^2 H^2 q_T} \,, \\
	\Upsilon =& \left[ 1 + W_8 - \frac{q_V}{4q_T^2} \frac{\phi^2}{M_{\text{Pl}}^2} \frac{\rho_M + P_M}{M_{\text{Pl}}^2 H^2} \left( 2W_9 (1-W_9) + (W_8+W_9)^2 \right) \right] \\
	& \times \left[ 1 + W_8 + \frac{q_V}{4q_T^2} \frac{\phi^2}{M_{\text{Pl}}^2} \frac{\rho_M + P_M}{M_{\text{Pl}}^2 H^2} \left( W_9 (-2 + W_9) + W_8^2 \right) \right] \,. \nonumber
\end{align}
One can see that the positivity of $c_{S,\psi}^{\text{(GP)},\, 2} $ necessarily implies $\omega_7 < 0$, whereas this condition is relaxed to be $\omega_7 / \Upsilon < 0$ in the EPN case.

One can now turn to the masses and derive their expressions in all generality, however once again their expressions are not particularly illuminating.  However under the same limiting choice of coefficients as we did previously, one can check explicitly that the scale of the mass of the fluid is set by the Hubble parameter $H$. In principle, we would require the fluid's mass to be positive to avoid tachyonic instabilities but as we have already discussed in the cosmological context, a negative square mass of order $H^2$ is not worrisome. As for the mass of the $\psi$-mode (or vector helicity-0 mode), it happens to vanish for that particular choice of parameters, however relaxing this choice (for instance choosing a non-zero $\omega_6$), one can check that the mass of this mode is also of order $H$, and there is therefore no risk of a faster than $H^2$ tachyonic instability in the scalar sector.

The stability of the matter fluid is easy to analyze. The condition $c_{S,M}^2>0$ is equivalent to $\Delta<1$, which in turn implies the null energy condition, $\rho_M + P_M>0$, in order to have $Q_{S,M}>0$. For the Proca mode $\psi$ to be stable we similarly require $Q_{S,\psi}>0$ and $c_{S,\psi}^2>0$. While these conditions are difficult to dissect given the long expressions, it is worth remembering that they include the results of GP theory as a particular case, in which context it has been shown that stability can be achieved for a wide range of parameters \cite{DeFelice:2016uil}.


\section{Discussion}
\label{sec:discussion}

Our aim in this paper was to explore further generalizations of the standard Einstein-Proca theory of a massive spin-1 field coupled to gravity, beyond those given by the GP class of models, motivated by the recent discovery of PN theory. In spite of being qualitatively different in their constraint structure, we have shown that PN can be extended simply by adding a subset of the GP operators, at least in flat spacetime. The full result for the flat-space Lagrangian is given in eq.\ \eqref{eq:LextPNflat}. This ``extended PN'' model is interesting already from a formal point of view in that it provides a link between GP and PN, both of which may be recovered by a particular choice (in a limiting sense in the case of GP) of the coefficients that define the theory.

Aside from generalized massive gravity \cite{deRham:2014lqa,deRham:2014gla}, the existence of another complete covariantization of EPN (and of PN) remains an open question. Here we have taken a first step toward its solution by proposing a covariant model (eq.\ \eqref{eq:SextPN}). With this covariantization in mind, one can find a null eigenvector (given in \eqref{eq:full EPN NEV}) for the full Hessian of the first family of operators (namely $\L_1(\X)$) proving that it enjoys a constraint at all orders. For the other family of operators (namely $\L_{n\ge 2}(\X)$), we showed that the same ansatz for the null eigenvector correctly annihilates the Hessian matrix at leading order in an expansion in the strong coupling scale $\Lambda$ but the process fails when pushing it to higher order. The result is nevertheless non-trivial and provides a hint that a full covariantization is in principle feasible. We note also that the failure of the constraint only occurs from mixing with the gravitational degrees of freedom and is thus Planck scale-suppressed. Moreover, we show that our proposed eigenvector remains a null one for the Hessian of the full theory (including the gravitational degrees of freedom), on any background where the tensor $\nabla_\mu A_\nu$ is symmetric. This directly implies the presence of a constraint that would remove the unwanted ghostly additional degree of freedom at linear order in perturbations about any such backgrounds, including on FLRW.

These results for the covariant EPN theory are by themselves sufficient to motivate the study of the predictions of the model in the context of cosmology. This is so because the theory describes the correct number of degrees of freedom at the level of cosmological backgrounds, defined by eqs.\ \eqref{eq:lineFLRW2} and \eqref{eq:vector bkgd2}, as well as at the level of linear perturbations about these solutions. In addition to establishing this result, in Sec.\ \ref{sec:fulltheory} we also derived the dispersion relations for the propagating variables in the presence of perfect fluid matter. Interestingly, EPN has some qualitative differences relative to GP in the dynamics of perturbations. Two particular results to highlight are that the Proca vector mode exhibits a non-linear dispersion relation (cf.\ eq.\ \eqref{eq:gen model vec disp rel}) and that the sound speed of the longitudinal matter perturbation (the phonon) is modified in the EPN set-up relative to its GR value (cf.\ eq.\ \eqref{eq:gen model phonon speed}). We also found that the kinetic coefficient of the phonon differs in EPN from its GR and GP values, an effect which may in principle percolate to higher-point interactions and hence be potentially observable. While we did not explore explicit solutions in this general model, we remark again that EPN contains GP as a particular case, in which set-up consistent cosmological solutions do exist. It would be interesting to perform a dedicated study of solutions and comparison with data within the complete theory.

In addition to investigating the possibility of covariantizing the full EPN theory, we have also considered the option that a subclass of the theory may admit a simpler covariantization, even if only a partial one in the sense we have described. Our so-called special model of Sec.\ \ref{sec:SpecEx} shows that this is the case, providing a particularly neat set-up with few unspecified functions and which has the virtue that the Proca field interacts with gravity only through minimal coupling terms. To our knowledge, this is the first instance of a generalized Einstein-Proca theory (i.e.\ models with non-trivial derivative self-interactions beyond those given by contractions of the Maxwell field strength and/or the undifferentiated field) with this property. As with the general model, the caveat is that the covariantization scheme is only a partial one, but it is again sufficient for cosmological applications as long as one is interested in linear perturbations about homogeneous and isotropic backgrounds. Our results of Sec.\ \ref{sec:SpecEx} show that the special model indeed describes the expected dynamical degrees of freedom. Moreover, we have shown that explicit solutions exist such that all the dynamical variables are ghost-free, gradient-stable and subluminal. We believe that these results motivate further scrutiny of the set-up.

\section*{Acknowledgements}

The work of CdR is supported by STFC grants ST/P000762/1 and ST/T000791/1. CdR thanks the Royal Society for support at ICL through a Wolfson Research Merit Award. SGS and CdR's work at ICL were supported by the European Union's Horizon 2020 Research Council grant 724659 MassiveCosmo ERC--2016--COG, and by a Simons Foundation award ID 555326 under the Simons Foundation's Origins of the Universe initiative, `\textit{Cosmology Beyond Einstein's Theory}'. CdR is also supported by a Simons Investigator award 690508.
LH is supported by funding from the European Research Council (ERC) under the European Unions Horizon 2020 research and innovation programme grant agreement No 801781 and by the Swiss National Science Foundation grant 179740. VP is funded by the Imperial College President's Fellowship.

\appendix

\section{Recovering GP from EPN}
\label{sec:GPfromEPN}

In this Appendix, we will show how one can recover most of the  GP from the Extended Proca-Nuevo Lagrangian given in \eqref{eq:LextPNflat}, in the limit where $\tilde \Lambda \to \infty$ while keeping the scale $\Lambda$ finite and the vector field mass finite.

We start with the extended PN Lagrangian \eqref{eq:LextPNflat} written in the form
\beq
 \label{eq:LextPNflatc}
	\L_{\text{EPN}} =\tilde\L_{\rm PN} + \L_{\rm GP}\,,
\eeq
with
\ba
\tilde\L_{\rm PN} &=& \tilde{\Lambda}^4 \sum_{n=0}^4 \alpha_n(\tilde{X}) \L_n[\tilde{\K}[A]] \,,
\ea
and where  $\L_{\rm GP}$ includes all the GP interactions aside from the generic function $f(F\mn,\tilde F\mn, X)$,
\ba
\L_{\rm GP}&=& \Lambda^4 \sum_{n=1}^4 d_n(X) \frac{\L_n[\p A]}{\Lambda^{2n}} \,.
\ea
Assuming analyticity of the functions $\alpha_n$ and $d_n$,
the mass term of the vector field is given by
\ba
m^2=\tilde \Lambda^2 \alpha_0'(0)\,.
\ea
Keeping the mass of the vector field finite in the $\tilde \Lambda \to \infty$ limit therefore requires scaling the coefficient $\alpha_0'(0)$ as $\alpha_0'(0)\to m^2 / \tilde \Lambda^2$. Since $\L_n(\tilde K)\sim \mathcal{O}\left( (\p A)^{n}/\tilde \Lambda^{2n} \right)$, we see that in the limit $\tilde \Lambda \to \infty$, keeping the scale $m$ fixed, the only relevant terms of PN origin are, up to irrelevant constant and total derivatives
\ba
\tilde\L_{\rm PN}
 \xrightarrow[\tilde \Lambda \rightarrow \Lambda]{} -\frac 12 m^2 A^2+\alpha_0''(0)A^4
 -\frac 12 \alpha_1'(0) A^2 \p A
 + \frac 14 \alpha_1(0) F\mn^2-\frac 12 \alpha_2(0)F\mn^2\,.
\ea
These are all of GP nature, so added to $\L_{\rm GP}$, we directly deduce that in the limit $\tilde \Lambda \to \infty$, keeping the scales $m$ and $\Lambda$ fixed, the extended PN Lagrangian given in  \eqref{eq:LextPNflatc} includes all the GP interactions aside from the generic function $f(F\mn,\tilde F\mn, X)$.

\section{Null eigenvector}
\label{sec:NEV}

\subsection{Null eigenvector on a fixed background}
\label{ssec:NEVFix}

In this section, we shall prove that the EPN theory defined in  \eqref{eq:covEPN} admits a constraint about any fixed background (no matter how curved and spacetime-dependent). To prove this, we simply need to show that the  vector $V_{\mu}$ defined in \eqref{eq:NEVcovariant} is indeed a null eigenvector for this EPN theory on any background. To be more precise, we have defined  $\mathcal{H}_{\alpha \beta}^{(n)}$ in eq.~\eqref{eq:HessianOrderN} to be the Hessian matrix corresponding to the Lagrangian $\L_n[\X]$ for $n=1,...,4$, as expressed in \eqref{eq:L0}--\eqref{eq:L3}. In what follows, we shall show that $V^{\alpha}$ is a null eigenvector for each $\mathcal{H}_{\alpha \beta}^{(n)}$ for $n=0,...,3$. The case $n=0$ is trivial since it is purely a potential term. For the other non-trivial Lagrangians, it turns out to be easier to consider them as functions of $\X$ rather than $\K$. This change of variable is always possible since the set $\{ \L_n[\X] \}$ is linearly related the set $\{ \L_n[\K] \}$ as long as one considers $\textit{all}$ interactions, i.e. including $\L_4$. In what follows we shall thus simply prove that
\begin{equation}
	\mathcal{H}_{\alpha \beta}^{(n)} V^{\alpha} = 0, \quad \text{for } n=1,...,4 \,.
\end{equation}
The case $n=1$ was proven in the main text \eqref{eq:HV1equals0}. Let us now turn to the proof that $V$ is indeed the NEV for $\L_n[\X]$, with $n=2,3,4$. To begin with, we make use of the following identity
\begin{equation}
	\frac{\p [\X^n] }{\p \dot{A}^{\alpha}} = n \Lambda^{-2} \left( \X^{n-2} \right)^0_{\phantom{0} \mu} \left( \delta^{\mu}_{\alpha} + \Lambda^{-2} \nabla^{\mu} A_{\alpha} \right) \,.
\end{equation}

\begin{itemize}
	\item \textbf{$\L_2[\X]$}

One can start by showing that the momentum associated with this Lagrangian reads
\begin{equation}
	p^{(2)}_{\alpha} = 2 \Lambda^{2} \left\lbrace [\X] V_{\alpha} - \left( \delta^{0}_{\alpha} + \Lambda^{-2} \nabla^{0} A_{\alpha} \right) \right\rbrace \,,
\end{equation}
it is then easy to prove that $V$ is indeed the correct null eigenvector,
\ba
	\mathcal{H}^{(2)}_{\alpha \beta} V^{\alpha} = 2 \Lambda^2 \left\lbrace \frac{\p [\X]}{\p \dot{A}^{\beta}} V_{\alpha} V^{\alpha} + [\X] \frac{\p V_{\alpha}}{\p \dot{A}^{\beta}} V^{\alpha} - \frac{1}{\Lambda^{2}} g^{00} g_{\alpha \beta} V^{\alpha} \right\rbrace = 2 \Lambda^2 g^{00} \left( \frac{\p [\X]}{\p \dot{A}^{\beta}} -  \frac{V_{\beta}}{\Lambda^2} \right) = 0 \,.\qquad
\ea

	\item \textbf{$\L_3[\X]$}
	
The momentum associated with $\L_3[\X]$ is given by
\begin{equation}
	\Lambda^{-2} p^{(3)}_{\alpha} = 3 \left( [\X]^2 - [\X^2] \right) V_{\alpha} - 6 [\X] \left( \delta^0_{\alpha} + \Lambda^{-2} \nabla^0 A_{\alpha} \right) + 6 \X^0_{\phantom{0} \mu} \left( \delta^{\mu}_{\alpha} + \Lambda^{-2} \nabla^{\mu} A_{\alpha} \right) \,.
\end{equation}
We will make use of the following identities
\begin{align}
	&V^{\alpha} \left( \delta^{\mu}_{\alpha} + \Lambda^{-2} \nabla^{\mu} A_{\alpha} \right) = \X^{0 \mu} \,, \\
	&\frac{\p \X^{0\mu}}{\p \dot{A}^{\beta}} \X^0_{\phantom{0} \mu} = \Lambda^{-2} g^{00} \left( \delta^0_{\beta} + \Lambda^{-2} \nabla^0 A_{\beta} \right) \,,
\end{align}
so as to derive the following matrix product between the Hessian and the vector $V$,
\begin{align}
	\mathcal{H}_{\alpha \beta}^{(3)} V^{\alpha} =& 6 \left\lbrace [\X] V_{\beta} - \left( \delta^0_{\beta} + \Lambda^{-2} \nabla^0 A_{\beta} \right) \right\rbrace V_{\alpha} V^{\alpha} + \frac{3}{2} \Lambda^2 \left( [\X]^2 - [\X^2] \right) \frac{\p \left( V_{\alpha} V^{\alpha} \right)}{\p \dot{A}^{\beta}} \nonumber \\
	& - 6 V_{\beta} \left( \delta^0_{\alpha} + \Lambda^{-2} \nabla^0 A_{\alpha} \right) V^{\alpha} - 6 [\X] g^{00} V_{\beta} + 6 \Lambda^2 \frac{\p \X^0_{\phantom{0} \mu}}{\p \dot{A}^{\beta}} \left( \delta^{\mu}_{\alpha} + \Lambda^{-2} \nabla^{\mu} A_{\alpha} \right) V^{\alpha} \nonumber \\
	& + 6 \X^{00} V_{\beta} \nonumber \\
	=& 6 \left[ - g^{00} \left( \delta^0_{\beta} + \Lambda^{-2} \nabla^0 A_{\beta} \right) - \X^{00} V_{\beta} + \Lambda^2_2 \frac{\p \X^0_{\phantom{0} \mu}}{\p \dot{A}^{\beta}} \X^{0 \mu} + \X^{00} V_{\beta} \right] \nonumber \\
	=& 0  \,.
\end{align}

\item \textbf{$\L_4[\X]$}

The canonical momentum coming from the Lagrangian at order 4 reads
\begin{equation}
	\Lambda^{-2} p_{\alpha}^{(4)} = 4 \L_3[\X] V_{\alpha} - 12 (\L_2[\X] \delta^0_{\mu} -2 ([\X] \X^0_{\phantom{0} \mu} - f^0_{\phantom{0} \mu})) \left( \delta^{\mu}_{\alpha} + \Lambda^{-2} \nabla^{\mu} A_{\alpha} \right) \,,
\end{equation}
and the eigenvalue equation follows directly
\begin{align}
	\mathcal{H}_{\alpha \beta}^{(4)} V^{\alpha} =& 4 \Lambda^{-2} p_{\beta}^{(3)} g^{00} + 4 \Lambda^2 \L_3[\X] \frac{\p V_{\alpha}}{\p \dot{A}^{\beta}} V_{\alpha} \nonumber \\
	& - 12 \left[ \Lambda^{-2} p_{\beta}^{(2)} \delta^0_{\mu} - 2 \Lambda^{-2} p_{\beta}^{(1)} \X^0_{\phantom{0} \mu} - 2 \Lambda^2 [\X] \frac{\p \X^0_{\phantom{0} \mu}}{\p \dot{A}^{\beta}} + 2 g^{00}  \left( g_{\mu \beta} + \Lambda^{-2} \nabla_{\mu} A_{\beta} \right) \right. \nonumber \\
	& \qquad \quad \left. \vphantom{\frac{\p \X^{0\mu}}{\p \dot{A}^{\beta}}} + 2 \delta^0_{\mu}  \left( \delta^{0}_{\beta} + \Lambda^{-2} \nabla^{0} A_{\beta} \right) \right] \X^{0 \mu} \nonumber \\
	&- 12 (\L_2[\X] \delta^0_{\mu} -2 ([\X] \X^0_{\phantom{0} \mu} - f^0_{\phantom{0} \mu})) g^{0 \mu} V_{\beta} \nonumber \\
	=& 4 g^{00} \left[ 3 \L_2[\X] V_{\beta} - 6 [\X] \left( \delta^{0}_{\beta} + \Lambda^{-2} \nabla^{0} A_{\beta} \right) + 6 \X^0_{\phantom{0} \mu} \left( \delta^{\mu}_{\beta} + \Lambda^{-2} \nabla^{\mu} A_{\beta} \right) \right] \nonumber \\
	&-24 \left[ \left\lbrace [\X] V_{\beta} - \left( \delta^{0}_{\beta} + \Lambda^{-2} \nabla^{0} A_{\beta} \right) \right\rbrace \X^{00} - f^{00} V_{\beta} - g^{00} [\X] \left( \delta^{0}_{\beta} + \Lambda^{-2} \nabla^{0} A_{\beta} \right) \right. \nonumber \\
	& \qquad \quad \left. + g^{00} \X^{0 \mu} \left( \delta_{\mu \beta} + \Lambda^{-2} \nabla_{\mu} A_{\beta} \right) + \X^{00} \left( \delta^{0}_{\beta} + \Lambda^{-2} \nabla^{0} A_{\beta} \right) \right] \nonumber \\
	&- 12 \left( g^{00} \L_2[\X] - 2 [\X] \X^{00} + 2 f^{00} \right) V_{\beta} \nonumber \\
	=& 0 \,.
\end{align}
This concludes the proof that $V$ is the common null eigenvector to $\L_1$, $\L_2$, $\L_3$ and $\L_4$,
\begin{equation}
	\Rightarrow \mathcal{H}^{(n)}_{\alpha \beta} V^{\alpha} = 0, \quad \text{for} \quad n=1,2,3,4 \,.
\end{equation}
	
\end{itemize}

\subsection{Null eigenvector on a dynamical background}
\label{ssec:NEVDyn}

In this section, we will consider the background to be dynamical, and hence extend the dynamical phase space including those contained in the gravitational sector. The NEV $V_{\mu}^{\ast}$ defined in \eqref{eq:Vast} is now embedded in a higher-dimensional vector $\bm{\mathcal{V}}=(V_{\mu}^{\ast},0)$ where the null entries run through the metric components. We have proven that $\mathcal{H}^{(n)}_{\alpha \beta} V^{\alpha} = 0$ for $n=1,...,3$ on a non-dynamical background. When coupling EPN to gravitational degrees of freedom, the vector $V_{\mu}^{\ast}$ is related to $V_{\mu}$ by the linear transformation $V_{\mu}^{\ast}=(M^{-1})_{\mu}^{\phantom{\mu}\nu} V_{\nu}$ and thus it is immediate to see that $\mathcal{H}^{\ast,(n)}_{\alpha \beta} V^{\ast,\alpha} = 0$ for $n=1,...,3$, i.e. $\bm{\mathcal{V}}$ annihilates the pure vector sector. This is trivially true for the pure metric sector. In order to prove that the higher-dimensional vector $\bm{\mathcal{V}}$ is the correct NEV on a dynamical background, one has to check that it also annihilates the mixed vector-metric sector. The Hessian matrix for the mixed vector-metric sector is defined to be
\ba
\mathcal{H}^{*(n)}_{\mu,ij} = \frac{\p p^{*(n)}_{\mu}}{\p \dot{\gamma}^{ij}}=
\Lambda^4 \frac{\p^2 \L_n[\X]}{\p \dot{A}^{\ast\mu}\p \dot{\gamma}^{ij}}\,.
\ea
We have previously shown that $V_{\mu}^{\ast}$ is indeed a NEV for the Hessian $\mathcal{H}^{*(1)}_{\mu,ij}$ and hence   $\bm{\mathcal{V}}$ is a NEV of the full Hessian associated with $\L_1[\X]$.
We will now prove that even though $\bm{\mathcal{V}}$ fails to remain a NEV for $\L_2$ (and $\L_3$), it is possible to add non-minimal couplings to $\L_2$ such that symbolically $\mathcal{H} \bm{\mathcal{V}}$ vanishes in all sectors at leading order in $(\nabla A) / \Lambda^2$. This seems to indicate that one could possibly add further non-minimal couplings to push the constraint to the next order and so on in an infinite series. However, this is only  postulated at this stage and proving such a statement in generality is beyond the scope of this work. Nevertheless, our results are  interesting in their own right and we will further show that it immediately follows that $\bm{\mathcal{V}}$ is the NEV of $\L_2^{(\text{non-min})}$ on any background such that $\nabla_{\mu} A_{\nu}$ is symmetric, e.g. FLRW. Even though the covariantization fails on a generic dynamical background, this is a proof that EPN can be considered for cosmology. An estimation for the mass of the resulting ghost on background where the field strength tensor acquired a non-vanishing vev is given in the main text.

\paragraph{$\L_2$ without non-minimal couplings.}

We start by computing the Hessian matrix associated with $\L_2$ in the mixed vector-metric sector. First note that with the covariantization introduced in \eqref{eq:SextPN}, the time-derivatives of the spatial metric do not only enter through the curvature, but also through the covariant derivative of the vector field. To include their contributions, we first consider the following derivatives
\begin{equation}
 \frac{\p \Gamma^{\beta}_{\mu \alpha}}{\p \dot{\gamma}_{kl}} = \frac14 g^{\beta \lambda} \left( \delta^0_{\mu} \delta^k_{\lambda} \delta^l_{\alpha} + \delta^0_{\alpha} \delta^k_{\lambda} \delta^l_{\mu} - \delta^0_{\lambda} \delta^k_{\mu} \delta^l_{\alpha} \right) + \left\{ k \leftrightarrow l \right\} \,.
\end{equation}
Throughout the rest of this appendix, we will consider $A^{\mu}$ (with upper index) to be constant with respect to $\gamma^{ij}$ and as a result $\p_{\mu} A_{\nu}$ will also contribute when differentiating with respect to $\dot{\gamma}^{ij}$. The derivative of $\nabla_\mu A^\nu$ is hence given by
\begin{align}
	\frac{\p (\nabla_{\mu} A^{\nu})}{\p \dot{\gamma}^{ij}} &= - \frac14 \gamma_{ik} \gamma_{jl} g^{\nu \lambda} \left( ( \delta^0_{\mu} \delta^k_{\lambda} - \delta^0_{\lambda} \delta^k_{\mu} ) A^l + \delta^k_{\mu} \delta^l_{\lambda} A^0 \right) + \left\lbrace i \leftrightarrow j \right\rbrace \nonumber \\
	&= - \frac14 \left[ (\delta^0_{\mu} \delta^{\nu}_i - g_{\mu i} g^{0 \nu}) A_j + g_{\mu i} \delta^{\nu}_j A^0 - 2 \delta^0_{\mu} \delta^{\nu}_i N_j A^0 + \delta^0_{\mu} g^{0 \nu} N_i N_j A^0 \right] + \left\lbrace i \leftrightarrow j \right\rbrace \,,
\end{align}
while that of $\nabla_{\mu} A_{\nu}$ follows trivially. Now, in order to compute the derivative of the momentum $p^{(2)}$ with respect to $\dot{\gamma}_{ij}$, we need
\begin{align}
	\frac{\p [\X]}{\p \dot{\gamma}^{ij}} &= \frac12 \frac{\p f\mn}{\p \dot{\gamma}^{ij}} \left(\X^{-1}\right)\mnup
	= \left[ \frac{\p(\nabla_{\mu} A_{\alpha})}{\p \dot{\gamma^{ij}}} + \Lambda^{-2} \frac{\p(\nabla_{\mu} A_{\nu})}{\p \dot{\gamma^{ij}}} \nabla_{\alpha} A^{\nu} \right] \left(\X^{-1}\right)^{\mu\alpha} \nonumber \\
	&= \Lambda^{-2} \frac{\p(\nabla_{\mu} A_{\nu})}{\p \dot{\gamma^{ij}}} \left(\X^{-1}\right)^{\mu\alpha} \left( \delta^{\nu}_{\alpha} + \Lambda^{-2} \nabla_{\alpha} A^{\nu} \right)
	= \Lambda^{-2} \frac{\p(\nabla_{\mu} A_{\nu})}{\p \dot{\gamma^{ij}}} V\mnup \nonumber \\
	&= \frac14 \Lambda^{-2} \left\lbrace A_i \left( V_j^{\phantom{j} 0} - V_j \right) + A^0 \left( 2 V_i N_j - V_{ij} - N_i N_j V^0 \right) \right\rbrace + \left\{ i \leftrightarrow j \right\}  \,,
\end{align}
where we have introduced
\begin{equation}
	V^{\mu \nu} = \left( \X^{-1} \right)^{\mu \alpha} \left( \delta^{\nu}_{\alpha} + \Lambda^{-2} \nabla_{\alpha} A^{\nu} \right) \,,
\end{equation}
such that
\begin{equation}
	V^{\mu} = V^{0 \mu} \,.
\end{equation}
On the hand, we have
\begin{equation}
	\frac{\p \left(\nabla^0 A_{\alpha} \right)}{\p \dot{\gamma}^{ij}} V^{\alpha} = \frac14 g^{00} \left( A^0 ( 2 V_i N_j - N_i N_j V^0 ) - A_i V_j \right) + \left\{ i \leftrightarrow j \right\} \,.
\end{equation}
Putting everything together we find that the contraction of the Hessian of $\alpha_{2,X} \L_2$ with the null eigenvector for $\L_1$ is now
\begin{align}
	\mathcal{H}^{(2)}_{\alpha,ij} V^{\alpha} &= \frac{\p p^{(2)}_{\alpha}}{\p \dot{\gamma}^{ij}} V^{\alpha} \nonumber \\
	&= 2 \alpha_{2,X} \Lambda^2 \left\lbrace \frac{\p [\X]}{\p \dot{\gamma}^{ij}} V_{\alpha} V^{\alpha} + [\X] \frac{\p V_{\alpha}}{\p \dot{\gamma}^{ij}} V^{\alpha} - \Lambda^{-2} \frac{\p \left(\nabla^0 A_{\alpha} \right)}{\p \dot{\gamma}^{ij}} V^{\alpha} \right\rbrace \nonumber \\
	&= 2 \alpha_{2,X} \Lambda^2 \left\lbrace g^{00} \frac{\p [\X]}{\p \dot{\gamma}^{ij}} - \Lambda^{-2} \frac{\p \left(\nabla^0 A_{\alpha} \right)}{\p \dot{\gamma}^{ij}} V^{\alpha} \right\rbrace \nonumber \\
	&= \frac12 \alpha_{2,X} g^{00} \left( V_i^{\phantom{i} 0} A_j + V_j^{\phantom{j} 0}  A_i - V_{ij} A^0 - V_{ji} A^0 \right) \nonumber \\
	&= - \alpha_{2,X} g^{00} \gamma_{ij} A^0 + \frac{1}{2\Lambda^2} \alpha_{2,X} g^{00} F_{(i}^{\phantom{(i}0}A_{j)} + \mathcal{O}((\nabla A)^{2}/\Lambda^{4}) \,,
	\label{eq:HV2}
\end{align}
which does not generically vanish. Considering this result in an operator expansion, or power expansion in $\nabla A/\Lambda^2$, we see that to leading order in that expansion, we get
\begin{equation}
	V\mn = g\mn + \frac{F\mn}{2\Lambda^2} - \frac{1}{8\Lambda^4} \left( F_{\mu}^{\phantom{\mu} \alpha} F_{\nu \alpha} - 4 \nabla_{\alpha} A_{[\mu} \nabla_{\nu]} A^{\alpha} \right) + \mathcal{O}((\nabla A)^{3}/\Lambda^{6}) \,,
\end{equation}
and it is clear that \eqref{eq:HV2} does not vanish at leading order in the operator expansion.
The previous result is surprising in itself and indeed the same occurs for GP at precisely the same level. The resolution in that case is the introduction of non-minimal couplings to gravity as already provided in \cite{Heisenberg:2014rta}.

\paragraph{Addition of non-minimal couplings}

In the context of EPN, the generalization of those non-minimal couplings is however much more challenging to find, particularly due to the fact that the constraint has to be satisfied non-linearly through mixing of orders. At this stage, there is no candidate for a straightforward and natural non-minimal coupling, however for lack of a better insight, we consider the inclusion of the following non-minimal coupling:
\begin{align}
	\sqrt{-g} \Lambda^2 \alpha_2[X] R &= \sqrt{-g} \Lambda^2 \alpha_2[X] g\mnup \left( \p_{\alpha} \Gamma^{\alpha}\mn - \p_{\nu} \Gamma^{\alpha}_{\alpha \mu} + \Gamma^{\alpha}\mn \Gamma^{\beta}_{\alpha \beta} - \Gamma^{\beta}_{\mu \alpha} \Gamma^{\alpha}_{\nu \beta} \right) \nonumber \\
	&= \sqrt{-g} \Lambda^2 \alpha_2[X] g\mnup \left( \nabla_{\alpha} \Gamma^{\alpha}\mn - \nabla_{\nu} \Gamma^{\alpha}_{\alpha \mu} + \mathcal{O}(\Gamma^2) \right) \nonumber \\
	&= \sqrt{-g} \Lambda^2 \alpha_2[X] \left( \nabla_{\alpha} \left( g\mnup \Gamma^{\alpha}\mn \right) - \nabla^{\mu} \Gamma^{\alpha}_{\alpha \mu} + \mathcal{O}(\Gamma^2) \right) \nonumber \\
	&= \sqrt{-g} \Lambda^2 \left( \Gamma^{\alpha}_{\alpha \mu} \nabla^{\mu} \alpha_2[X] - g\mnup \Gamma^{\alpha}\mn \nabla_{\alpha} \alpha_2[X] + \mathcal{O}(\Gamma^2) \right) \nonumber \\
	&= \sqrt{-g} \alpha_{2,X}[X] A_{\beta} \left( g\mnup \Gamma^{\alpha}\mn \nabla_{\alpha} A^{\beta} - \Gamma^{\alpha}_{\alpha \mu} \nabla^{\mu}A^{\beta} + \mathcal{O}(\Gamma^2) \right) \,.
\end{align}
Now defining $p^{(2,R)}$ as the momentum conjugate to $A$ with respect to the non-minimal coupling part of the Lagrangian at order 2,

\begin{equation}
	\Rightarrow p^{(2,R)}_{\alpha} = \frac{\p \left( \Lambda^2 \alpha_2[X] R \right)}{\p \dot{A}^{\alpha}} = \alpha_{2,X} A_{\alpha} \left( g\mnup \Gamma^0\mn - g^{\mu 0} \Gamma^{\nu}\mn \right) \,,
\end{equation}
and $\mathcal{H}^{(2,R)}_{\alpha,ij}$ as the contribution to the second-order Hamiltonian purely coming from the second-order non-minimal coupling to gravity,

\begin{align}
	\mathcal{H}^{(2,R)}_{\alpha,ij} &= \frac{\p p^{(2,R)}_{\alpha}}{\p \dot{\gamma}^{ij}} \nonumber \\
	&= - \gamma_{ik} \gamma_{jl} \alpha_{2,X} A_{\alpha} \left( g\mnup \frac{\p \Gamma^0\mn}{\p \dot{\gamma}_{kl}} - g^{\mu 0} \frac{\p \Gamma^{\nu}\mn}{\p \dot{\gamma}_{kl}} \right) \nonumber \\
	&= - \frac14 \gamma_{ik} \gamma_{jl} \alpha_{2,X} A_{\alpha} \left( \left(4 g^{0k} g^{0l} - 2 g^{00} g^{kl} \right) - 2 g^{00} g^{kl} \right) \nonumber \\
	&= \gamma_{ik} \gamma_{jl} \alpha_{2,X} A_{\alpha} \left( g^{00} g^{kl} - g^{0k} g^{0l} \right) \nonumber \\
	&= \alpha_{2,X} A_{\alpha} \left( g^{00} \left( \gamma_{ij} + g^{00} N_i N_j \right) - \left( - g^{00} N_i \right) \left( - g^{00} N_j \right) \right) \nonumber \\
	&= g^{00} \gamma_{ij} \alpha_{2,X} A_{\alpha} \,,
\end{align}
then, we get the following eigenstate equation

\begin{align}
	\Rightarrow \mathcal{H}^{(2,R)}_{\alpha,ij} V^{\alpha} &= g^{00} \gamma_{ij} \alpha_{2,X} A_{\alpha} V^{\alpha} \nonumber \\
	&= \alpha_{2,X} g^{00} \gamma_{ij} \left( A^0 + \frac{1}{2\Lambda^2} F^{0 \alpha} A_{\alpha} + \mathcal{O}((\nabla A)^{2}/\Lambda^{4}) \right) \,.
	\label{eq:HV2R}
\end{align}
Separately, neither \eqref{eq:HV2} nor \eqref{eq:HV2R} vanish at leading order in $(\nabla A)/\Lambda^{2}$. However, when adding these two contributions, we get a cancellation at leading order,
\begin{align}
	\Rightarrow \left( \mathcal{H}^{(2)}_{\alpha,ij} + \mathcal{H}^{(2,R)}_{\alpha,ij} \right) V^{\alpha} &= \frac12 g^{00} \alpha_{2,X} \left( 2 \gamma_{ij} A_{\alpha} V^{\alpha} + V_i^{\phantom{i} 0} A_j + V_j^{\phantom{j} 0}  A_i - V_{ij} A^0 - V_{ji} A^0 \right) \nonumber \\
	&= 0 + \frac{1}{2\Lambda^2} \alpha_{2,X} g^{00} \left( F_{(i}^{\phantom{(i}0} A_{j)} + \gamma_{ij} F^{0 \alpha} A_{\alpha} \right) + \mathcal{O}((\nabla A)^{2}/\Lambda^{4}) \,.
	\label{eq:sumHcoupl}
\end{align}
Now, this equation is vanishing at leading order in $(\nabla A)/\Lambda^{2}$ but not to higher order. From this we  conclude that by itself the minimal coupling $\alpha_2[X] R$ does help with the pushing the breaking of the constraint to a higher order but is not sufficient to ensure that the constraint will be satisfied to all orders. Other more general non-minimal couplings are currently under investigations but those are kept to another study since for what interests us in the context of cosmology is to ensure the absence of ghosts on cosmological backgrounds. In this context, the tensor $\nabla_{\mu}A_\nu$ is symmetric and the right hand side of \eqref{eq:sumHcoupl} then vanishes.  Indeed, if $\nabla A$ is symmetric then $f$ is nothing other than $(1 + \nabla A / \Lambda^2)^2$, meaning that $\chi$ reduces to the simple form $1 + \nabla A / \Lambda^2$. Finally, we have
\begin{equation}
	V\mnup = \left[ \left( 1 + \nabla A / \Lambda^2 \right)^{-1} \right]^{\mu \alpha} \left[ 1 + \nabla A / \Lambda^2 \right]_{\alpha}^{\phantom{\alpha}\nu} = g\mnup \,,
\end{equation}
which is simply the zeroth-order of the general formula, proving that the right hand side of \eqref{eq:sumHcoupl} vanishes for any configurations where the field strength tensor $F\mn$ vanishes.

\subsection{Special example with no non-minimal coupling}
\label{ssec:NEVspec}

We now establish whether the special example considered in Sec.\ \ref{sec:SpecEx} enjoys a constraint when coupled to gravity. We start by defining the momenta associated with the GP operators as
\begin{equation}
	p^{(n,{\rm GP})}_{\alpha} =\frac{\p \L_n[\nabla A]}{\p \dot{A}^{\alpha}} \,,
\end{equation}
so that we have
\begin{align}
	p^{(1,{\rm GP})}_{\alpha} &= \delta^0_{\alpha} \,, \\
	p^{(2,{\rm GP})}_{\alpha} &= 2 \left( \delta^0_{\alpha} \nabla_{\mu} A^{\mu} - \nabla_{\alpha} A^0 \right) \,, \\
	p^{(3,{\rm GP})}_{\alpha} &= 3 \delta^0_{\alpha} \left( (\nabla_{\mu} A^{\mu})^2 - \nabla_{\mu} A^{\nu} \nabla_{\nu} A^{\mu} \right) + 6 \left( \nabla_{\alpha} A^{\mu} \nabla_{\mu} A^{0} - \nabla_{\alpha} A^{0} \nabla_{\mu} A^{\mu} \right) \,.
\end{align}
From there we can immediately check that by themselves, the GP type of terms constructed out of symmetric polynomials of $(\nabla A)$  do not contribute to the Hessian matrix, namely
\begin{equation}
	\mathcal{H}^{(n,{\rm GP})}_{\alpha \beta} = 0, \qquad n=1,2,3 \,.
\end{equation}
Is is then clear that $V^{\alpha}$ (the covariantization of the Minkowski NEV for EPN) is still satisfying

\begin{equation}
	\mathcal{H}_{\alpha \beta} ( \hat{\L} ) V^{\alpha} = 0 \,.
\end{equation}
Now focusing on the part of the Hessian matrix that probes the mixing between $\dot{A}$ and $\dot{\gamma}$ we find
\begin{equation}
	\mathcal{H}_{\alpha,ij}^{(2,{\rm GP})} V^{\alpha} = - \frac12 \left[ \gamma_{ij} A^0 V^0 + g^{00} A_i V_j \right] + \left\{ i \leftrightarrow j \right\} \,,
\end{equation}
leading to
\begin{align}
	\mathcal{H}_{\alpha,ij}(\hat{\L}^{(2)}) V^{\alpha} &= \frac12 \alpha_{2,X} \left[ A^0 (\gamma_{ij} V^0 - V_{ij} g^{00} ) + g^{00} A_i ( V_j + V_j^{\phantom{j} 0} ) \right] + \left\{ i \leftrightarrow j \right\}  \\
	&= 0 + \frac{0}{\Lambda^2} - \frac{1}{8\Lambda^4} \left[ \gamma_{ij} A^0 F^{0 \alpha} F_{0 \alpha} + g^{00} \left( 2 F^{0 \alpha} A_{(i} F_{j) \alpha} - A^0 F_{i}^{\phantom{i}\alpha} F_{j \alpha} \right) \right] + \mathcal{O}((\nabla A)^{3}/\Lambda^{6}) \,,\nonumber
\end{align}
which again fails to vanish at all orders but vanishes at leading and next-to-leading order in the operator expansion and vanishes on any background for which the field strength tensor vanishes, $F\mn=0$, as is the case for cosmology.

\section{Definition of some coefficients in the perturbed quadratic actions}
\label{sec:DefCoefs}

\subsection{Coefficients of the scalar perturbations in the special model}
\label{ssec:DefCoefsExScal}

We define here the $7$ coefficients entering the quadratic scalar action of the the special model \eqref{eq:SSnocoupl2},

\begin{align}
	\hat{\omega}_1 &= - 2 M_{\text{Pl}}^2 H - \phi^3 \left( \alpha_{1,X} + d_{1,X} \right) \,, \nonumber \\
	\hat{\omega}_2 &= \hat{\omega}_1 + 2 M_{\text{Pl}}^2 H \,, \nonumber \\
	\hat{\omega}_3 &= -2 \phi^2 \hat{q}_V \,, \nonumber \\
	\hat{\omega}_4 &=  -3 M_{\text{Pl}}^2 H^2 + \frac{1}{2} \phi^4 \alpha_{0,XX} - \frac{3}{2} H \phi^3 \left[ ( \alpha_{1,X} +  d_{1,X} ) - \frac{\phi^2}{\Lambda^2} \left( \alpha_{1,XX} +  d_{1,XX} \right) \right] \,, \nonumber \\
	\hat{\omega}_5 &= \hat{\omega}_4 - \frac32 H ( \hat{\omega}_1 + \hat{\omega}_2 ) \,, \nonumber \\
	\hat{\omega}_6 &= - \phi^2 \left( \alpha_{1,X} + d_{1,X} \right) \,, \nonumber \\
	\hat{\omega}_7 &= - \dot{\phi} \left( \alpha_{1,X} + d_{1,X} \right) \,.
	\label{eq:coefsomegaspec}
\end{align}

\subsection{Masses of the scalar modes in the special model}
\label{ssec:MassesExScal}

In this Appendix, we present the results for the square masses of both scalar modes, the matter perturbation $\delta \rho_M / k$ and the scalar $\psi$. These masses are inferred from the dispersion relation \eqref{eq:SpecExdispSM} and hence are canonically normalized. To begin with, the mass of the matter field is given by

\begin{equation}
	\hat{m}_{S,M}^2 = \frac{\hat{\Theta}}{2 \phi^2 \hat{q}_V (\hat{\omega}_1 - 2 \hat{\omega}_2)^2} \,,
\end{equation}
where

\begin{align}
	\hat{\Theta} = 2 (\rho_M + P_M) \hat{\omega}_2^2 + \hat{\omega}_3 &\left[ (\rho_M + P_M)^2 - H (\rho_M + P_M) \left\lbrace (\hat{\omega}_1 - 2 \hat{\omega}_2 ) \left( 1 + 6 c_M^2 \right) + \frac{\dot{\hat{\omega}}_1 - 2 \dot{\hat{\omega}}_2}{H} \right\rbrace \right.  \\
	& \quad \left. + 3 H^2 (\hat{\omega}_1 - 2 \hat{\omega}_2 )^2 \left\lbrace 3 c_M^4 + c_M^2 - 2 - (1 + c_M^2) \left( \frac{\dot{H}}{H^2} - \frac{\p_t ( \rho_M + P_M)}{H ( \rho_M + P_M )} \right\rbrace \right) \right] \,.\nonumber
\end{align}
Avoiding a tachyonic instability is achieved by requiring

\begin{equation}
	\hat{\Theta} > 0 \,.
\end{equation}
On another hand, one can also derive the mass of the scalar field $\psi$,

\begin{equation}
	\hat{m}_{S,\psi}^2 = \frac{1}{(\hat{\omega}_1-2\hat{\omega}_2)^2 \hat{\omega}_3^2} \frac{\hat{\Xi}_1}{\hat{\Xi}_2} \,,
\end{equation}
where

\begin{align}
	\hat{\Xi}_1 =& - (\hat{\omega}_1-2\hat{\omega}_2)^2 \hat{\omega}_3 \hat{\Xi}_2 \phi \dot{\hat{\omega}}_6 \nonumber \\
	& + 2 (\hat{\omega}_1-2\hat{\omega}_2)(\hat{\omega}_1-\hat{\omega}_2) \hat{\omega}_3 \phi \left[ \hat{\omega}_2 (\hat{\omega}_1^2 - \hat{\omega}_1 \hat{\omega}_2 + (P_M + \rho_M) \hat{\omega}_3) + (\hat{\omega}_1 - 2 \hat{\omega}_2)(\hat{\omega}_1-\hat{\omega}_2) \hat{\omega}_6 \phi \right] \dot{\hat{\omega}}_4 \nonumber \\
	& + (\hat{\omega}_1 - 2 \hat{\omega}_2) ( \hat{\omega}_1 \hat{\omega}_2 + (\hat{\omega}_1 - 2 \hat{\omega}_2) \hat{\omega}_6 \phi) \hat{\Xi}_2 \dot{\hat{\omega}}_3 \nonumber \\
	&- \hat{\omega}_3 \phi \left[ 3 H \hat{\omega}_1^2 \left\lbrace \hat{\omega}_1 (\hat{\omega}_1 (\hat{\omega}_1 +2 \hat{\omega}_2) - 2\hat{\omega}_2^2) + (\hat{\omega}_1-2\hat{\omega}_2)(3\hat{\omega}_1-2\hat{\omega}_2)\hat{\omega}_6 \phi) \right. \right. \nonumber \\
	&\phantom{- \hat{\omega}_3 \phi} \qquad \qquad \left. \left. + (P_M + \rho_M)(\hat{\omega}_1 + 4 \hat{\omega}_2) \hat{\omega}_3 \right\rbrace \right. \nonumber \\
	&\phantom{- \hat{\omega}_3 \phi} \quad \left. - 2 \hat{\omega}_4 \left\lbrace \hat{\omega}_1 (\hat{\omega}_1 - \hat{\omega}_2) (\hat{\omega}_1 (\hat{\omega}_1 + \hat{\omega}_2) + 2(\hat{\omega}_1-2\hat{\omega}_2)\hat{\omega}_6 \phi) \right. \right. \nonumber \\
	&\phantom{- \hat{\omega}_3 \phi} \qquad \qquad \left. \left. + (P_M + \rho_M) (\hat{\omega}_1 (\hat{\omega}_1 +2 \hat{\omega}_2) - 2\hat{\omega}_2^2) \hat{\omega}_3 \right\rbrace \right] \dot{\hat{\omega}}_2 \nonumber \\
	&- \hat{\omega}_3 \phi \left[ 3 H \hat{\omega}_1 \left\lbrace \hat{\omega}_1 \hat{\omega}_2 (\hat{\omega}_1^2 - 8 \hat{\omega}_1 \hat{\omega}_2 + 6\hat{\omega}_2^2) + (\hat{\omega}_1 - 2 \hat{\omega}_2) (\hat{\omega}_1^2 - 6\hat{\omega}_1 \hat{\omega}_2 + 4\hat{\omega}_2^2) \hat{\omega}_6 \phi \right. \right. \nonumber \\
	&\phantom{- \hat{\omega}_3 \phi} \qquad \qquad \left. - (P_M + \rho_M) (\hat{\omega}_1 + 4\hat{\omega}_2) \hat{\omega}_2 \hat{\omega}_3 \right\rbrace \nonumber \\
	&\phantom{- \hat{\omega}_3 \phi} \quad \left. + 2 \hat{\omega}_2 \hat{\omega}_4 \left\lbrace (2\hat{\omega}_1- \hat{\omega}_2)(2(\hat{\omega}_1-\hat{\omega}_2)\hat{\omega}_2 + (P_M + \rho_M)\hat{\omega}_3) + 2(\hat{\omega}_1-2\hat{\omega}_2) (\hat{\omega}_1-\hat{\omega}_2) \hat{\omega}_6 \phi \right\rbrace \right] \dot{\hat{\omega}}_1 \nonumber \\
	& + 3\hat{\omega}_1^2 (\hat{\omega}_1 - 2\hat{\omega}_2) \hat{\omega}_3 \left[ \hat{\omega}_2 (\hat{\omega}_1 (\hat{\omega}_1-\hat{\omega}_2) + \hat{\omega}_3 (P_M + \rho_M)) + (\hat{\omega}_1-2\hat{\omega}_2)(\hat{\omega}_1-\hat{\omega}_2)\hat{\omega}_6 \phi \right] \phi \dot{H} \nonumber \\
	& - (\hat{\omega}_1 - 2\hat{\omega}_2) \hat{\omega}_2 \hat{\omega}_3^2 \left[ 3H \hat{\omega}_1^2 - 2 (\hat{\omega}_1-\hat{\omega}_2)\hat{\omega}_4 \right] \phi \p_t (P_M + \rho_M) \nonumber \\
	& + (\hat{\omega}_1-2\hat{\omega}_2)\hat{\omega}_3 (3 H \hat{\omega}_1^2 -2(\hat{\omega}_1-\hat{\omega}_2)\hat{\omega}_4) \nonumber \\
	& \qquad \qquad \times \left[ 2 \hat{\omega}_2(\hat{\omega}_1(\hat{\omega}_1-\hat{\omega}_2)+ \hat{\omega}_3(P_M+\rho_M)) + (\hat{\omega}_1-2\hat{\omega}_2)(\hat{\omega}_1-\hat{\omega}_2)\hat{\omega}_6 \phi \right] \dot{\phi} \nonumber \\
	& - (\hat{\omega}_1 - 2\hat{\omega}_2) \left[ 2 \hat{\omega}_2 \hat{\omega}_3 (3 H \hat{\omega}_2 + (P_M + \rho_M)) + \hat{\omega}_1^2 (2 \hat{\omega}_2 + 3 H \hat{\omega}_3) - \hat{\omega}_1 \hat{\omega}_2 (2 \hat{\omega}_2 + 9H \hat{\omega}_3) \right] \nonumber \\
	& \qquad \qquad \times (3 H \hat{\omega}_1^2 -2 (\hat{\omega}_1 -\hat{\omega}_2) \hat{\omega}_4) \phi^2 \hat{\omega}_6 \nonumber \\
	&+ \left[ 2 (\hat{\omega}_1-2\hat{\omega}_2)^2(\hat{\omega}_1-\hat{\omega}_2)^2 \hat{\omega}_6^2 \phi^3 + 2 \hat{\omega}_2 ( \hat{\omega}_1(\hat{\omega}_1-\hat{\omega}_2) + \hat{\omega}_3 (P_M + \rho_M) ) \right. \nonumber \\
	&\phantom{+} \qquad \qquad \qquad \left. \times \left\lbrace \hat{\omega}_1 \hat{\omega}_2 (\hat{\omega}_1 - \hat{\omega}_2) + \hat{\omega}_3 ( 3H(\hat{\omega}_1-2\hat{\omega}_2)(\hat{\omega}_1-\hat{\omega}_2) + \hat{\omega}_2 (P_M + \rho_M)) \right\rbrace \phi \right] \hat{\omega}_4 \nonumber \\
	&- 12 H \phi \hat{\omega}_2^3 \hat{\omega}_3^2 (P_M + \rho_M)^2 \nonumber \\
	&-3 H \hat{\omega}_1^2 \phi \left[ \hat{\omega}_2 \hat{\omega}_3 (2\hat{\omega}_1 \hat{\omega}_2 + 3H (\hat{\omega}_1 -2\hat{\omega}_2) \hat{\omega}_3) (P_M + \rho_M) \right. \nonumber \\
	&\phantom{-3 H \hat{\omega}_1^2 \phi} \quad \left.+ (\hat{\omega}_1-\hat{\omega}_2) \left\lbrace 3 H \hat{\omega}_1 (\hat{\omega}_1 - 2 \hat{\omega}_2) \hat{\omega}_2 \hat{\omega}_3 + \hat{\omega}_1^2 \hat{\omega}_2^2 + (\hat{\omega}_1 - 2 \hat{\omega}_2)^2 \hat{\omega}_6^2 \phi^2 \right\rbrace \right] \,,
\end{align}
and

\begin{equation}
	\hat{\Xi}_2 = (\hat{\omega}_1- \hat{\omega}_2) (3H \hat{\omega}_1^2 - 2(\hat{\omega}_1-\hat{\omega}_2)\hat{\omega}_4) \phi \,.
\end{equation}
The absence of tachyonic instabilities for the scalar mode $\psi$ is ensured by the positivity of $\hat{m}_{S,\psi}^2$, which is equivalent to require $\hat{\Xi}_1 / \hat{\Xi}_2 > 0$.

\subsection{Coefficients of the vector perturbations in the general model}
\label{ssec:DefCoefsVec}

In this Appendix, we will define the coefficients entering the quadratic action for the vector field in the general model \eqref{eq:SV2}.

\begin{align}
	q_V &= 1 - \frac{1}{2\mu} \alpha_1 + \frac{1}{\mu} \left( 1 - 2 \frac{H \phi}{\Lambda^2} \right) \alpha_{2,X} \nonumber \\
	&\qquad \qquad + \frac{1}{2\mu^2} \left\lbrace  \frac{3H^2 \mu - \left( 1 + H \phi / \Lambda^2 \right) \dot{H}}{\Lambda^2} \alpha_3 - \mu  \frac{H \phi}{\Lambda^2} \left( 2 - \frac{H \phi}{\Lambda^2} \right) \alpha_{3,X} \right\rbrace  \,, \nonumber \\
	\mathcal{C}_1 &= 1 + \frac{1}{2 \left( 1 + H \phi / \Lambda^2 \right)} \left[ - \alpha_1 + 2 \left( 1 - \frac{H \phi + \dot{\phi}}{\Lambda^2} \right) \alpha_{2,X} + \frac{3H^2 + 2 \dot{H}}{\Lambda^2} \alpha_3 \right. \nonumber \\
	& \qquad \qquad \qquad \qquad \qquad \qquad \left. - \left( \frac{H \phi}{\Lambda^2} + \left( 1 - \frac{H \phi}{\Lambda^2} \right) \frac{\dot{\phi}}{\Lambda^2} \right) \alpha_{3,X} \right] \,, \nonumber \\
	\mathcal{C}_2 &= \left( \alpha_{1,X} + d_{1,X} \right) \frac{\dot{\phi}}{H^2} + \frac{\dot{H}}{\Lambda^2 \mu} \alpha_3 + \frac{1}{H^2} \p_t \left( H \left[ 4 (\alpha_{2,X} + d_{2,X} ) + \frac{\dot{H}}{\Lambda^2 \mu} \alpha_3 - \frac{H \phi}{\Lambda^2} \left( \alpha_{3,X} + d_{3,X} \right) \right] \right) \nonumber \\
	& \qquad + 2 q_V + \frac{\p_t (q_V H)}{H^2} \,, \nonumber \\
	\mathcal{C}_3 &= 2 \frac{\phi}{M_{\text{Pl}}} \left( \alpha_{2,X} + d_{2,X} \right) - \frac{H}{M_{\text{Pl}}} \left( \frac{H \phi - \dot{\phi}}{2 \Lambda^2 \mu} \alpha_3 + \frac{\phi^2}{\Lambda^2} \left( \alpha_{3,X} + d_{3,X} \right) \right) \,, \nonumber \\
	\mathcal{C}_4 &= \frac{\dot{\phi} - H \phi}{2 \Lambda^2 M_{\text{Pl}} \mu} \alpha_3 \,,
\end{align}
where we have used

\begin{equation}
	\mu = 1 + \frac{H \phi + \dot{\phi}}{2\Lambda} \,.
\end{equation}

\subsection{Coefficients of the scalar perturbations in the general model}
\label{ssec:DefCoefsScal}

Finally, we define the coefficients entering the quadratic action for the scalar sector of the general model \eqref{eq:SS2}.

\begin{align}
	\omega_1 &= -2 M_{\text{Pl}}^2 H - \phi^3 ( \alpha_{1,X} + d_{1,X} )- 4 \Lambda^2 H \left[ (\alpha_2 + d_2 ) + \frac{\phi^4}{\Lambda^4} ( \alpha_{2,XX} + d_{2,XX} ) \right]  \\
	&\qquad + \frac{H^2 \phi^3}{\Lambda^2} \left[ \left(\alpha_{3,X} + d_{3,X} \right) + \frac{\phi^2}{\Lambda^2} \left(\alpha_{3,XX} + d_{3,XX} \right) \right] \,, \nonumber \\
	\omega_2 &= \omega_1 + 2 M_{\text{Pl}}^2 H q_T \,, \nonumber \\
	\omega_3 &= -2 \phi^2 q_V \,, \nonumber \\
	\omega_4 &= - 3 M_{\text{Pl}}^2 H^2 + \frac12 \phi^4 \alpha_{0,XX} - \frac32 H \phi^3 \left[ \left(\alpha_{1,X} + d_{1,X} \right) - \frac{\phi^2}{\Lambda^2} \left(\alpha_{1,XX} + d_{1,XX} \right) \right] \nonumber \\
	&\qquad - 3 \Lambda^2 H^2 \left[ 2 ( \alpha_2 + d_2 ) + 2 \frac{\phi^2}{\Lambda^2} (\alpha_{2,X} + d_{2,X}) + \frac{\phi^4}{\Lambda^4} (\alpha_{2,XX} + d_{2,XX}) - \frac{\phi^6}{\Lambda^6} (\alpha_{2,XXX} + d_{2,XXX}) \right] \nonumber \\
	&\qquad + \frac12 \frac{H^3 \phi^3}{\Lambda^2} \left[ 9 \left(\alpha_{3,X} + d_{3,X} \right) - \frac{\phi^4}{\Lambda^4} \left(\alpha_{3,XXX} + d_{3,XXX} \right) \right] \,, \nonumber \\
	\omega_5 &= \omega_4 - \frac32 H (\omega_1 + \omega_2) \,, \nonumber
\end{align}
\begin{align}
	\omega_6 &= - \phi^2 ( \alpha_{1,X} + d_{1,X} ) + 4 H \phi \left[ (\alpha_{2,X} + d_{2,X} ) - \frac{\phi^2}{\Lambda^2} ( \alpha_{2,XX} + d_{2,XX} ) \right]  \\
	&\qquad - \frac{H^2 \phi^2 }{\Lambda^2} \left[ \left(\alpha_{3,X} + d_{3,X} \right) - \frac{\phi^2}{\Lambda^2} \left(\alpha_{3,XX} + d_{3,XX} \right) \right] + 2 \frac{\dot{H} H \phi}{\mu \Lambda^2} \alpha_3 \,, \nonumber \\
	\omega_7 &= - \dot{\phi} ( \alpha_{1,X} + d_{1,X} ) - 4 \left[ \dot{H}(\alpha_{2,X} + d_{2,X} ) + \frac{H \phi \dot{\phi}}{\Lambda^2} ( \alpha_{2,XX} + d_{2,XX} ) \right] \nonumber \\
	&\qquad + \frac{H( 2 \dot{H} \phi + H \dot{\phi})}{\Lambda^2} \left(\alpha_{3,X} + d_{3,X} \right) + \frac{H^2 \phi^2 \dot{\phi}}{\Lambda^4} \left(\alpha_{3,XX} + d_{3,XX} \right) - \frac{\dot{H} H^2}{\mu \Lambda^2} \alpha_3 - \p_t \left( \frac{\dot{H} H}{\mu \Lambda^2} \alpha_3 \right) \,, \nonumber \\
	\omega_8 &= \frac{\dot{H} \phi^2}{\mu \Lambda^2} \alpha_3 \,, \nonumber \\
	\omega_9 &= \frac{H \phi (H \phi - \dot{\phi})}{\mu \Lambda^2} \alpha_3 \,.\nonumber
\end{align}
On a side note, these coefficients do not reproduce the ones of the special model depicted in \eqref{eq:coefsomegaspec} in the limit $\alpha_2 = - d_2$ and $\alpha_3 = - d_3$, showing the non-trivial relation between the special and the general models. This highlights the importance of either the tuning in the special model or the non-minimal couplings in the general model to construct a healthy theory.

\section{Operator relevance about the de Sitter point}
\label{sec:OperatordS}

We have seen in Sec.~\ref{sssec:PertSpec} that the scalar and vector velocities vanish about the dS point. This could potentially suggest the presence of strong coupling issues associated with the breakdown of perturbative unitarity as we reach those fix points. We will show that this is not the case. Schematically, the action for the  mode $\psi$ reads
\begin{equation}
	S^{(2)}_{S,\psi} = \int \d t \d^3 x a^3 \left( - \frac12 Z\mnup \p_{\mu} \psi \p_{\nu} \psi \right) = \int \d t \d^3 x \frac{a^3}{2} \hat{Q}(t) \left( \dot{\psi}^2 - \frac{\hat{c}(t)^2}{a^2} (\p_i \psi)^2 \right) \,,
\end{equation}
defining an effective metric $Z\mnup$.  To determine the strong coupling scale, or the scale at which perturbative unitarity gets broken, we first normalise the field and the spacetime coordinates as performed for instance in \cite{PhysRevD.95.123523},
\ba
\tilde t=\int \hat c(t)\d t\,, \quad \tilde x^i=x^i\,, \quad {\rm and}\quad \psi =(\hat Q \hat c)^{-1/2}\tilde \psi\,,
\ea
so that in terms of these variables the field is canonically normalized
\ba
S^{(2)}_{S,\psi} = \int \d \tilde t \d^3 \tilde x \frac{a^3}{2} \left( \(\p_{\tilde t}\tilde \psi\)^2 - \frac{1}{a^2} \(\p_{\tilde x^i} \tilde\psi\)^2 \right) \,.
\ea
Treated as an EFT, we would expect the theory to include an infinite number of irrelevant operators of the form $a^{3-2L} \psi^N \dot{\psi}^M (\p_i \psi)^{2L}$ entering  at the respective  scale $\Lambda_{NML}$, where $\Lambda_{NML}$ is expected to be at least of order  $\Lambda$ and where $N$, $M$ and $L$ are positive integers respecting $N + 2M + 4L > 4$.
In terms of the normalized fields and coordinates, this operator enters at the physical scale $\tilde{\mu}_{NML}$ given by
\begin{equation}
	\tilde{\mu}_{NML}
= \hat{Q}^{\frac{N + M + 2L}{2(N+2M+4L-4)}} \left( \hat{c}^2 \right)^{\frac{N - M + 2L + 2}{4(N+2M+4L-4)}} \Lambda_{NML}\,.
\end{equation}
Now, the kinetic term $\hat{Q}$ and the velocity $\hat{c}^2$ are power laws of the scale factor $a(t)$ when approaching the dS point, for both the vector and the scalar mode. Let us define
\begin{equation}
	p_Q = \frac{\text{ln}(\hat{Q})}{\text{ln}(a)}, \qquad p_c = \frac{\text{ln}(\hat{c}^2)}{\text{ln}(a)} \,,
\end{equation}
so that
\begin{equation}
p_{NML}	\equiv \frac{\text{ln} ( \tilde{\mu}_{NML} / \Lambda_{NML} )}{\text{ln} (a)} = \frac{(N+2L)(2p_Q+p_c) + M(2p_Q-p_c) +2p_c}{4(N+2M+4L-4)} \,.
\end{equation}
The validity of the EFT is preserved as long as no scale $\tilde{\mu}_{NML}$ has a much smaller value than the corresponding scale $\Lambda_{NML}$, for any non-negative integers $N$, $M$, $L$ such that $N + 2M +4L > 4$. This translates into
\begin{equation}
	p_{NML} > 0, \quad \text{for all } N, M, L \ge 0, \quad N + 2M +4L > 4 \,.
\end{equation}
The scaling of the kinetic terms and velocities of the scalar and vector modes in terms of the scale factor $a(t)$ can be deduced from Figs.~\ref{fig:velocities} and \ref{fig:kin} and is summarised in the Table \ref{tab:apowers} below.
\begin{table}[h!]
\begin{center}
\begin{tabular}{ c | c | c }
  & vector & scalar $\psi$ \\ \hline
  $\hat{Q}$ & $a^3$ & $a^6$ \\
  $\hat{c}^2$ & $a^{-3}$ & $a^{-3}$
\end{tabular}
\caption{Power-law behaviour of the vector and scalar kinetic terms and velocities as functions of the scale factor $a(t)$ about the dS point.}
\label{tab:apowers}
\end{center}
\end{table}

\noindent It follows that the value of $p_{NML}$ for the vector and scalar $\psi$ modes about the dS point read
\begin{equation}
	\begin{cases}
		(p_{NML})_{V} &= \frac{3}{4} \frac{N + 3M + 2L - 2}{N+2M+4L-4} \,, \\
		(p_{NML})_{S,\psi} &= \frac{3}{4} \frac{3N + 5M + 6L - 2}{N+2M+4L-4} \,.
	\end{cases}
\end{equation}
From here, it is easy to prove that $p_{NML}$ is always strictly positive for any of the allowed values of $N$, $M$ and $L$, for both the scalar and the vector. This concludes the proof that our EFT does not suffer any strong coupling issue due to the vanishing of the velocities about the de Sitter point (in link with the divergence of the associated kinetic terms).

\pagebreak

\bibliographystyle{JHEP}
\bibliography{references}

\end{document}